\documentclass[aip, jcp, reprint]{revtex4-2}
\usepackage{graphicx}
\usepackage{xcolor}
\usepackage{amsmath, amsfonts, amssymb}
\usepackage{bm}
\usepackage{gensymb}
\usepackage{xr}
\usepackage{chemformula}
\usepackage{soul}
\setstcolor{red}

\makeatletter
\newcommand{\revision}[1]{{\color{black}#1}}

\newcommand{\mc}[3]{\multicolumn{#1}{#2}{#3}}

\begin{document}

\title{Performant Automatic Differentiation of Local Coupled Cluster Theories: Response Properties and Ab Initio Molecular Dynamics}


\author{Xing Zhang}
\author{Chenghan Li}
\affiliation
{Division of Chemistry and Chemical Engineering,
California Institute of Technology, Pasadena, CA 91125, USA}
\author{Hong-Zhou Ye}
\author{Timothy C. Berkelbach}
\affiliation
{Department of Chemistry, Columbia University, New York, NY 10027, USA}
\author{Garnet Kin-Lic Chan}
\affiliation
{Division of Chemistry and Chemical Engineering,
California Institute of Technology, Pasadena, CA 91125, USA}
\email{gkc1000@gmail.com}



\begin{abstract}
In this work, we introduce a differentiable implementation of
the local natural orbital coupled cluster (LNO-CC) method
within the automatic differentiation framework of the \textsc{PySCFAD} package.
The implementation is comprehensively tuned for enhanced performance,
which enables the calculation of first-order static
response properties on medium-sized molecular systems using coupled cluster theory
with single, double, and perturbative triple excitations [CCSD(T)].
We evaluate the accuracy of our method by benchmarking it against the canonical CCSD(T) reference for nuclear gradients, dipole moments, and geometry optimizations.
In addition, we demonstrate the possibility of property calculations for chemically interesting systems through the
computation of bond orders and M{\"o}ssbauer spectroscopy parameters
for a [NiFe]-hydrogenase active site model, along with the simulation of infrared (IR) spectra via \textit{ab initio} LNO-CC molecular dynamics for a protonated water hexamer.
\end{abstract}

\maketitle
\section{Introduction}
Since the pioneering work of Pulay and S{\ae}b{\o},
\cite{Pulay1983,Saebo1985,Pulay1986,Saebo1993,Boughton1993}
local electron correlation methods have seen significant advances
over the past two decades.
\cite{Hampel1996,Schutz2000,Schutz2001,Schutz2002a,Schutz2002b,
Li2002,Li2006,Li2009,
Neese2009a,Neese2009b,
Werner2011,
Rolik2011,
Yang2011, Kurashige2012, Yang2012, Schutz2013,
Rolik2013,
Riplinger2013a,Riplinger2013b,Sparta2014,
Schmitz2014,
Kallay2015,
Liakos2015a,Liakos2015b,Riplinger2016,
Schmitz2016,
Guo2016,Pavovsevic2017,
Ma2017,Ma2018, Guo2018,
Liakos2019,
Nagy2018,Nagy2019,
Ni2019,Wang2019,Ni2021,Wang2022,Li2023,
Ye2023}
With modern developments, calculations employing local approximations to
coupled cluster theory with single, double,
and perturbative triple excitations\cite{Raghavachari1989} [CCSD(T)]
can now be routinely performed for large molecular systems
\cite{Liakos2015b,Guo2018,Nagy2018,Nagy2019,Ni2019,Ni2021}
and solids.\cite{Wang2022,Ye2023,Ye2023b}
However, the majority of these calculations are focused on ground-state electronic energetics,
while the application of local correlation methods to molecular properties remains relatively underexplored.
This may be attributed to the inherent complexity of these methods,
which makes the implementation of analytic derivatives or response theory more challenging, as compared to their canonical counterparts.
Nevertheless, notable efforts have been made on this subject.
Analytic nuclear gradients for projected atomic orbital (PAO) based local
correlation methods,\cite{Hampel1996,Schutz2000,Schutz2001}
including local second-order M{\o}ller--Plesset perturbation theory (LMP2) and
local coupled cluster theory with single and double excitations (LCCSD),
have been developed by Werner and co-workers.
\cite{El1998,Rauhut2001,Schutz2004,Dornbach2019}
Meanwhile, Sch{\"u}tz and co-workers have implemented analytic nuclear gradients
and dipole moments for the local CC2 method, applied to both ground and excited states.
\cite{Ledermuller2013,Ledermuller2014}
In addition, calculations of nuclear magnetic resonance (NMR) shieldings\cite{Gauss2000,Loibl2012}
and magnetizabilities\cite{Loibl2014} have been reported
at the LMP2 level.
More recently, Neese and co-workers extended their domain-based
local pair natural orbital (DLPNO) approaches\cite{Neese2009a,Neese2009b} for computing static
response properties.
These include first\cite{Pinski2018,Pinski2019} and second\cite{Stoychev2021} derivatives of DLPNO-MP2 and orbital-unrelaxed first derivatives of DLPNO-CCSD.\cite{Datta2016}
Similarly, Yang and co-workers published analytic nuclear gradient implementations for the orbital-specific virtual (OSV) MP2 method.\cite{Zhou2019}
Finally, Crawford and co-workers studied dynamic (frequency dependent) response properties,
such as polarizabilities and optical rotations,
by applying the local coupled cluster linear response theory. \cite{Russ2004,Russ2008,McAlexander2012,McAlexander2016,DCunha2021}

Typically, static response properties are calculated as energy (or Lagrangian)
derivatives with respect to the perturbations.
Computing analytic derivatives simply involves applying chain rules to the objective function.
However, manually tracking the entire workflow of a complex calculation can be tedious and error-prone.
Thankfully, modern automatic differentiation (AD) tools\cite{Autograd,Jax2018,Pytorch} offload the task of tracing the program's
execution to the computer,\cite{Frostig2018,TorchFx} greatly simplifying the implementation of analytic derivatives.
In recent years, there has been a growing effort to integrate AD techniques into quantum chemistry computations.
\cite{Tamayo2018,Song2021,Abbott2021,Kasim2022,pyscfad,Vargas2023,Mahajan2023,MCasares2024}
Among these, the \textsc{PySCFAD} package is distinguished by its extensive functionality and its flexibility.\cite{pyscfad}
In an earlier publication,\cite{pyscfad} we demonstrated a proof-of-concept showing that \textsc{PySCFAD} could be utilized for the rapid prototyping of new methodologies. Here, we further illustrate that \textsc{PySCFAD} is now also effective for production-level calculations.

In this work, we present a differentiable implementation of the local natural
orbital coupled cluster (LNO-CC) method, first introduced by Rolik and K{\'a}llay,\cite{Rolik2011} within the AD framework of \textsc{PySCFAD}.
The modular design of \textsc{PySCFAD} allows for the integration of new methods with virtually no changes to the existing components of the software.
Furthermore, performance optimizations can be targeted to a small segment of the program, e.g., the tensor contractions for computing the triple excitation correction in the CCSD(T) method, without compromising the differentiability of the entire computational workflow.
We demonstrate the resulting differentiable LNO-CC method in some non-trivial applications: calculating the bond orders and M{\"o}ssbauer spectroscopy parameters for a model system of the [NiFe]-hydrogenase active site, as well as obtaining the anharmonic infrared (IR) spectra of a protonated water \revision{hexamer} from \textit{ab initio} molecular dynamics.

\section{Notations}
Throughout the paper, we use Greek letters ($\mu$, $\nu$, ...) to denote atomic orbitals.
Occupied and virtual canonical Hartree-Fock (HF) orbitals are labeled
using lowercase letters ($i$, $j$, ... for occupied; $a$, $b$, ... for virtual),
while uppercase letters ($I$, $J$, ...) signify local orbitals\revision{, whether semi-canonicalized or not}.
The energies corresponding to canonical HF orbitals and semi-canonical local orbitals
are denoted by $\varepsilon$ and $\mathcal{E}$, respectively.
Orbitals that span a local active space are distinguished by tildes ($\tilde{i}$, $\tilde{j}$, ...).
Unless specified otherwise, equations are expressed in the spin orbital formalism, although our implementation is spin-adapted,
and we adopt Dirac's notation for the electron integrals.

\section{Methodology}
Our differentiable LNO-CC method is built upon the recent implementation of Ye and Berkelbach,\cite{Ye2023} with the latter limited to calculating ground-state energies.
In this section, we elaborate on this realization of the LNO-CC method within the \textsc{PySCFAD} framework, highlighting adjustments made for efficient differentiation,
and also discuss the potential issues that arise during the computation of response properties.

The procedure starts by constructing a set of occupied orthonormal local orbitals.
In the original scheme by Rolik and K\'{a}llay,\cite{Rolik2011}
each local orbital along with the corresponding local natural orbitals
(LNOs, see the definition below) defines a local active space,
within which the local electron correlation calculation is carried out.
We name this the \textit{one-orbital} scheme. However,
the size of the computational graph grows linearly with the number of
local electron correlation calculations.
For systems containing hundreds of electrons, it becomes time-consuming
to trace the computation and to compile the program in a just-in-time\cite{Frostig2018} (JIT) fashion.
A workaround, which we call the \textit{multi-orbital} scheme,
is to group the local orbitals to form fragments,
similar to the strategy employed by the fragment molecular orbital (FMO) approach.\cite{FMO}
In particular, we designate each heavy atom and
its surrounding hydrogen atoms as a fragment, and the local orbitals
are assigned to the corresponding fragments based on
a L\"{o}wdin population analysis.
The resulting fragments are sufficiently small to allow for efficient high-level local electron correlation calculations [e.g., at the CCSD(T) level],
while maintaining a moderate number of local calculations,
facilitating tractable computation tracing, JIT compilation, and gradient backpropagation.

With the local orbitals and fragments, we determine the local active space as follows.
Suppose $|\phi_I^{\Omega} \rangle$ are the semi-canonical local orbitals on a fragment $\Omega$,
then the corresponding local active space
is spanned by $|\phi_I^{\Omega} \rangle$ plus a set of LNOs
determined from diagonalizing the fragment contribution to the MP2 density matrix.
Following Ref.~\citenum{Rolik2011},
the occupied-occupied (OO) and virtual-virtual (VV) blocks of this density matrix read as
\begin{equation} \label{eq:dmoo}
  D_{jk}^{\Omega} = \sum_{mn} \mathcal{P}^{\Omega \top}_{jm} \left(\frac{1}{2} \sum_{I \in \Omega} \sum_{ab} t_{Imab} t_{Inab} \right) \mathcal{P}^{\Omega}_{nk} \;,
\end{equation}
and
\begin{equation} \label{eq:dmvv}
  D_{ab}^{\Omega} = \frac{1}{2} \sum_{I \in \Omega} \sum_{jc} t_{Ijac} t_{Ijbc} \;,
\end{equation}
respectively, where
\begin{equation}
  t_{Ijab} = \frac{\langle ab || Ij \rangle}{\mathcal{E}_I + \varepsilon_j - \varepsilon_a - \varepsilon_b} \;,
\end{equation}
and $\bm{\mathcal{P}}^{\Omega}$ is a projection matrix that removes the contributions
of $|\phi_I^{\Omega} \rangle$ from the density matrix.
Note that if $|\phi_I^{\Omega} \rangle$ overlaps with the virtual space,
e.g., when intrinsic atomic orbitals (IAOs) are taken as the local orbitals,
a similar projection is performed in Eq.~\ref{eq:dmvv} as well.
Diagonalizing the OO (Eq.~\ref{eq:dmoo}) and
VV (Eq.~\ref{eq:dmvv}) blocks of the density matrix gives the
occupied and virtual LNOs associated with fragment $\Omega$, respectively.
In practice, we truncate the local active space by
discarding those LNOs whose occupation numbers are smaller than a
predefined threshold $\zeta$. The retained LNOs along with $|\phi_I^{\Omega} \rangle$
are further semi-canonicalized to simplify the subsequent local electron correlation calculations on each fragment.

Taking MP2 as an example, the local correlation energy on a fragment can
be expressed as
\begin{equation} \label{eq:emp2}
  E^{\Omega}_{\text{MP2}} = \sum_{I \in \Omega} \sum _{\tilde{m}\tilde{n} \in \Omega} U^{\Omega \top}_{I\tilde{m}} \left(\frac{1}{4} \sum_{\tilde{j}\tilde{a}\tilde{b} \in \Omega} \langle \tilde{a}\tilde{b} || \tilde{m}\tilde{j} \rangle  t_{\tilde{n}\tilde{j}\tilde{a}\tilde{b}} \right) U^{\Omega}_{\tilde{n}I} \;,
\end{equation}
where $\mathbf{U}^{\Omega}$ transforms the semi-canonical orbitals
that span a local active space to the corresponding
local orbitals on fragment $\Omega$.
This orbital transformation restores the fragment-based energy partitioning.
Therefore, the total correlation energy of the system can be computed by summing over
the fragments
\begin{equation}
  E_{\text{LNO-MP2}} = \sum_{\Omega} E^{\Omega}_{\text{MP2}} \;.
\end{equation}

The energy expression for the LNO-CC method can be derived similarly, and
is detailed in the supporting information (see Sec.~\ref{sec:cc_energy}).
One caveat is that the orbital transformation in Eq.~\ref{eq:emp2}
may break certain permutation symmetries of electron integrals
and CC amplitudes.
In particular, this happens when computing the triple excitation
correction within the LNO-CCSD(T) method,\cite{Rolik2013}
which increases the computational cost by a factor of at most three
for closed-shell systems,
compared to the canonical CCSD(T) calculation with the same correlation domain size.
Finally, it is often beneficial to
perform a global electron correlation calculation at a lower level
(such as MP2 in this work) to correct for the correlation effects
due to the weak pair interactions.\cite{Saebo1993}
This results in the correlation energy expression for the LNO-CC method being
\begin{equation}
  E_{\text{LNO-CC}} = \sum_{\Omega} \left( E^{\Omega}_{\text{CC}} - E^{\Omega}_{\text{MP2}} \right) + E_{\text{MP2}} \;.
\end{equation}

In the course of implementing the LNO-CC method, several enhancements have been applied to the core components of \textsc{PySCFAD}, leading to improved performance. These include:
(i) incorporating permutation symmetries for electron integrals and coupled cluster amplitudes,
(ii) a manually optimized implementation for the gradient of tensor contractions associated with the triple excitation correction in CCSD(T),
and (iii) minimizing the memory footprint for gradient calculations by recomputing intermediate quantities during backpropagation.
(More details can be found in the supporting information Sec.~\ref{sec:opt}.)
The optimized \textsc{PySCFAD} exhibits efficiency on par with its parent program \textsc{PySCF},\cite{pyscf} especially for MP2 and CC methods
(see Fig.~\ref{fig:timing}).
Additionally, the LNO-CC calculations have been parallelized using the Message Passing Interface (MPI),\cite{mpi4py} whereby the computations on distinct fragments are distributed across multiple processes.
(The mean-field part of the calculation is duplicated
within each process for straightforward backpropagation.)

Within the LNO-CC method, local correlation domains are truncated
according to a simple cutoff for the LNOs, which neither yields
a continuous energy function across the potential energy surface,
nor preserves the molecular point-group symmetry.
As such, nuclear gradients or any other response properties computed using the LNO-CC method may inherently contain errors stemming from
the energy discontinuity or the symmetry breaking.
Nonetheless, it is observed in practice that these errors tend to be small,
provided that the correlation domains are properly converged with respect to the total energy (see next section).

Finally, there are situations where the evaluation of the orbital localization response becomes ill-defined.\cite{Pinski2019}
This occurs when a continuum of solutions that fulfill
the localization criteria exists.\cite{Scheurer2000a, Scheurer2000b}
(In other words, the orbital rotation Hessian is rank deficient.)
A comprehensive exploration of this problem falls beyond
the scope of our current work.
Instead, we offer a practical remedy to prevent singular gradients,
which closely follows the strategy introduced in Ref.~\citenum{Pinski2019}
(see Sec.~\ref{sec:local_orb_autodiff}).

\section{Results and Discussions}
In this section, we present calculations of
nuclear gradients, dipole moments, geometry optimizations,
bond orders, quadrupole splittings (via electric field gradients), and \textit{ab initio} molecular dynamics, using the LNO-CCSD(T) method.
All calculations employ the density fitting\cite{Whitten1973,Dunlap1979} approximation for the two electron repulsion integral.

\subsection{Nuclear Gradient and Dipole Moment}
\begin{figure*}
\includegraphics{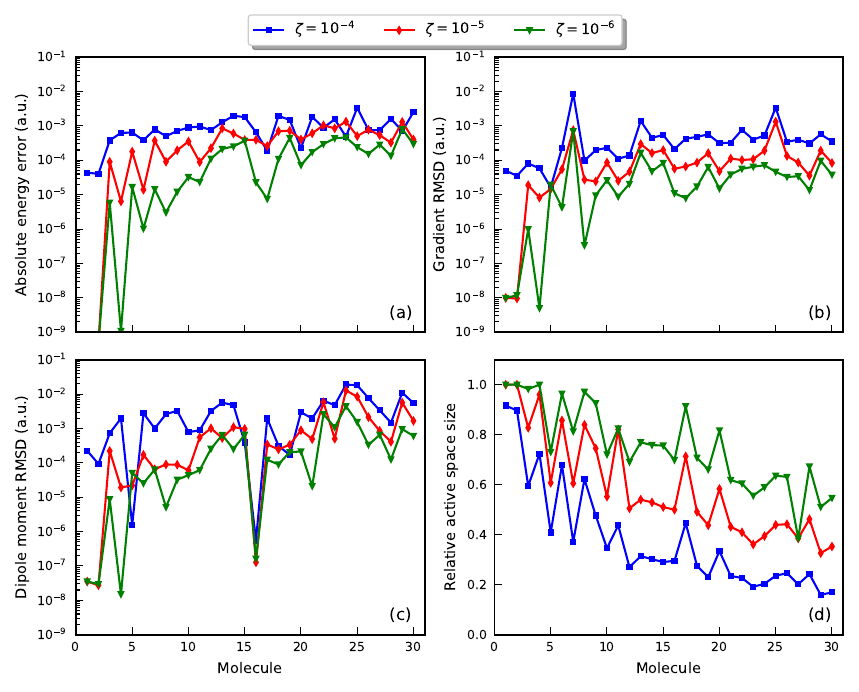}
\centering
\caption{Absolute energy error, nuclear gradient RMSD, dipole moment RMSD, and relative active space size computed at the
LNO-CCSD(T)/PM/cc-pVDZ\cite{ccpvtz} level for the Baker test set.
The reference data were obtained at the CCSD(T)/cc-pVDZ level,
and the geometries were optimized at the MP2/cc-pVDZ level.}
\label{fig:grad_pm}
\end{figure*}

First, we examine the correctness and convergence with threshold of nuclear gradients and dipole moments computed using the LNO-CCSD(T) method.
The Pipek-Mezey (PM) procedure\cite{pipek1989fast} was employed to determine the local orbitals.
The Baker test set, a set of 30 molecules with main-group elements of the first three rows ranging in size from 3 to 29 atoms,\cite{Baker} is considered,
and the reference data were obtained at the canonical CCSD(T) level of theory (where AD was employed to compute the nuclear gradients and the dipole moments).
In Fig.~\ref{fig:grad_pm}, we plot, for each molecule in the test set,
the absolute energy error, the nuclear gradient root mean square deviation (RMSD),
the dipole moment RMSD, and the relative active space size.
(The relative active space size is defined as the ratio of the mean number of orbitals within the local active space on each fragment to the total number of orbitals.)

It can be seen that, on average, the errors in nuclear gradients and dipole moments
are of the same order as those in energies.
\revision{Enlarging the basis set from double zeta to triple zeta does not lead to an increase in the errors (as shown in Fig.~\ref{fig:grad_pm_TZ}).}
We note that both nuclear gradients and dipole moments are types of first-order response properties, which can be determined using the zeroth-order wavefunction (or density matrix), according to Wigner's $2n+1$ rule.
However, for higher order response properties,
such as polarizabilities, which depend on the first-order wavefunction,
it may be necessary to use larger correlation domains to achieve the desired accuracy,
as suggested by Crawford and co-workers.\cite{McAlexander2016}

There are also some outliers in Fig.~\ref{fig:grad_pm}, such as benzene (molecule 7),
for which the large errors are likely due to symmetry breaking
(note the non-zero dipole moment).
For these systems, the coupled perturbed localization (CPL) equations
(Eq.~\ref{eq:local_zvec}) may not have solutions
unless both the fragmentation and the truncation of LNOs
are performed in ways that preserve the point-group symmetry.
These kinds of errors, however, can be significantly reduced by using
projection-based local orbitals, such as intrinsic atomic orbitals\cite{IAO} (IAOs),
which eliminate the need to solve the CPL equations.
(See results in Fig.~\ref{fig:grad_iao}.)

Finally, increasing the size of the local correlation domain generally improves the accuracy of energies, nuclear gradients, and dipole moments.
Many of the molecules in this test set are small (10 of them have fewer than 10 atoms), but
for the majority of systems assessed, an accuracy of $10^{-3}$ a.u. may be achieved with a local active space containing no more than $50\%$ of the total orbitals, with a smaller fraction of orbitals needed in the larger molecules.

\subsection{Geometry Optimization}
Geometry optimizations were also performed for molecules in the Baker test set,
through an interface with the \textsc{GeomeTRIC} library.\cite{geometric}
The default optimization convergence criteria were employed, including a nuclear gradient root mean square less than $3 \times 10^{-4}$ hartree/Bohr and an absolute energy change less than $1 \times 10^{-6}$ hartree.

We compare the LNO-CCSD(T) results to the canonical CCSD(T) reference,
and present the errors in Table~\ref{tab:geom_opt_pm}.
The observed errors are small for both bond lengths and various types of angles,
even with the loose LNO cutoff of $\zeta=10^{-4}$.
It is worth mentioning that the geometry optimization for benzene with
$\zeta=10^{-4}$ did not converge due to large gradient fluctuations caused by symmetry breaking.
Nevertheless, such an issue can be avoided by using a smaller $\zeta$ or
by employing the IAO local orbitals.

The present implementation of LNO-CCSD(T) has a formal scaling of
$O(N^5+\tilde{N}^7)$, where $N$ is the number of basis functions, and
$\tilde{N}$ represents the dimensions of the local correlation domains.
The $O(N^5)$ part of the computational complexity originates from the global MP2 correction and the LNO construction scheme,
while the $O(\tilde{N}^7)$ part is due to the local CCSD(T) calculations.
For the chemical systems for which one would use CCSD(T), $\tilde{N}$ does not need to grow with the system size for a given energy accuracy per atom,
making LNO-CCSD(T) orders of magnitude more efficient than canonical CCSD(T),
especially for large $N$
\revision{(see Fig.~\ref{fig:time_pm_TZ} and the corresponding discussion)}.
However, employing a looser LNO cutoff (and thus a smaller $\tilde{N}$)
might introduce more severe discontinuities into the potential energy surface, leading to more iterations to converge the geometry optimization.
Overall, our tests suggest that our LNO-CCSD(T) method is capable of accurate geometry optimizations
with computational scaling comparable to that of canonical MP2, albeit potentially with a large prefactor.

\begin{table*}
  \centering
  \caption{Mean absolute error (MAE)
and maximum error (max) in bond lengths,
angles, dihedral angles and out-of-plane angles of
the geometries optimized at the LNO-CCSD(T)/PM/cc-pVDZ level
for the Baker test set. The reference geometries were
optimized at the canonical CCSD(T) level with the same basis set.}
  \begin{tabular*}{1.\textwidth}{@{\extracolsep{\fill}}llcccc}
    \hline\hline
    $\zeta$   &     & bond length (\AA) & angle (\degree) & dihedral angle (\degree) & out-of-plane angle (\degree) \\
    \hline
    $10^{-4}$ \revision{$^a$} & MAE & 0.0008          & 0.038 & 0.059         & 0.008 \\
              & max & 0.0082          & 0.363 & 0.803         & 0.152 \\
    $10^{-5}$ & MAE & 0.0002          & 0.014 & 0.025         & 0.002 \\
              & max & 0.0019          & 0.085 & 0.774         & 0.028 \\
    \hline
  \end{tabular*}
  \label{tab:geom_opt_pm}
  \begin{flushleft}\footnotesize
  \revision{$^a$ The geometry used for benzene was obtained from iteration 100 of the geometry optimization.}
  \end{flushleft}
\end{table*}

\subsection{Relaxed Density Matrix}
Most first-order static response properties can be computed with the zeroth-order density matrix, which is readily available through AD as the energy derivative with respect to the unperturbed Hamiltonian.
This is especially useful for the LNO-CC method, which never calculates a global wavefunction that could be used to calculate the density matrix.
Specifically, the one-electron reduced density matrix may be expressed as
\begin{equation}
  D_{\mu\nu} = \frac{\partial E_{\text{LNO-CC}}}{\partial h_{\mu\nu}^{\text{core}}} \;,
\end{equation}
where $\mathbf{h}^{\text{core}}$ is the one-electron core Hamiltonian matrix.
Note that orbital relaxation is incorporated here by using the Hamiltonian in the atomic orbital basis.
In the following, we show an example of computing bond orders and
M\"{o}ssbauer spectroscopy parameters for a [NiFe]-hydrogenase
active site model system using the LNO-CCSD(T) method.

\begin{figure}[h]
\includegraphics{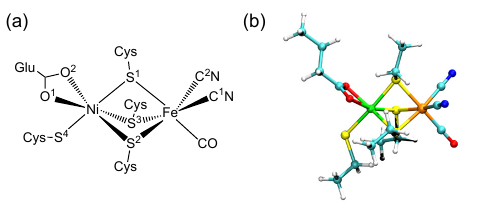}
\centering
\caption{
  (a) The chemical structure of the active site of \textit{Ht}-SH
  in its fully oxidized state. (b) The corresponding model system studied in this work.
}
\label{fig:NiFe_chemical}
\end{figure}

The crystal structures\cite{Shomura2017} of
the NAD$^{+}$-reducing soluble [NiFe]-hydrogenase (SH)
from \textit{Hydrogenophilus thermoluteolus} (\textit{Ht})
reveal an unusual arrangement at the oxidized active site
[see Fig.~\ref{fig:NiFe_chemical} (a)].
There, the nickel center adopts a distorted octahedral six-coordinate configuration,
featuring three bridging cysteines, one terminal cysteine,
and a bidentate glutamate coordination.
Meanwhile, the iron site is coordinated by two cyanide
and one carbon monoxide ligand. Notably, the IR spectrum\cite{Preissler2018,Kulka-Peschke2022}
of the oxidized state exhibits a unique CO vibration band at 1993 cm$^{-1}$,
which is distinct from all other [NiFe]-hydrogenases.
Such spectral features were proposed to originate from
a biologically unprecedented Ni(IV)/Fe(II) ground state,
supported by density functional theory (DFT) calculations.\cite{Kulka-Peschke2022}
However, a subsequent study, also conducted at the DFT level,
suggests that the spectral properties and the coordination geometry
of the oxidized \textit{Ht}-SH may be attributed to
an open-shell singlet Ni(III)/Fe(III) state instead.\cite{Kumar2023}

It would be interesting to study the structural and spectral properties of
the \textit{Ht}-SH active site beyond mean-field theory.
Here, we employ the LNO-CCSD(T) method to calculate
bond orders and M\"{o}ssbauer spectroscopy parameters,
offering a comparative analysis with DFT results.
Note that the primary aim of the current work is to demonstrate the
feasibility of our LNO-CC method, while a comprehensive investigation of the
electronic structure of the \textit{Ht}-SH system will be deferred to future studies.

The active site of \textit{Ht}-SH in the fully oxidized state
is represented by the ``model 20'' structure [see Fig.~\ref{fig:NiFe_chemical} (b)] from Ref.~\citenum{Kulka-Peschke2022}.
Its geometry, optimized at the DFT/TPSSh\cite{TPSSH}/def2-TZVP\cite{def2TZVP} level,
is taken from Ref.~\citenum{Kumar2023}.
Only the closed-shell singlet ground state is considered below.

In Table \ref{tab:NiFe_bond_order1}, the orbital resolved bond orders
(defined in Eq.~\ref{eq:bond_order})
are presented for the two metal centers.
Minor contributions to bonding are omitted,
and only representative bonds among those with similar chemical environments are shown.
\revision{(See Tables~\ref{tab:NiFe_bond_order1_full}
and \ref{tab:NiFe_bond_order} for the complete data.)}
We compare the results computed by DFT with those obtained by the LNO-CCSD(T) method.
For the latter, IAOs were employed as the local orbitals,
and an LNO cutoff of $\zeta = 2\times 10^{-5}$ was adopted, leading to fragments with correlation domains comprising
approximately $80$ occupied and $200$ virtual orbitals at most.
(Such choices are discussed in the supporting information Sec.~\ref{sec:HtSH}.)
The key findings are the following:
(i) Bond orders calculated by the two methods qualitatively agree with each other.
(ii) The $4s$ orbitals of both Fe and Ni do not contribute significantly to bonding.
(iii) Relatively strong covalent bonding is observed between Fe and CO.
(iv) The three bridging cysteines exhibit weak bonding interactions with the two metal centers, characterized by the stronger Ni--S bonds compared to the Fe--S bonds.
Interestingly, the LNO-CCSD(T) method tends to predict non-bonding characters for the Fe--S bonds.

\begin{table}[h]
  \centering
  \caption{
  Bond orders for the \textit{Ht}-SH active site model system,
  computed at the DFT/TPSSh and LNO-CCSD(T)/IAO levels
  with the def2-TZVP basis set.}
  \begin{tabular*}{0.48\textwidth}{@{\extracolsep{\fill}}lcc}
    \hline\hline
    Bond & TPSSh & LNO-CCSD(T) \\
    \hline
    Fe(3d)--C$^1$(2p)N & 0.24 & 0.17 \\
    Fe(3d)--C(2p)O & 0.57 & 0.87 \\
    Fe(3d)--S$^1$(3p)Cys & 0.21 & 0.09 \\
    Fe(3d)--S$^2$(3p)Cys & 0.17 & 0.07 \\
    Ni(3d)--S$^1$(3p)Cys & 0.26 & 0.31 \\
    Ni(3d)--S$^2$(3p)Cys & 0.34 & 0.14 \\
    Ni(3d)--S$^4$(3p)Cys & 0.33 & 0.32 \\
    Ni(4s)--O$^1$(2p) & 0.12 & 0.09 \\
    \hline
  \end{tabular*}
  \label{tab:NiFe_bond_order1}
\end{table}

We have also computed the quadrupole splittings $\Delta_v$
(defined in Eq.~\ref{eq:Delta_v}) for both metal centers,
as shown in Table~\ref{tab:NiFe_Delta_v}.
It can be seen that DFT and LNO-CCSD(T) give consistent results for Fe,
whereas noticeable discrepancies emerge for Ni.
This suggests that the electron density distribution around the Ni atom is described differently by the two methods, warranting further investigation.

\begin{table}[h]
  \centering
  \caption{
  Quadrupole splittings for the iron and nickel centers of the \textit{Ht}-SH active site model system,
  computed at the DFT/TPSSh and LNO-CCSD(T)/IAO levels
  with the def2-TZVP basis set.}
  \begin{tabular*}{0.48\textwidth}{@{\extracolsep{\fill}}lcc}
    \hline\hline
    Method & \mc{2}{c}{$\Delta_v$ (mm/s)} \\
    \cline{2-3}
    & $^{57}$Fe & $^{61}$Ni \\
    \hline
    TPSSh  & 0.46 & 0.17 \\
    LNO-CCSD(T) & 0.43 & 0.35 \\
    \hline
  \end{tabular*}
  \label{tab:NiFe_Delta_v}
\end{table}

\subsection{Ab Initio Molecular Dynamics}
As the final example, we present the calculation of the IR spectrum for a protonated water hexamer, \ch{H^+(H2O)6}, via \textit{ab initio} molecular dynamics (AIMD).
The IR intensity ($A$) is proportional to the dipole-dipole correlation function.
In the frequency domain, it can be expressed as
\begin{equation}
    A(\omega) \propto \int \mathrm{d}t \langle \boldsymbol{\Dot{\mu}}(0)\cdot\boldsymbol{\Dot{\mu}}(t)\rangle e^{-i\omega t} ,
    \label{eq:ir}
\end{equation}
where the dipole moment, $\bm{\mu}$, was computed through AD
as the energy derivative with respect to the external electric field,
and its time derivative was computed via finite difference.
Additionally, the bracket in Eq.~\ref{eq:ir} indicates an ensemble average over the MD trajectories,
which were computed at the LNO-CCSD(T) level using the cc-pVTZ\cite{ccpvtz} basis set.
The initial geometries were prepared by Avogadro\cite{hanwell2012avogadro} to represent likely conformers of the protonated water cluster,
and then loosely relaxed at the DFT/$\omega$B97X\cite{wB97X}/cc-pVDZ level.
This results in four distinct structures from which we further seed the dynamics:
one Zundel-like (\ch{H5O2^+}) cation, one Eigen-like [\ch{H3O^+(H2O)3}] cation,
and two with a similar four-water-ring [\ch{H^+(H2O)4}] structure (see Fig.~\ref{fig:water}).
The LNO-CCSD(T) calculations employed the PM local orbitals and an LNO cutoff of $\zeta=5\times 10^{-6}$.
Such a setup was found to reproduce the canonical CCSD(T)/cc-pVTZ energies of the DFT-relaxed geometries to well within chemical accuracy
(with a mean absolute deviation of 0.04 kcal/mol and with a largest deviation of 0.06 kcal/mol).
The MD equilibration in the canonical ($NVT$) ensemble was then performed starting from the four DFT-relaxed geometries.
The temperature was controlled by a Langevin thermostat at 50 K and the nuclear dynamics was integrated using a time step of 1 fs.
The production runs using the microcanonical ($NVE$) ensemble were performed for 10 ps following 1 ps of $NVT$ equilibration.
For the $NVE$ simulations, we employed a longer time step and adopted the multiple time-stepping (MTS) integrator to reduce the time step error of the AIMD. For this we utilized a 2 fs outer MD time step with LNO-CCSD(T) forces, with a 0.5 fs inner time step with machine-learned forces, following the machine-learning (ML) accelerated AIMD approach\cite{li2022using}.
We chose the Allegro framework\cite{musaelian2023learning} to train the ML reference potential on the $NVT$-sampled configurations (1001 configurations for each conformer) and their energies and forces.
To benchmark the error of the MTS integrator,
a separate $NVE$ run (without applying the MTS method) using a time step of 0.5 fs was compared with the MTS dynamics.
Agreement between the two resulting IR spectra validates our MTS approach (see Fig.~\ref{fig:mts_compare}).
As discussed earlier, the potential energy surface (PES) calculated by the LNO-CC method is not strictly continuous,
and thus an energy drift is expected in the $NVE$ MD.
However, the observed energy drift in our simulations is negligible ($<0.05$ kcal/mol/ps),
as the result of using a small enough $\zeta$.

\begin{figure*}
\includegraphics{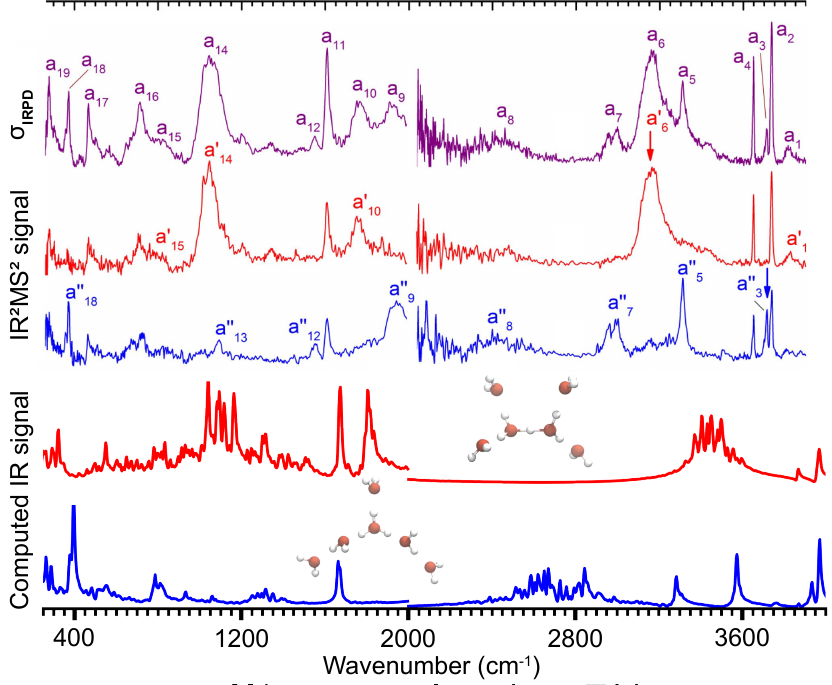}
\centering
\caption{IR spectra of the protonated water cluster. The first panel shows the experimental gas-phase \ch{H2}-predissociation spectrum of \ch{H^+(H2O)6.H2}. The second and third panels show the experimental IR$^2$MS$^2$ spectra of \ch{H^+(H2O)6.H2}, probing the transitions at 3159 cm$^{-1}$ and 3715 cm$^{-1}$ respectively (indicated by the red and blue arrows).
(The experimental spectra were reprinted with permission from Ref.~\citenum{heine2013isomer}. Copyright {2013} American Chemical Society.)
The last two panels show the computed IR spectra for the Zundel-like and Eigen-like conformers. The intensity under 2000 cm$^{-1}$ in the computed spectra is multiplied by 3 for clarity, and the spectra are convoluted using a Gaussian kernel with a width of 1 cm$^{-1}$.}
\label{fig:ir}
\end{figure*}

By computing the IR spectrum using Eq.~\ref{eq:ir} with MD simulations on the LNO-CCSD(T) PES,
we have explicitly accounted for the electron correlation at the CCSD(T) level (within the local approximation) and included anharmonic vibrational effects. Assuming the electronic structure and sampling to be converged,
the only remaining physical component to be included is the nuclear quantum effect (NQE).
While NQEs can in principle be incorporated through
path integral MD\cite{rossi2014remove,paesani2010quantitative} or by explicitly solving
the vibrational Schr\"{o}dinger equation,
\cite{kjaergaard2008calculation,yu2019classical,carpenter2020decoding}
herein we restrict ourselves to classical nuclear dynamics,
and thus quantitative agreement with the experiment is not expected.
As shown in Figure~\ref{fig:ir}, the computed IR spectra of the Zundel- and Eigen-like conformers agree qualitatively with the experimental IR$^2$MS$^2$ spectra,
while blue-shifted signals are observed compared to the experimental peak positions.
We attribute the blue shifting to the missing NQEs instead of the basis set incompleteness error,
since we found that enlarging the basis set from cc-pVTZ to aug-cc-pVQZ
red shifts the OH stretching bands by only $\sim$20 cm$^{-1}$
in a gas-phase harmonic analysis for a single water molecule at the canonical CCSD(T) level.
The experimental IR$^2$MS$^2$ signals\cite{heine2013isomer} were obtained by
probing the bands at 3159 cm$^{-1}$ and 3715 cm$^{-1}$
(denoted as $a_6$ and $a_3$, respectively),
which are assigned to the OH stretches in the Zundel and Eigen cations, respectively.
The absence of $a_6$/$a_3$ signals in the calculated IR spectra
of the Eigen/Zundel-like conformers clearly suggests that
the two frequencies are unique features of the Zundel/Eigen cations.
Moreover, the IR spectra computed for the four-water-ring-like conformers show distinct signatures (see Fig.~\ref{fig:ir_all}),
unambiguously validating that the IR$^2$MS$^2$ measurements correspond to the signals of interest from the Zundel- and Eigen-like conformers.

\section{Conclusions}
In this work, we introduced a differentiable implementation of
a local coupled cluster theory, utilizing the newly improved
AD framework provided by the \textsc{PySCFAD} package.
Calculations of first-order response properties, including nuclear gradients, dipole moments, electric field gradients and relaxed density matrices,
demonstrate the feasibility of our framework for analytic derivative computations involving complex computational workflows.
Moreover, the application of our method to geometry optimizations and AIMD simulations for medium-sized molecular systems further indicates that \textsc{PySCFAD} is not only a useful
platform to rapidly prototype new methodologies but also effective for production-level calculations.

While significant progress has been made, there remain several challenges to address.
Firstly, computing higher-order response properties with the current LNO-CC method can become prohibitively expensive.
Our implementation has a formal scaling of $O(N^5)$,
where $N$ is the number of basis functions.
Moving to each higher order of response increases both computational and memory complexities
by a factor equal to the dimension of the perturbation variable.
This may be manageable for low-dimensional perturbations such as the electric field,
which has a dimension of three.
However, for variables like nuclear coordinates, the resultant cost elevation could be substantial.
One potential solution involves employing the various domain truncation algorithms
\cite{Rolik2013,Riplinger2016} to mitigate the overall computational expense.
These algorithms can be seamlessly integrated into our differentiable LNO-CC implementation,
assuming a well defined energy expression exists.

Secondly, only static response properties are readily computable at the moment,
whereas a straightforward formalism for calculating dynamic response properties
via analytic derivatives is still lacking.
As discussed before,\cite{pyscfad} dynamic response properties may be
computed as the derivatives of the quasienergy,
which is defined as the time-averaged expectation value of
$\hat{H}-i\partial/\partial t$ over the time-dependent wavefunction.\cite{Christiansen1998}
However, a time-dependent implementation of local correlation methods remains to be developed.


\begin{acknowledgements}
This work was primarily supported by the United States Department of
Energy, Office of Science, Basic Energy Sciences, Chemical Sciences, Geosciences,
and Biosciences Division, FWP LANLE3F2 awarded to Los Alamos National
Laboratory under Triad National Security, LLC (‘Triad’) contract grant no.
89233218CNA000001, subaward C2448 to the California Institute of Technology.
Additional support for GKC was provided by the Camille and Henry Dreyfus Foundation via a grant from the program "Machine Learning in the Chemical Sciences
and Engineering"
Some of the calculations were performed at the National Energy Research Scientific Computing Center (NERSC),
a U.S. Department of Energy, Office of Science User Facility located at the Lawrence Berkeley National Laboratory.
\end{acknowledgements}

\section*{Author Contributions}
X. Z. and C. L. contributed equally to this work.
X. Z., C. L., and G. K.-L. C. designed this project.
H.-Z. Y. and T. C. B. contributed to the original implementation of the LNO-CC method,
based on which X. Z. developed the differentiable version used here.
H.-Z. Y. and T. C. B. also provided help and advice with the methodology development in the early part of the project.
C. L. developed the AIMD simulation workflow.
X. Z. and C. L. performed the calculations and wrote the original draft.
X. Z., C. L., and G. K.-L. C. validated the data,
and all contributed to the reviewing and editing of the manuscript.

\section*{Data Availability}
The data that supports the findings of this study are available within the article,
and/or from the corresponding author upon reasonable request.
The \textsc{PySCFAD} source code can be found at \url{https://github.com/fishjojo/pyscfad}.

\section*{References}
\bibliography{ref}

\begin{thebibliography}{108}%
\makeatletter
\providecommand \@ifxundefined [1]{%
 \@ifx{#1\undefined}
}%
\providecommand \@ifnum [1]{%
 \ifnum #1\expandafter \@firstoftwo
 \else \expandafter \@secondoftwo
 \fi
}%
\providecommand \@ifx [1]{%
 \ifx #1\expandafter \@firstoftwo
 \else \expandafter \@secondoftwo
 \fi
}%
\providecommand \natexlab [1]{#1}%
\providecommand \enquote  [1]{``#1''}%
\providecommand \bibnamefont  [1]{#1}%
\providecommand \bibfnamefont [1]{#1}%
\providecommand \citenamefont [1]{#1}%
\providecommand \href@noop [0]{\@secondoftwo}%
\providecommand \href [0]{\begingroup \@sanitize@url \@href}%
\providecommand \@href[1]{\@@startlink{#1}\@@href}%
\providecommand \@@href[1]{\endgroup#1\@@endlink}%
\providecommand \@sanitize@url [0]{\catcode `\\12\catcode `\$12\catcode
  `\&12\catcode `\#12\catcode `\^12\catcode `\_12\catcode `\%12\relax}%
\providecommand \@@startlink[1]{}%
\providecommand \@@endlink[0]{}%
\providecommand \url  [0]{\begingroup\@sanitize@url \@url }%
\providecommand \@url [1]{\endgroup\@href {#1}{\urlprefix }}%
\providecommand \urlprefix  [0]{URL }%
\providecommand \Eprint [0]{\href }%
\providecommand \doibase [0]{https://doi.org/}%
\providecommand \selectlanguage [0]{\@gobble}%
\providecommand \bibinfo  [0]{\@secondoftwo}%
\providecommand \bibfield  [0]{\@secondoftwo}%
\providecommand \translation [1]{[#1]}%
\providecommand \BibitemOpen [0]{}%
\providecommand \bibitemStop [0]{}%
\providecommand \bibitemNoStop [0]{.\EOS\space}%
\providecommand \EOS [0]{\spacefactor3000\relax}%
\providecommand \BibitemShut  [1]{\csname bibitem#1\endcsname}%
\let\auto@bib@innerbib\@empty
\bibitem [{\citenamefont {Pulay}(1983)}]{Pulay1983}%
  \BibitemOpen
  \bibfield  {author} {\bibinfo {author} {\bibfnamefont {P.}~\bibnamefont
  {Pulay}},\ }\bibfield  {title} {\enquote {\bibinfo {title} {Localizability of
  dynamic electron correlation},}\ }\href@noop {} {\bibfield  {journal}
  {\bibinfo  {journal} {Chemical physics letters}\ }\textbf {\bibinfo {volume}
  {100}},\ \bibinfo {pages} {151--154} (\bibinfo {year} {1983})}\BibitemShut
  {NoStop}%
\bibitem [{\citenamefont {S{\ae}b{\o}}\ and\ \citenamefont
  {Pulay}(1985)}]{Saebo1985}%
  \BibitemOpen
  \bibfield  {author} {\bibinfo {author} {\bibfnamefont {S.}~\bibnamefont
  {S{\ae}b{\o}}}\ and\ \bibinfo {author} {\bibfnamefont {P.}~\bibnamefont
  {Pulay}},\ }\bibfield  {title} {\enquote {\bibinfo {title} {Local
  configuration interaction: An efficient approach for larger molecules},}\
  }\href@noop {} {\bibfield  {journal} {\bibinfo  {journal} {Chemical physics
  letters}\ }\textbf {\bibinfo {volume} {113}},\ \bibinfo {pages} {13--18}
  (\bibinfo {year} {1985})}\BibitemShut {NoStop}%
\bibitem [{\citenamefont {Pulay}\ and\ \citenamefont
  {Saeb{\o}}(1986)}]{Pulay1986}%
  \BibitemOpen
  \bibfield  {author} {\bibinfo {author} {\bibfnamefont {P.}~\bibnamefont
  {Pulay}}\ and\ \bibinfo {author} {\bibfnamefont {S.}~\bibnamefont
  {Saeb{\o}}},\ }\bibfield  {title} {\enquote {\bibinfo {title}
  {Orbital-invariant formulation and second-order gradient evaluation in
  {M{\o}ller-Plesset} perturbation theory},}\ }\href@noop {} {\bibfield
  {journal} {\bibinfo  {journal} {Theoretica chimica acta}\ }\textbf {\bibinfo
  {volume} {69}},\ \bibinfo {pages} {357--368} (\bibinfo {year}
  {1986})}\BibitemShut {NoStop}%
\bibitem [{\citenamefont {S{\ae}b{\o}}\ and\ \citenamefont
  {Pulay}(1993)}]{Saebo1993}%
  \BibitemOpen
  \bibfield  {author} {\bibinfo {author} {\bibfnamefont {S.}~\bibnamefont
  {S{\ae}b{\o}}}\ and\ \bibinfo {author} {\bibfnamefont {P.}~\bibnamefont
  {Pulay}},\ }\bibfield  {title} {\enquote {\bibinfo {title} {Local treatment
  of electron correlation},}\ }\href@noop {} {\bibfield  {journal} {\bibinfo
  {journal} {Annual Review of Physical Chemistry}\ }\textbf {\bibinfo {volume}
  {44}},\ \bibinfo {pages} {213--236} (\bibinfo {year} {1993})}\BibitemShut
  {NoStop}%
\bibitem [{\citenamefont {Boughton}\ and\ \citenamefont
  {Pulay}(1993)}]{Boughton1993}%
  \BibitemOpen
  \bibfield  {author} {\bibinfo {author} {\bibfnamefont {J.~W.}\ \bibnamefont
  {Boughton}}\ and\ \bibinfo {author} {\bibfnamefont {P.}~\bibnamefont
  {Pulay}},\ }\bibfield  {title} {\enquote {\bibinfo {title} {Comparison of the
  boys and {Pipek--Mezey} localizations in the local correlation approach and
  automatic virtual basis selection},}\ }\href@noop {} {\bibfield  {journal}
  {\bibinfo  {journal} {Journal of computational chemistry}\ }\textbf {\bibinfo
  {volume} {14}},\ \bibinfo {pages} {736--740} (\bibinfo {year}
  {1993})}\BibitemShut {NoStop}%
\bibitem [{\citenamefont {Hampel}\ and\ \citenamefont
  {Werner}(1996)}]{Hampel1996}%
  \BibitemOpen
  \bibfield  {author} {\bibinfo {author} {\bibfnamefont {C.}~\bibnamefont
  {Hampel}}\ and\ \bibinfo {author} {\bibfnamefont {H.-J.}\ \bibnamefont
  {Werner}},\ }\bibfield  {title} {\enquote {\bibinfo {title} {Local treatment
  of electron correlation in coupled cluster theory},}\ }\href@noop {}
  {\bibfield  {journal} {\bibinfo  {journal} {The Journal of chemical physics}\
  }\textbf {\bibinfo {volume} {104}},\ \bibinfo {pages} {6286--6297} (\bibinfo
  {year} {1996})}\BibitemShut {NoStop}%
\bibitem [{\citenamefont {Sch{\"u}tz}\ and\ \citenamefont
  {Werner}(2000)}]{Schutz2000}%
  \BibitemOpen
  \bibfield  {author} {\bibinfo {author} {\bibfnamefont {M.}~\bibnamefont
  {Sch{\"u}tz}}\ and\ \bibinfo {author} {\bibfnamefont {H.-J.}\ \bibnamefont
  {Werner}},\ }\bibfield  {title} {\enquote {\bibinfo {title} {Local
  perturbative triples correction {(T)} with linear cost scaling},}\
  }\href@noop {} {\bibfield  {journal} {\bibinfo  {journal} {Chemical Physics
  Letters}\ }\textbf {\bibinfo {volume} {318}},\ \bibinfo {pages} {370--378}
  (\bibinfo {year} {2000})}\BibitemShut {NoStop}%
\bibitem [{\citenamefont {Sch{\"u}tz}\ and\ \citenamefont
  {Werner}(2001)}]{Schutz2001}%
  \BibitemOpen
  \bibfield  {author} {\bibinfo {author} {\bibfnamefont {M.}~\bibnamefont
  {Sch{\"u}tz}}\ and\ \bibinfo {author} {\bibfnamefont {H.-J.}\ \bibnamefont
  {Werner}},\ }\bibfield  {title} {\enquote {\bibinfo {title} {Low-order
  scaling local electron correlation methods. {IV}. linear scaling local
  coupled-cluster {(LCCSD)}},}\ }\href@noop {} {\bibfield  {journal} {\bibinfo
  {journal} {The Journal of Chemical Physics}\ }\textbf {\bibinfo {volume}
  {114}},\ \bibinfo {pages} {661--681} (\bibinfo {year} {2001})}\BibitemShut
  {NoStop}%
\bibitem [{\citenamefont {Sch{\"u}tz}(2002{\natexlab{a}})}]{Schutz2002a}%
  \BibitemOpen
  \bibfield  {author} {\bibinfo {author} {\bibfnamefont {M.}~\bibnamefont
  {Sch{\"u}tz}},\ }\bibfield  {title} {\enquote {\bibinfo {title} {Low-order
  scaling local electron correlation methods. {V}. connected triples beyond
  {(T)}: Linear scaling local {CCSDT-1b}},}\ }\href@noop {} {\bibfield
  {journal} {\bibinfo  {journal} {The Journal of chemical physics}\ }\textbf
  {\bibinfo {volume} {116}},\ \bibinfo {pages} {8772--8785} (\bibinfo {year}
  {2002}{\natexlab{a}})}\BibitemShut {NoStop}%
\bibitem [{\citenamefont {Sch{\"u}tz}(2002{\natexlab{b}})}]{Schutz2002b}%
  \BibitemOpen
  \bibfield  {author} {\bibinfo {author} {\bibfnamefont {M.}~\bibnamefont
  {Sch{\"u}tz}},\ }\bibfield  {title} {\enquote {\bibinfo {title} {A new, fast,
  semi-direct implementation of linear scaling local coupled cluster theory},}\
  }\href@noop {} {\bibfield  {journal} {\bibinfo  {journal} {Physical Chemistry
  Chemical Physics}\ }\textbf {\bibinfo {volume} {4}},\ \bibinfo {pages}
  {3941--3947} (\bibinfo {year} {2002}{\natexlab{b}})}\BibitemShut {NoStop}%
\bibitem [{\citenamefont {Li}, \citenamefont {Ma},\ and\ \citenamefont
  {Jiang}(2002)}]{Li2002}%
  \BibitemOpen
  \bibfield  {author} {\bibinfo {author} {\bibfnamefont {S.}~\bibnamefont
  {Li}}, \bibinfo {author} {\bibfnamefont {J.}~\bibnamefont {Ma}},\ and\
  \bibinfo {author} {\bibfnamefont {Y.}~\bibnamefont {Jiang}},\ }\bibfield
  {title} {\enquote {\bibinfo {title} {Linear scaling local correlation
  approach for solving the coupled cluster equations of large systems},}\
  }\href@noop {} {\bibfield  {journal} {\bibinfo  {journal} {Journal of
  computational chemistry}\ }\textbf {\bibinfo {volume} {23}},\ \bibinfo
  {pages} {237--244} (\bibinfo {year} {2002})}\BibitemShut {NoStop}%
\bibitem [{\citenamefont {Li}\ \emph {et~al.}(2006)\citenamefont {Li},
  \citenamefont {Shen}, \citenamefont {Li},\ and\ \citenamefont
  {Jiang}}]{Li2006}%
  \BibitemOpen
  \bibfield  {author} {\bibinfo {author} {\bibfnamefont {S.}~\bibnamefont
  {Li}}, \bibinfo {author} {\bibfnamefont {J.}~\bibnamefont {Shen}}, \bibinfo
  {author} {\bibfnamefont {W.}~\bibnamefont {Li}},\ and\ \bibinfo {author}
  {\bibfnamefont {Y.}~\bibnamefont {Jiang}},\ }\bibfield  {title} {\enquote
  {\bibinfo {title} {An efficient implementation of the
  {``cluster-in-molecule''} approach for local electron correlation
  calculations},}\ }\href@noop {} {\bibfield  {journal} {\bibinfo  {journal}
  {The Journal of chemical physics}\ }\textbf {\bibinfo {volume} {125}},\
  \bibinfo {pages} {074109} (\bibinfo {year} {2006})}\BibitemShut {NoStop}%
\bibitem [{\citenamefont {Li}\ \emph {et~al.}(2009)\citenamefont {Li},
  \citenamefont {Piecuch}, \citenamefont {Gour},\ and\ \citenamefont
  {Li}}]{Li2009}%
  \BibitemOpen
  \bibfield  {author} {\bibinfo {author} {\bibfnamefont {W.}~\bibnamefont
  {Li}}, \bibinfo {author} {\bibfnamefont {P.}~\bibnamefont {Piecuch}},
  \bibinfo {author} {\bibfnamefont {J.~R.}\ \bibnamefont {Gour}},\ and\
  \bibinfo {author} {\bibfnamefont {S.}~\bibnamefont {Li}},\ }\bibfield
  {title} {\enquote {\bibinfo {title} {Local correlation calculations using
  standard and renormalized coupled-cluster approaches},}\ }\href@noop {}
  {\bibfield  {journal} {\bibinfo  {journal} {The Journal of chemical physics}\
  }\textbf {\bibinfo {volume} {131}},\ \bibinfo {pages} {114109} (\bibinfo
  {year} {2009})}\BibitemShut {NoStop}%
\bibitem [{\citenamefont {Neese}, \citenamefont {Wennmohs},\ and\ \citenamefont
  {Hansen}(2009)}]{Neese2009a}%
  \BibitemOpen
  \bibfield  {author} {\bibinfo {author} {\bibfnamefont {F.}~\bibnamefont
  {Neese}}, \bibinfo {author} {\bibfnamefont {F.}~\bibnamefont {Wennmohs}},\
  and\ \bibinfo {author} {\bibfnamefont {A.}~\bibnamefont {Hansen}},\
  }\bibfield  {title} {\enquote {\bibinfo {title} {Efficient and accurate local
  approximations to coupled-electron pair approaches: An attempt to revive the
  pair natural orbital method},}\ }\href@noop {} {\bibfield  {journal}
  {\bibinfo  {journal} {The Journal of chemical physics}\ }\textbf {\bibinfo
  {volume} {130}},\ \bibinfo {pages} {114108} (\bibinfo {year}
  {2009})}\BibitemShut {NoStop}%
\bibitem [{\citenamefont {Neese}, \citenamefont {Hansen},\ and\ \citenamefont
  {Liakos}(2009)}]{Neese2009b}%
  \BibitemOpen
  \bibfield  {author} {\bibinfo {author} {\bibfnamefont {F.}~\bibnamefont
  {Neese}}, \bibinfo {author} {\bibfnamefont {A.}~\bibnamefont {Hansen}},\ and\
  \bibinfo {author} {\bibfnamefont {D.~G.}\ \bibnamefont {Liakos}},\ }\bibfield
   {title} {\enquote {\bibinfo {title} {Efficient and accurate approximations
  to the local coupled cluster singles doubles method using a truncated pair
  natural orbital basis},}\ }\href@noop {} {\bibfield  {journal} {\bibinfo
  {journal} {The Journal of chemical physics}\ }\textbf {\bibinfo {volume}
  {131}},\ \bibinfo {pages} {064103} (\bibinfo {year} {2009})}\BibitemShut
  {NoStop}%
\bibitem [{\citenamefont {Werner}\ and\ \citenamefont
  {Sch{\"u}tz}(2011)}]{Werner2011}%
  \BibitemOpen
  \bibfield  {author} {\bibinfo {author} {\bibfnamefont {H.-J.}\ \bibnamefont
  {Werner}}\ and\ \bibinfo {author} {\bibfnamefont {M.}~\bibnamefont
  {Sch{\"u}tz}},\ }\bibfield  {title} {\enquote {\bibinfo {title} {An efficient
  local coupled cluster method for accurate thermochemistry of large
  systems},}\ }\href@noop {} {\bibfield  {journal} {\bibinfo  {journal} {The
  Journal of Chemical Physics}\ }\textbf {\bibinfo {volume} {135}},\ \bibinfo
  {pages} {144116} (\bibinfo {year} {2011})}\BibitemShut {NoStop}%
\bibitem [{\citenamefont {Rolik}\ and\ \citenamefont
  {K{\'a}llay}(2011)}]{Rolik2011}%
  \BibitemOpen
  \bibfield  {author} {\bibinfo {author} {\bibfnamefont {Z.}~\bibnamefont
  {Rolik}}\ and\ \bibinfo {author} {\bibfnamefont {M.}~\bibnamefont
  {K{\'a}llay}},\ }\bibfield  {title} {\enquote {\bibinfo {title} {A
  general-order local coupled-cluster method based on the cluster-in-molecule
  approach},}\ }\href@noop {} {\bibfield  {journal} {\bibinfo  {journal} {The
  Journal of chemical physics}\ }\textbf {\bibinfo {volume} {135}},\ \bibinfo
  {pages} {104111} (\bibinfo {year} {2011})}\BibitemShut {NoStop}%
\bibitem [{\citenamefont {Yang}\ \emph {et~al.}(2011)\citenamefont {Yang},
  \citenamefont {Kurashige}, \citenamefont {Manby},\ and\ \citenamefont
  {Chan}}]{Yang2011}%
  \BibitemOpen
  \bibfield  {author} {\bibinfo {author} {\bibfnamefont {J.}~\bibnamefont
  {Yang}}, \bibinfo {author} {\bibfnamefont {Y.}~\bibnamefont {Kurashige}},
  \bibinfo {author} {\bibfnamefont {F.~R.}\ \bibnamefont {Manby}},\ and\
  \bibinfo {author} {\bibfnamefont {G.~K.~L.}\ \bibnamefont {Chan}},\
  }\bibfield  {title} {\enquote {\bibinfo {title} {{Tensor factorizations of
  local second-order {M{\o}ller–Plesset} theory}},}\ }\href
  {https://doi.org/10.1063/1.3528935} {\bibfield  {journal} {\bibinfo
  {journal} {The Journal of Chemical Physics}\ }\textbf {\bibinfo {volume}
  {134}},\ \bibinfo {pages} {044123} (\bibinfo {year} {2011})}\BibitemShut
  {NoStop}%
\bibitem [{\citenamefont {Kurashige}\ \emph {et~al.}(2012)\citenamefont
  {Kurashige}, \citenamefont {Yang}, \citenamefont {Chan},\ and\ \citenamefont
  {Manby}}]{Kurashige2012}%
  \BibitemOpen
  \bibfield  {author} {\bibinfo {author} {\bibfnamefont {Y.}~\bibnamefont
  {Kurashige}}, \bibinfo {author} {\bibfnamefont {J.}~\bibnamefont {Yang}},
  \bibinfo {author} {\bibfnamefont {G.~K.-L.}\ \bibnamefont {Chan}},\ and\
  \bibinfo {author} {\bibfnamefont {F.~R.}\ \bibnamefont {Manby}},\ }\bibfield
  {title} {\enquote {\bibinfo {title} {{Optimization of orbital-specific
  virtuals in local {M{\o}ller-Plesset} perturbation theory}},}\ }\href
  {https://doi.org/10.1063/1.3696962} {\bibfield  {journal} {\bibinfo
  {journal} {The Journal of Chemical Physics}\ }\textbf {\bibinfo {volume}
  {136}},\ \bibinfo {pages} {124106} (\bibinfo {year} {2012})}\BibitemShut
  {NoStop}%
\bibitem [{\citenamefont {Yang}\ \emph {et~al.}(2012)\citenamefont {Yang},
  \citenamefont {Chan}, \citenamefont {Manby}, \citenamefont {Sch{\"u}tz},\
  and\ \citenamefont {Werner}}]{Yang2012}%
  \BibitemOpen
  \bibfield  {author} {\bibinfo {author} {\bibfnamefont {J.}~\bibnamefont
  {Yang}}, \bibinfo {author} {\bibfnamefont {G.~K.-L.}\ \bibnamefont {Chan}},
  \bibinfo {author} {\bibfnamefont {F.~R.}\ \bibnamefont {Manby}}, \bibinfo
  {author} {\bibfnamefont {M.}~\bibnamefont {Sch{\"u}tz}},\ and\ \bibinfo
  {author} {\bibfnamefont {H.-J.}\ \bibnamefont {Werner}},\ }\bibfield  {title}
  {\enquote {\bibinfo {title} {{The orbital-specific-virtual local coupled
  cluster singles and doubles method}},}\ }\href
  {https://doi.org/10.1063/1.3696963} {\bibfield  {journal} {\bibinfo
  {journal} {The Journal of Chemical Physics}\ }\textbf {\bibinfo {volume}
  {136}},\ \bibinfo {pages} {144105} (\bibinfo {year} {2012})}\BibitemShut
  {NoStop}%
\bibitem [{\citenamefont {Sch{\"u}tz}\ \emph {et~al.}(2013)\citenamefont
  {Sch{\"u}tz}, \citenamefont {Yang}, \citenamefont {Chan}, \citenamefont
  {Manby},\ and\ \citenamefont {Werner}}]{Schutz2013}%
  \BibitemOpen
  \bibfield  {author} {\bibinfo {author} {\bibfnamefont {M.}~\bibnamefont
  {Sch{\"u}tz}}, \bibinfo {author} {\bibfnamefont {J.}~\bibnamefont {Yang}},
  \bibinfo {author} {\bibfnamefont {G.~K.}\ \bibnamefont {Chan}}, \bibinfo
  {author} {\bibfnamefont {F.~R.}\ \bibnamefont {Manby}},\ and\ \bibinfo
  {author} {\bibfnamefont {H.-J.}\ \bibnamefont {Werner}},\ }\bibfield  {title}
  {\enquote {\bibinfo {title} {The orbital-specific virtual local triples
  correction: {OSV-L (T)}},}\ }\href@noop {} {\bibfield  {journal} {\bibinfo
  {journal} {The Journal of Chemical Physics}\ }\textbf {\bibinfo {volume}
  {138}},\ \bibinfo {pages} {054109} (\bibinfo {year} {2013})}\BibitemShut
  {NoStop}%
\bibitem [{\citenamefont {Rolik}\ \emph {et~al.}(2013)\citenamefont {Rolik},
  \citenamefont {Szegedy}, \citenamefont {Ladj{\'a}nszki}, \citenamefont
  {Lad{\'o}czki},\ and\ \citenamefont {K{\'a}llay}}]{Rolik2013}%
  \BibitemOpen
  \bibfield  {author} {\bibinfo {author} {\bibfnamefont {Z.}~\bibnamefont
  {Rolik}}, \bibinfo {author} {\bibfnamefont {L.}~\bibnamefont {Szegedy}},
  \bibinfo {author} {\bibfnamefont {I.}~\bibnamefont {Ladj{\'a}nszki}},
  \bibinfo {author} {\bibfnamefont {B.}~\bibnamefont {Lad{\'o}czki}},\ and\
  \bibinfo {author} {\bibfnamefont {M.}~\bibnamefont {K{\'a}llay}},\ }\bibfield
   {title} {\enquote {\bibinfo {title} {An efficient linear-scaling {CCSD (T)}
  method based on local natural orbitals},}\ }\href@noop {} {\bibfield
  {journal} {\bibinfo  {journal} {The Journal of chemical physics}\ }\textbf
  {\bibinfo {volume} {139}},\ \bibinfo {pages} {094105} (\bibinfo {year}
  {2013})}\BibitemShut {NoStop}%
\bibitem [{\citenamefont {Riplinger}\ and\ \citenamefont
  {Neese}(2013)}]{Riplinger2013a}%
  \BibitemOpen
  \bibfield  {author} {\bibinfo {author} {\bibfnamefont {C.}~\bibnamefont
  {Riplinger}}\ and\ \bibinfo {author} {\bibfnamefont {F.}~\bibnamefont
  {Neese}},\ }\bibfield  {title} {\enquote {\bibinfo {title} {An efficient and
  near linear scaling pair natural orbital based local coupled cluster
  method},}\ }\href@noop {} {\bibfield  {journal} {\bibinfo  {journal} {The
  Journal of chemical physics}\ }\textbf {\bibinfo {volume} {138}},\ \bibinfo
  {pages} {034106} (\bibinfo {year} {2013})}\BibitemShut {NoStop}%
\bibitem [{\citenamefont {Riplinger}\ \emph {et~al.}(2013)\citenamefont
  {Riplinger}, \citenamefont {Sandhoefer}, \citenamefont {Hansen},\ and\
  \citenamefont {Neese}}]{Riplinger2013b}%
  \BibitemOpen
  \bibfield  {author} {\bibinfo {author} {\bibfnamefont {C.}~\bibnamefont
  {Riplinger}}, \bibinfo {author} {\bibfnamefont {B.}~\bibnamefont
  {Sandhoefer}}, \bibinfo {author} {\bibfnamefont {A.}~\bibnamefont {Hansen}},\
  and\ \bibinfo {author} {\bibfnamefont {F.}~\bibnamefont {Neese}},\ }\bibfield
   {title} {\enquote {\bibinfo {title} {Natural triple excitations in local
  coupled cluster calculations with pair natural orbitals},}\ }\href@noop {}
  {\bibfield  {journal} {\bibinfo  {journal} {The Journal of chemical physics}\
  }\textbf {\bibinfo {volume} {139}},\ \bibinfo {pages} {134101} (\bibinfo
  {year} {2013})}\BibitemShut {NoStop}%
\bibitem [{\citenamefont {Sparta}\ and\ \citenamefont
  {Neese}(2014)}]{Sparta2014}%
  \BibitemOpen
  \bibfield  {author} {\bibinfo {author} {\bibfnamefont {M.}~\bibnamefont
  {Sparta}}\ and\ \bibinfo {author} {\bibfnamefont {F.}~\bibnamefont {Neese}},\
  }\bibfield  {title} {\enquote {\bibinfo {title} {Chemical applications
  carried out by local pair natural orbital based coupled-cluster methods},}\
  }\href@noop {} {\bibfield  {journal} {\bibinfo  {journal} {Chemical Society
  Reviews}\ }\textbf {\bibinfo {volume} {43}},\ \bibinfo {pages} {5032--5041}
  (\bibinfo {year} {2014})}\BibitemShut {NoStop}%
\bibitem [{\citenamefont {Schmitz}, \citenamefont {H{\"a}ttig},\ and\
  \citenamefont {Tew}(2014)}]{Schmitz2014}%
  \BibitemOpen
  \bibfield  {author} {\bibinfo {author} {\bibfnamefont {G.}~\bibnamefont
  {Schmitz}}, \bibinfo {author} {\bibfnamefont {C.}~\bibnamefont
  {H{\"a}ttig}},\ and\ \bibinfo {author} {\bibfnamefont {D.~P.}\ \bibnamefont
  {Tew}},\ }\bibfield  {title} {\enquote {\bibinfo {title} {Explicitly
  correlated {PNO-MP2} and {PNO-CCSD} and their application to the {S66} set
  and large molecular systems},}\ }\href@noop {} {\bibfield  {journal}
  {\bibinfo  {journal} {Physical Chemistry Chemical Physics}\ }\textbf
  {\bibinfo {volume} {16}},\ \bibinfo {pages} {22167--22178} (\bibinfo {year}
  {2014})}\BibitemShut {NoStop}%
\bibitem [{\citenamefont {K{\'a}llay}(2015)}]{Kallay2015}%
  \BibitemOpen
  \bibfield  {author} {\bibinfo {author} {\bibfnamefont {M.}~\bibnamefont
  {K{\'a}llay}},\ }\bibfield  {title} {\enquote {\bibinfo {title}
  {Linear-scaling implementation of the direct random-phase approximation},}\
  }\href@noop {} {\bibfield  {journal} {\bibinfo  {journal} {The Journal of
  chemical physics}\ }\textbf {\bibinfo {volume} {142}},\ \bibinfo {pages}
  {204105} (\bibinfo {year} {2015})}\BibitemShut {NoStop}%
\bibitem [{\citenamefont {Liakos}\ \emph {et~al.}(2015)\citenamefont {Liakos},
  \citenamefont {Sparta}, \citenamefont {Kesharwani}, \citenamefont {Martin},\
  and\ \citenamefont {Neese}}]{Liakos2015a}%
  \BibitemOpen
  \bibfield  {author} {\bibinfo {author} {\bibfnamefont {D.~G.}\ \bibnamefont
  {Liakos}}, \bibinfo {author} {\bibfnamefont {M.}~\bibnamefont {Sparta}},
  \bibinfo {author} {\bibfnamefont {M.~K.}\ \bibnamefont {Kesharwani}},
  \bibinfo {author} {\bibfnamefont {J.~M.}\ \bibnamefont {Martin}},\ and\
  \bibinfo {author} {\bibfnamefont {F.}~\bibnamefont {Neese}},\ }\bibfield
  {title} {\enquote {\bibinfo {title} {Exploring the accuracy limits of local
  pair natural orbital coupled-cluster theory},}\ }\href@noop {} {\bibfield
  {journal} {\bibinfo  {journal} {Journal of chemical theory and computation}\
  }\textbf {\bibinfo {volume} {11}},\ \bibinfo {pages} {1525--1539} (\bibinfo
  {year} {2015})}\BibitemShut {NoStop}%
\bibitem [{\citenamefont {Liakos}\ and\ \citenamefont
  {Neese}(2015)}]{Liakos2015b}%
  \BibitemOpen
  \bibfield  {author} {\bibinfo {author} {\bibfnamefont {D.~G.}\ \bibnamefont
  {Liakos}}\ and\ \bibinfo {author} {\bibfnamefont {F.}~\bibnamefont {Neese}},\
  }\bibfield  {title} {\enquote {\bibinfo {title} {Is it possible to obtain
  coupled cluster quality energies at near density functional theory cost?
  domain-based local pair natural orbital coupled cluster vs modern density
  functional theory},}\ }\href@noop {} {\bibfield  {journal} {\bibinfo
  {journal} {Journal of chemical theory and computation}\ }\textbf {\bibinfo
  {volume} {11}},\ \bibinfo {pages} {4054--4063} (\bibinfo {year}
  {2015})}\BibitemShut {NoStop}%
\bibitem [{\citenamefont {Riplinger}\ \emph {et~al.}(2016)\citenamefont
  {Riplinger}, \citenamefont {Pinski}, \citenamefont {Becker}, \citenamefont
  {Valeev},\ and\ \citenamefont {Neese}}]{Riplinger2016}%
  \BibitemOpen
  \bibfield  {author} {\bibinfo {author} {\bibfnamefont {C.}~\bibnamefont
  {Riplinger}}, \bibinfo {author} {\bibfnamefont {P.}~\bibnamefont {Pinski}},
  \bibinfo {author} {\bibfnamefont {U.}~\bibnamefont {Becker}}, \bibinfo
  {author} {\bibfnamefont {E.~F.}\ \bibnamefont {Valeev}},\ and\ \bibinfo
  {author} {\bibfnamefont {F.}~\bibnamefont {Neese}},\ }\bibfield  {title}
  {\enquote {\bibinfo {title} {Sparse maps—a systematic infrastructure for
  reduced-scaling electronic structure methods. {II}. linear scaling domain
  based pair natural orbital coupled cluster theory},}\ }\href@noop {}
  {\bibfield  {journal} {\bibinfo  {journal} {The Journal of chemical physics}\
  }\textbf {\bibinfo {volume} {144}},\ \bibinfo {pages} {024109} (\bibinfo
  {year} {2016})}\BibitemShut {NoStop}%
\bibitem [{\citenamefont {Schmitz}\ and\ \citenamefont
  {H{\"a}ttig}(2016)}]{Schmitz2016}%
  \BibitemOpen
  \bibfield  {author} {\bibinfo {author} {\bibfnamefont {G.}~\bibnamefont
  {Schmitz}}\ and\ \bibinfo {author} {\bibfnamefont {C.}~\bibnamefont
  {H{\"a}ttig}},\ }\bibfield  {title} {\enquote {\bibinfo {title} {Perturbative
  triples correction for local pair natural orbital based explicitly correlated
  {CCSD (F12*)} using laplace transformation techniques},}\ }\href@noop {}
  {\bibfield  {journal} {\bibinfo  {journal} {The Journal of Chemical Physics}\
  }\textbf {\bibinfo {volume} {145}},\ \bibinfo {pages} {234107} (\bibinfo
  {year} {2016})}\BibitemShut {NoStop}%
\bibitem [{\citenamefont {Guo}\ \emph {et~al.}(2016)\citenamefont {Guo},
  \citenamefont {Sivalingam}, \citenamefont {Valeev},\ and\ \citenamefont
  {Neese}}]{Guo2016}%
  \BibitemOpen
  \bibfield  {author} {\bibinfo {author} {\bibfnamefont {Y.}~\bibnamefont
  {Guo}}, \bibinfo {author} {\bibfnamefont {K.}~\bibnamefont {Sivalingam}},
  \bibinfo {author} {\bibfnamefont {E.~F.}\ \bibnamefont {Valeev}},\ and\
  \bibinfo {author} {\bibfnamefont {F.}~\bibnamefont {Neese}},\ }\bibfield
  {title} {\enquote {\bibinfo {title} {Sparsemaps—a systematic infrastructure
  for reduced-scaling electronic structure methods. {III}. linear-scaling
  multireference domain-based pair natural orbital {N}-electron valence
  perturbation theory},}\ }\href@noop {} {\bibfield  {journal} {\bibinfo
  {journal} {The Journal of chemical physics}\ }\textbf {\bibinfo {volume}
  {144}},\ \bibinfo {pages} {094111} (\bibinfo {year} {2016})}\BibitemShut
  {NoStop}%
\bibitem [{\citenamefont {Pavo{\v{s}}evi{\'c}}\ \emph
  {et~al.}(2017)\citenamefont {Pavo{\v{s}}evi{\'c}}, \citenamefont {Peng},
  \citenamefont {Pinski}, \citenamefont {Riplinger}, \citenamefont {Neese},\
  and\ \citenamefont {Valeev}}]{Pavovsevic2017}%
  \BibitemOpen
  \bibfield  {author} {\bibinfo {author} {\bibfnamefont {F.}~\bibnamefont
  {Pavo{\v{s}}evi{\'c}}}, \bibinfo {author} {\bibfnamefont {C.}~\bibnamefont
  {Peng}}, \bibinfo {author} {\bibfnamefont {P.}~\bibnamefont {Pinski}},
  \bibinfo {author} {\bibfnamefont {C.}~\bibnamefont {Riplinger}}, \bibinfo
  {author} {\bibfnamefont {F.}~\bibnamefont {Neese}},\ and\ \bibinfo {author}
  {\bibfnamefont {E.~F.}\ \bibnamefont {Valeev}},\ }\bibfield  {title}
  {\enquote {\bibinfo {title} {Sparsemaps—a systematic infrastructure for
  reduced scaling electronic structure methods. {V}. linear scaling explicitly
  correlated coupled-cluster method with pair natural orbitals},}\ }\href@noop
  {} {\bibfield  {journal} {\bibinfo  {journal} {The Journal of chemical
  physics}\ }\textbf {\bibinfo {volume} {146}},\ \bibinfo {pages} {174108}
  (\bibinfo {year} {2017})}\BibitemShut {NoStop}%
\bibitem [{\citenamefont {Ma}\ \emph {et~al.}(2017)\citenamefont {Ma},
  \citenamefont {Schwilk}, \citenamefont {K{\"o}ppl},\ and\ \citenamefont
  {Werner}}]{Ma2017}%
  \BibitemOpen
  \bibfield  {author} {\bibinfo {author} {\bibfnamefont {Q.}~\bibnamefont
  {Ma}}, \bibinfo {author} {\bibfnamefont {M.}~\bibnamefont {Schwilk}},
  \bibinfo {author} {\bibfnamefont {C.}~\bibnamefont {K{\"o}ppl}},\ and\
  \bibinfo {author} {\bibfnamefont {H.-J.}\ \bibnamefont {Werner}},\ }\bibfield
   {title} {\enquote {\bibinfo {title} {Scalable electron correlation methods.
  4. parallel explicitly correlated local coupled cluster with pair natural
  orbitals ({PNO-LCCSD-F12})},}\ }\href@noop {} {\bibfield  {journal} {\bibinfo
   {journal} {Journal of chemical theory and computation}\ }\textbf {\bibinfo
  {volume} {13}},\ \bibinfo {pages} {4871--4896} (\bibinfo {year}
  {2017})}\BibitemShut {NoStop}%
\bibitem [{\citenamefont {Ma}\ and\ \citenamefont {Werner}(2018)}]{Ma2018}%
  \BibitemOpen
  \bibfield  {author} {\bibinfo {author} {\bibfnamefont {Q.}~\bibnamefont
  {Ma}}\ and\ \bibinfo {author} {\bibfnamefont {H.-J.}\ \bibnamefont
  {Werner}},\ }\bibfield  {title} {\enquote {\bibinfo {title} {Scalable
  electron correlation methods. 5. parallel perturbative triples correction for
  explicitly correlated local coupled cluster with pair natural orbitals},}\
  }\href@noop {} {\bibfield  {journal} {\bibinfo  {journal} {Journal of
  Chemical Theory and Computation}\ }\textbf {\bibinfo {volume} {14}},\
  \bibinfo {pages} {198--215} (\bibinfo {year} {2018})}\BibitemShut {NoStop}%
\bibitem [{\citenamefont {Guo}, \citenamefont {Becker},\ and\ \citenamefont
  {Neese}(2018)}]{Guo2018}%
  \BibitemOpen
  \bibfield  {author} {\bibinfo {author} {\bibfnamefont {Y.}~\bibnamefont
  {Guo}}, \bibinfo {author} {\bibfnamefont {U.}~\bibnamefont {Becker}},\ and\
  \bibinfo {author} {\bibfnamefont {F.}~\bibnamefont {Neese}},\ }\bibfield
  {title} {\enquote {\bibinfo {title} {Comparison and combination of
  {``direct''} and fragment based local correlation methods: Cluster in
  molecules and domain based local pair natural orbital perturbation and
  coupled cluster theories},}\ }\href@noop {} {\bibfield  {journal} {\bibinfo
  {journal} {The Journal of Chemical Physics}\ }\textbf {\bibinfo {volume}
  {148}},\ \bibinfo {pages} {124117} (\bibinfo {year} {2018})}\BibitemShut
  {NoStop}%
\bibitem [{\citenamefont {Liakos}, \citenamefont {Guo},\ and\ \citenamefont
  {Neese}(2019)}]{Liakos2019}%
  \BibitemOpen
  \bibfield  {author} {\bibinfo {author} {\bibfnamefont {D.~G.}\ \bibnamefont
  {Liakos}}, \bibinfo {author} {\bibfnamefont {Y.}~\bibnamefont {Guo}},\ and\
  \bibinfo {author} {\bibfnamefont {F.}~\bibnamefont {Neese}},\ }\bibfield
  {title} {\enquote {\bibinfo {title} {Comprehensive benchmark results for the
  domain based local pair natural orbital coupled cluster method ({DLPNO-CCSD
  (T)}) for closed-and open-shell systems},}\ }\href@noop {} {\bibfield
  {journal} {\bibinfo  {journal} {The Journal of Physical Chemistry A}\
  }\textbf {\bibinfo {volume} {124}},\ \bibinfo {pages} {90--100} (\bibinfo
  {year} {2019})}\BibitemShut {NoStop}%
\bibitem [{\citenamefont {Nagy}, \citenamefont {Samu},\ and\ \citenamefont
  {K{\'a}llay}(2018)}]{Nagy2018}%
  \BibitemOpen
  \bibfield  {author} {\bibinfo {author} {\bibfnamefont {P.~R.}\ \bibnamefont
  {Nagy}}, \bibinfo {author} {\bibfnamefont {G.}~\bibnamefont {Samu}},\ and\
  \bibinfo {author} {\bibfnamefont {M.}~\bibnamefont {K{\'a}llay}},\ }\bibfield
   {title} {\enquote {\bibinfo {title} {Optimization of the linear-scaling
  local natural orbital {CCSD (T)} method: Improved algorithm and benchmark
  applications},}\ }\href@noop {} {\bibfield  {journal} {\bibinfo  {journal}
  {Journal of Chemical Theory and Computation}\ }\textbf {\bibinfo {volume}
  {14}},\ \bibinfo {pages} {4193--4215} (\bibinfo {year} {2018})}\BibitemShut
  {NoStop}%
\bibitem [{\citenamefont {Nagy}\ and\ \citenamefont
  {K{\'a}llay}(2019)}]{Nagy2019}%
  \BibitemOpen
  \bibfield  {author} {\bibinfo {author} {\bibfnamefont {P.~R.}\ \bibnamefont
  {Nagy}}\ and\ \bibinfo {author} {\bibfnamefont {M.}~\bibnamefont
  {K{\'a}llay}},\ }\bibfield  {title} {\enquote {\bibinfo {title} {Approaching
  the basis set limit of {CCSD (T)} energies for large molecules with local
  natural orbital coupled-cluster methods},}\ }\href@noop {} {\bibfield
  {journal} {\bibinfo  {journal} {Journal of Chemical Theory and Computation}\
  }\textbf {\bibinfo {volume} {15}},\ \bibinfo {pages} {5275--5298} (\bibinfo
  {year} {2019})}\BibitemShut {NoStop}%
\bibitem [{\citenamefont {Ni}, \citenamefont {Li},\ and\ \citenamefont
  {Li}(2019)}]{Ni2019}%
  \BibitemOpen
  \bibfield  {author} {\bibinfo {author} {\bibfnamefont {Z.}~\bibnamefont
  {Ni}}, \bibinfo {author} {\bibfnamefont {W.}~\bibnamefont {Li}},\ and\
  \bibinfo {author} {\bibfnamefont {S.}~\bibnamefont {Li}},\ }\bibfield
  {title} {\enquote {\bibinfo {title} {Fully optimized implementation of the
  cluster-in-molecule local correlation approach for electron correlation
  calculations of large systems},}\ }\href@noop {} {\bibfield  {journal}
  {\bibinfo  {journal} {Journal of Computational Chemistry}\ }\textbf {\bibinfo
  {volume} {40}},\ \bibinfo {pages} {1130--1140} (\bibinfo {year}
  {2019})}\BibitemShut {NoStop}%
\bibitem [{\citenamefont {Wang}\ \emph {et~al.}(2019)\citenamefont {Wang},
  \citenamefont {Ni}, \citenamefont {Li},\ and\ \citenamefont {Li}}]{Wang2019}%
  \BibitemOpen
  \bibfield  {author} {\bibinfo {author} {\bibfnamefont {Y.}~\bibnamefont
  {Wang}}, \bibinfo {author} {\bibfnamefont {Z.}~\bibnamefont {Ni}}, \bibinfo
  {author} {\bibfnamefont {W.}~\bibnamefont {Li}},\ and\ \bibinfo {author}
  {\bibfnamefont {S.}~\bibnamefont {Li}},\ }\bibfield  {title} {\enquote
  {\bibinfo {title} {Cluster-in-molecule local correlation approach for
  periodic systems},}\ }\href@noop {} {\bibfield  {journal} {\bibinfo
  {journal} {Journal of Chemical Theory and Computation}\ }\textbf {\bibinfo
  {volume} {15}},\ \bibinfo {pages} {2933--2943} (\bibinfo {year}
  {2019})}\BibitemShut {NoStop}%
\bibitem [{\citenamefont {Ni}\ \emph {et~al.}(2021)\citenamefont {Ni},
  \citenamefont {Guo}, \citenamefont {Neese}, \citenamefont {Li},\ and\
  \citenamefont {Li}}]{Ni2021}%
  \BibitemOpen
  \bibfield  {author} {\bibinfo {author} {\bibfnamefont {Z.}~\bibnamefont
  {Ni}}, \bibinfo {author} {\bibfnamefont {Y.}~\bibnamefont {Guo}}, \bibinfo
  {author} {\bibfnamefont {F.}~\bibnamefont {Neese}}, \bibinfo {author}
  {\bibfnamefont {W.}~\bibnamefont {Li}},\ and\ \bibinfo {author}
  {\bibfnamefont {S.}~\bibnamefont {Li}},\ }\bibfield  {title} {\enquote
  {\bibinfo {title} {Cluster-in-molecule local correlation method with an
  accurate distant pair correction for large systems},}\ }\href@noop {}
  {\bibfield  {journal} {\bibinfo  {journal} {Journal of Chemical Theory and
  Computation}\ }\textbf {\bibinfo {volume} {17}},\ \bibinfo {pages} {756--766}
  (\bibinfo {year} {2021})}\BibitemShut {NoStop}%
\bibitem [{\citenamefont {Wang}\ \emph {et~al.}(2022)\citenamefont {Wang},
  \citenamefont {Ni}, \citenamefont {Neese}, \citenamefont {Li}, \citenamefont
  {Guo},\ and\ \citenamefont {Li}}]{Wang2022}%
  \BibitemOpen
  \bibfield  {author} {\bibinfo {author} {\bibfnamefont {Y.}~\bibnamefont
  {Wang}}, \bibinfo {author} {\bibfnamefont {Z.}~\bibnamefont {Ni}}, \bibinfo
  {author} {\bibfnamefont {F.}~\bibnamefont {Neese}}, \bibinfo {author}
  {\bibfnamefont {W.}~\bibnamefont {Li}}, \bibinfo {author} {\bibfnamefont
  {Y.}~\bibnamefont {Guo}},\ and\ \bibinfo {author} {\bibfnamefont
  {S.}~\bibnamefont {Li}},\ }\bibfield  {title} {\enquote {\bibinfo {title}
  {Cluster-in-molecule method combined with the domain-based local pair natural
  orbital approach for electron correlation calculations of periodic
  systems},}\ }\href@noop {} {\bibfield  {journal} {\bibinfo  {journal}
  {Journal of Chemical Theory and Computation}\ }\textbf {\bibinfo {volume}
  {18}},\ \bibinfo {pages} {6510--6521} (\bibinfo {year} {2022})}\BibitemShut
  {NoStop}%
\bibitem [{\citenamefont {Li}\ \emph {et~al.}(2023)\citenamefont {Li},
  \citenamefont {Wang}, \citenamefont {Ni},\ and\ \citenamefont {Li}}]{Li2023}%
  \BibitemOpen
  \bibfield  {author} {\bibinfo {author} {\bibfnamefont {W.}~\bibnamefont
  {Li}}, \bibinfo {author} {\bibfnamefont {Y.}~\bibnamefont {Wang}}, \bibinfo
  {author} {\bibfnamefont {Z.}~\bibnamefont {Ni}},\ and\ \bibinfo {author}
  {\bibfnamefont {S.}~\bibnamefont {Li}},\ }\bibfield  {title} {\enquote
  {\bibinfo {title} {Cluster-in-molecule local correlation method for
  dispersion interactions in large systems and periodic systems},}\ }\href@noop
  {} {\bibfield  {journal} {\bibinfo  {journal} {Accounts of Chemical
  Research}\ }\textbf {\bibinfo {volume} {56}},\ \bibinfo {pages} {3462--3474}
  (\bibinfo {year} {2023})}\BibitemShut {NoStop}%
\bibitem [{\citenamefont {Ye}\ and\ \citenamefont
  {Berkelbach}(2024{\natexlab{a}})}]{Ye2023}%
  \BibitemOpen
  \bibfield  {author} {\bibinfo {author} {\bibfnamefont {H.-Z.}\ \bibnamefont
  {Ye}}\ and\ \bibinfo {author} {\bibfnamefont {T.~C.}\ \bibnamefont
  {Berkelbach}},\ }\href@noop {} {\enquote {\bibinfo {title} {Ab initio surface
  chemistry with chemical accuracy},}\ } (\bibinfo {year}
  {2024}{\natexlab{a}}),\ \Eprint {https://arxiv.org/abs/2309.14640}
  {arXiv:2309.14640 [cond-mat.mtrl-sci]} \BibitemShut {NoStop}%
\bibitem [{\citenamefont {Raghavachari}\ \emph {et~al.}(1989)\citenamefont
  {Raghavachari}, \citenamefont {Trucks}, \citenamefont {Pople},\ and\
  \citenamefont {{Head-Gordon}}}]{Raghavachari1989}%
  \BibitemOpen
  \bibfield  {author} {\bibinfo {author} {\bibfnamefont {K.}~\bibnamefont
  {Raghavachari}}, \bibinfo {author} {\bibfnamefont {G.~W.}\ \bibnamefont
  {Trucks}}, \bibinfo {author} {\bibfnamefont {J.~A.}\ \bibnamefont {Pople}},\
  and\ \bibinfo {author} {\bibfnamefont {M.}~\bibnamefont {{Head-Gordon}}},\
  }\bibfield  {title} {\enquote {\bibinfo {title} {A {fifth-order} perturbation
  comparison of electron correlation theories},}\ }\href@noop {} {\bibfield
  {journal} {\bibinfo  {journal} {Chemical Physics Letters}\ }\textbf {\bibinfo
  {volume} {157}},\ \bibinfo {pages} {479--483} (\bibinfo {year}
  {1989})}\BibitemShut {NoStop}%
\bibitem [{\citenamefont {Ye}\ and\ \citenamefont
  {Berkelbach}(2024{\natexlab{b}})}]{Ye2023b}%
  \BibitemOpen
  \bibfield  {author} {\bibinfo {author} {\bibfnamefont {H.-Z.}\ \bibnamefont
  {Ye}}\ and\ \bibinfo {author} {\bibfnamefont {T.~C.}\ \bibnamefont
  {Berkelbach}},\ }\href@noop {} {\enquote {\bibinfo {title} {Adsorption and
  vibrational spectroscopy of {CO} on the surface of {MgO} from periodic local
  coupled-cluster theory},}\ } (\bibinfo {year} {2024}{\natexlab{b}}),\ \Eprint
  {https://arxiv.org/abs/2309.14651} {arXiv:2309.14651 [cond-mat.mtrl-sci]}
  \BibitemShut {NoStop}%
\bibitem [{\citenamefont {El~Azhary}\ \emph {et~al.}(1998)\citenamefont
  {El~Azhary}, \citenamefont {Rauhut}, \citenamefont {Pulay},\ and\
  \citenamefont {Werner}}]{El1998}%
  \BibitemOpen
  \bibfield  {author} {\bibinfo {author} {\bibfnamefont {A.}~\bibnamefont
  {El~Azhary}}, \bibinfo {author} {\bibfnamefont {G.}~\bibnamefont {Rauhut}},
  \bibinfo {author} {\bibfnamefont {P.}~\bibnamefont {Pulay}},\ and\ \bibinfo
  {author} {\bibfnamefont {H.-J.}\ \bibnamefont {Werner}},\ }\bibfield  {title}
  {\enquote {\bibinfo {title} {Analytical energy gradients for local
  second-order {M{\o}ller--Plesset} perturbation theory},}\ }\href@noop {}
  {\bibfield  {journal} {\bibinfo  {journal} {The Journal of chemical physics}\
  }\textbf {\bibinfo {volume} {108}},\ \bibinfo {pages} {5185--5193} (\bibinfo
  {year} {1998})}\BibitemShut {NoStop}%
\bibitem [{\citenamefont {Rauhut}\ and\ \citenamefont
  {Werner}(2001)}]{Rauhut2001}%
  \BibitemOpen
  \bibfield  {author} {\bibinfo {author} {\bibfnamefont {G.}~\bibnamefont
  {Rauhut}}\ and\ \bibinfo {author} {\bibfnamefont {H.-J.}\ \bibnamefont
  {Werner}},\ }\bibfield  {title} {\enquote {\bibinfo {title} {Analytical
  energy gradients for local coupled-cluster methods},}\ }\href@noop {}
  {\bibfield  {journal} {\bibinfo  {journal} {Physical Chemistry Chemical
  Physics}\ }\textbf {\bibinfo {volume} {3}},\ \bibinfo {pages} {4853--4862}
  (\bibinfo {year} {2001})}\BibitemShut {NoStop}%
\bibitem [{\citenamefont {Sch{\"u}tz}\ \emph {et~al.}(2004)\citenamefont
  {Sch{\"u}tz}, \citenamefont {Werner}, \citenamefont {Lindh},\ and\
  \citenamefont {Manby}}]{Schutz2004}%
  \BibitemOpen
  \bibfield  {author} {\bibinfo {author} {\bibfnamefont {M.}~\bibnamefont
  {Sch{\"u}tz}}, \bibinfo {author} {\bibfnamefont {H.-J.}\ \bibnamefont
  {Werner}}, \bibinfo {author} {\bibfnamefont {R.}~\bibnamefont {Lindh}},\ and\
  \bibinfo {author} {\bibfnamefont {F.~R.}\ \bibnamefont {Manby}},\ }\bibfield
  {title} {\enquote {\bibinfo {title} {Analytical energy gradients for local
  second-order {M{\o}ller--Plesset} perturbation theory using density fitting
  approximations},}\ }\href@noop {} {\bibfield  {journal} {\bibinfo  {journal}
  {The Journal of chemical physics}\ }\textbf {\bibinfo {volume} {121}},\
  \bibinfo {pages} {737--750} (\bibinfo {year} {2004})}\BibitemShut {NoStop}%
\bibitem [{\citenamefont {Dornbach}\ and\ \citenamefont
  {Werner}(2019)}]{Dornbach2019}%
  \BibitemOpen
  \bibfield  {author} {\bibinfo {author} {\bibfnamefont {M.}~\bibnamefont
  {Dornbach}}\ and\ \bibinfo {author} {\bibfnamefont {H.-J.}\ \bibnamefont
  {Werner}},\ }\bibfield  {title} {\enquote {\bibinfo {title} {Analytical
  energy gradients for local second-order {M{\o}ller-Plesset} perturbation
  theory using intrinsic bond orbitals},}\ }\href@noop {} {\bibfield  {journal}
  {\bibinfo  {journal} {Molecular Physics}\ }\textbf {\bibinfo {volume}
  {117}},\ \bibinfo {pages} {1252--1263} (\bibinfo {year} {2019})}\BibitemShut
  {NoStop}%
\bibitem [{\citenamefont {Lederm{\"u}ller}, \citenamefont {Kats},\ and\
  \citenamefont {Sch{\"u}tz}(2013)}]{Ledermuller2013}%
  \BibitemOpen
  \bibfield  {author} {\bibinfo {author} {\bibfnamefont {K.}~\bibnamefont
  {Lederm{\"u}ller}}, \bibinfo {author} {\bibfnamefont {D.}~\bibnamefont
  {Kats}},\ and\ \bibinfo {author} {\bibfnamefont {M.}~\bibnamefont
  {Sch{\"u}tz}},\ }\bibfield  {title} {\enquote {\bibinfo {title} {Local {CC2}
  response method based on the laplace transform: Orbital-relaxed first-order
  properties for excited states},}\ }\href@noop {} {\bibfield  {journal}
  {\bibinfo  {journal} {The Journal of Chemical Physics}\ }\textbf {\bibinfo
  {volume} {139}},\ \bibinfo {pages} {084111} (\bibinfo {year}
  {2013})}\BibitemShut {NoStop}%
\bibitem [{\citenamefont {Lederm{\"u}ller}\ and\ \citenamefont
  {Sch{\"u}tz}(2014)}]{Ledermuller2014}%
  \BibitemOpen
  \bibfield  {author} {\bibinfo {author} {\bibfnamefont {K.}~\bibnamefont
  {Lederm{\"u}ller}}\ and\ \bibinfo {author} {\bibfnamefont {M.}~\bibnamefont
  {Sch{\"u}tz}},\ }\bibfield  {title} {\enquote {\bibinfo {title} {Local {CC2}
  response method based on the laplace transform: Analytic energy gradients for
  ground and excited states},}\ }\href@noop {} {\bibfield  {journal} {\bibinfo
  {journal} {The Journal of Chemical Physics}\ }\textbf {\bibinfo {volume}
  {140}},\ \bibinfo {pages} {164113} (\bibinfo {year} {2014})}\BibitemShut
  {NoStop}%
\bibitem [{\citenamefont {Gauss}\ and\ \citenamefont
  {Werner}(2000)}]{Gauss2000}%
  \BibitemOpen
  \bibfield  {author} {\bibinfo {author} {\bibfnamefont {J.}~\bibnamefont
  {Gauss}}\ and\ \bibinfo {author} {\bibfnamefont {H.-J.}\ \bibnamefont
  {Werner}},\ }\bibfield  {title} {\enquote {\bibinfo {title} {{NMR} chemical
  shift calculations within local correlation methods: the {GIAO-LMP2}
  approach},}\ }\href@noop {} {\bibfield  {journal} {\bibinfo  {journal}
  {Physical Chemistry Chemical Physics}\ }\textbf {\bibinfo {volume} {2}},\
  \bibinfo {pages} {2083--2090} (\bibinfo {year} {2000})}\BibitemShut {NoStop}%
\bibitem [{\citenamefont {Loibl}\ and\ \citenamefont
  {Sch{\"u}tz}(2012)}]{Loibl2012}%
  \BibitemOpen
  \bibfield  {author} {\bibinfo {author} {\bibfnamefont {S.}~\bibnamefont
  {Loibl}}\ and\ \bibinfo {author} {\bibfnamefont {M.}~\bibnamefont
  {Sch{\"u}tz}},\ }\bibfield  {title} {\enquote {\bibinfo {title} {{NMR}
  shielding tensors for density fitted local second-order {M{\o}ller-Plesset}
  perturbation theory using gauge including atomic orbitals},}\ }\href@noop {}
  {\bibfield  {journal} {\bibinfo  {journal} {The Journal of Chemical Physics}\
  }\textbf {\bibinfo {volume} {137}},\ \bibinfo {pages} {084107} (\bibinfo
  {year} {2012})}\BibitemShut {NoStop}%
\bibitem [{\citenamefont {Loibl}\ and\ \citenamefont
  {Sch{\"u}tz}(2014)}]{Loibl2014}%
  \BibitemOpen
  \bibfield  {author} {\bibinfo {author} {\bibfnamefont {S.}~\bibnamefont
  {Loibl}}\ and\ \bibinfo {author} {\bibfnamefont {M.}~\bibnamefont
  {Sch{\"u}tz}},\ }\bibfield  {title} {\enquote {\bibinfo {title}
  {Magnetizability and rotational g tensors for density fitted local
  second-order {M{\o}ller-Plesset} perturbation theory using gauge-including
  atomic orbitals},}\ }\href@noop {} {\bibfield  {journal} {\bibinfo  {journal}
  {The Journal of Chemical Physics}\ }\textbf {\bibinfo {volume} {141}},\
  \bibinfo {pages} {024108} (\bibinfo {year} {2014})}\BibitemShut {NoStop}%
\bibitem [{\citenamefont {Pinski}\ and\ \citenamefont
  {Neese}(2018)}]{Pinski2018}%
  \BibitemOpen
  \bibfield  {author} {\bibinfo {author} {\bibfnamefont {P.}~\bibnamefont
  {Pinski}}\ and\ \bibinfo {author} {\bibfnamefont {F.}~\bibnamefont {Neese}},\
  }\bibfield  {title} {\enquote {\bibinfo {title} {Communication: Exact
  analytical derivatives for the domain-based local pair natural orbital mp2
  method ({DLPNO-MP2})},}\ }\href@noop {} {\bibfield  {journal} {\bibinfo
  {journal} {The Journal of Chemical Physics}\ }\textbf {\bibinfo {volume}
  {148}},\ \bibinfo {pages} {031101} (\bibinfo {year} {2018})}\BibitemShut
  {NoStop}%
\bibitem [{\citenamefont {Pinski}\ and\ \citenamefont
  {Neese}(2019)}]{Pinski2019}%
  \BibitemOpen
  \bibfield  {author} {\bibinfo {author} {\bibfnamefont {P.}~\bibnamefont
  {Pinski}}\ and\ \bibinfo {author} {\bibfnamefont {F.}~\bibnamefont {Neese}},\
  }\bibfield  {title} {\enquote {\bibinfo {title} {Analytical gradient for the
  domain-based local pair natural orbital second order {M{\o}ller-Plesset}
  perturbation theory method ({DLPNO-MP2})},}\ }\href@noop {} {\bibfield
  {journal} {\bibinfo  {journal} {The Journal of Chemical Physics}\ }\textbf
  {\bibinfo {volume} {150}},\ \bibinfo {pages} {164102} (\bibinfo {year}
  {2019})}\BibitemShut {NoStop}%
\bibitem [{\citenamefont {Stoychev}\ \emph {et~al.}(2021)\citenamefont
  {Stoychev}, \citenamefont {Auer}, \citenamefont {Gauss},\ and\ \citenamefont
  {Neese}}]{Stoychev2021}%
  \BibitemOpen
  \bibfield  {author} {\bibinfo {author} {\bibfnamefont {G.~L.}\ \bibnamefont
  {Stoychev}}, \bibinfo {author} {\bibfnamefont {A.~A.}\ \bibnamefont {Auer}},
  \bibinfo {author} {\bibfnamefont {J.}~\bibnamefont {Gauss}},\ and\ \bibinfo
  {author} {\bibfnamefont {F.}~\bibnamefont {Neese}},\ }\bibfield  {title}
  {\enquote {\bibinfo {title} {{DLPNO-MP2} second derivatives for the
  computation of polarizabilities and {NMR} shieldings},}\ }\href@noop {}
  {\bibfield  {journal} {\bibinfo  {journal} {The Journal of Chemical Physics}\
  }\textbf {\bibinfo {volume} {154}},\ \bibinfo {pages} {164110} (\bibinfo
  {year} {2021})}\BibitemShut {NoStop}%
\bibitem [{\citenamefont {Datta}, \citenamefont {Kossmann},\ and\ \citenamefont
  {Neese}(2016)}]{Datta2016}%
  \BibitemOpen
  \bibfield  {author} {\bibinfo {author} {\bibfnamefont {D.}~\bibnamefont
  {Datta}}, \bibinfo {author} {\bibfnamefont {S.}~\bibnamefont {Kossmann}},\
  and\ \bibinfo {author} {\bibfnamefont {F.}~\bibnamefont {Neese}},\ }\bibfield
   {title} {\enquote {\bibinfo {title} {Analytic energy derivatives for the
  calculation of the first-order molecular properties using the domain-based
  local pair-natural orbital coupled-cluster theory},}\ }\href@noop {}
  {\bibfield  {journal} {\bibinfo  {journal} {The Journal of Chemical Physics}\
  }\textbf {\bibinfo {volume} {145}},\ \bibinfo {pages} {114101} (\bibinfo
  {year} {2016})}\BibitemShut {NoStop}%
\bibitem [{\citenamefont {Zhou}, \citenamefont {Liang},\ and\ \citenamefont
  {Yang}(2019)}]{Zhou2019}%
  \BibitemOpen
  \bibfield  {author} {\bibinfo {author} {\bibfnamefont {R.}~\bibnamefont
  {Zhou}}, \bibinfo {author} {\bibfnamefont {Q.}~\bibnamefont {Liang}},\ and\
  \bibinfo {author} {\bibfnamefont {J.}~\bibnamefont {Yang}},\ }\bibfield
  {title} {\enquote {\bibinfo {title} {Complete {OSV-MP2} analytical gradient
  theory for molecular structure and dynamics simulations},}\ }\href@noop {}
  {\bibfield  {journal} {\bibinfo  {journal} {Journal of Chemical Theory and
  Computation}\ }\textbf {\bibinfo {volume} {16}},\ \bibinfo {pages} {196--210}
  (\bibinfo {year} {2019})}\BibitemShut {NoStop}%
\bibitem [{\citenamefont {Russ}\ and\ \citenamefont
  {Crawford}(2004)}]{Russ2004}%
  \BibitemOpen
  \bibfield  {author} {\bibinfo {author} {\bibfnamefont {N.~J.}\ \bibnamefont
  {Russ}}\ and\ \bibinfo {author} {\bibfnamefont {T.~D.}\ \bibnamefont
  {Crawford}},\ }\bibfield  {title} {\enquote {\bibinfo {title} {Local
  correlation in coupled cluster calculations of molecular response
  properties},}\ }\href@noop {} {\bibfield  {journal} {\bibinfo  {journal}
  {Chemical physics letters}\ }\textbf {\bibinfo {volume} {400}},\ \bibinfo
  {pages} {104--111} (\bibinfo {year} {2004})}\BibitemShut {NoStop}%
\bibitem [{\citenamefont {Russ}\ and\ \citenamefont
  {Crawford}(2008)}]{Russ2008}%
  \BibitemOpen
  \bibfield  {author} {\bibinfo {author} {\bibfnamefont {N.~J.}\ \bibnamefont
  {Russ}}\ and\ \bibinfo {author} {\bibfnamefont {T.~D.}\ \bibnamefont
  {Crawford}},\ }\bibfield  {title} {\enquote {\bibinfo {title} {Local
  correlation domains for coupled cluster theory: optical rotation and
  magnetic-field perturbations},}\ }\href@noop {} {\bibfield  {journal}
  {\bibinfo  {journal} {Physical Chemistry Chemical Physics}\ }\textbf
  {\bibinfo {volume} {10}},\ \bibinfo {pages} {3345--3352} (\bibinfo {year}
  {2008})}\BibitemShut {NoStop}%
\bibitem [{\citenamefont {McAlexander}, \citenamefont {Mach},\ and\
  \citenamefont {Crawford}(2012)}]{McAlexander2012}%
  \BibitemOpen
  \bibfield  {author} {\bibinfo {author} {\bibfnamefont {H.~R.}\ \bibnamefont
  {McAlexander}}, \bibinfo {author} {\bibfnamefont {T.~J.}\ \bibnamefont
  {Mach}},\ and\ \bibinfo {author} {\bibfnamefont {T.~D.}\ \bibnamefont
  {Crawford}},\ }\bibfield  {title} {\enquote {\bibinfo {title} {Localized
  optimized orbitals, coupled cluster theory, and chiroptical response
  properties},}\ }\href@noop {} {\bibfield  {journal} {\bibinfo  {journal}
  {Physical Chemistry Chemical Physics}\ }\textbf {\bibinfo {volume} {14}},\
  \bibinfo {pages} {7830--7836} (\bibinfo {year} {2012})}\BibitemShut {NoStop}%
\bibitem [{\citenamefont {McAlexander}\ and\ \citenamefont
  {Crawford}(2016)}]{McAlexander2016}%
  \BibitemOpen
  \bibfield  {author} {\bibinfo {author} {\bibfnamefont {H.~R.}\ \bibnamefont
  {McAlexander}}\ and\ \bibinfo {author} {\bibfnamefont {T.~D.}\ \bibnamefont
  {Crawford}},\ }\bibfield  {title} {\enquote {\bibinfo {title} {A comparison
  of three approaches to the reduced-scaling coupled cluster treatment of
  non-resonant molecular response properties},}\ }\href
  {https://doi.org/10.1021/acs.jctc.5b00898} {\bibfield  {journal} {\bibinfo
  {journal} {Journal of Chemical Theory and Computation}\ }\textbf {\bibinfo
  {volume} {12}},\ \bibinfo {pages} {209--222} (\bibinfo {year}
  {2016})}\BibitemShut {NoStop}%
\bibitem [{\citenamefont {D'Cunha}\ and\ \citenamefont
  {Crawford}(2021)}]{DCunha2021}%
  \BibitemOpen
  \bibfield  {author} {\bibinfo {author} {\bibfnamefont {R.}~\bibnamefont
  {D'Cunha}}\ and\ \bibinfo {author} {\bibfnamefont {T.~D.}\ \bibnamefont
  {Crawford}},\ }\bibfield  {title} {\enquote {\bibinfo {title} {{PNO++}:
  Perturbed pair natural orbitals for coupled cluster linear response
  theory},}\ }\href {https://doi.org/10.1021/acs.jctc.0c01086} {\bibfield
  {journal} {\bibinfo  {journal} {Journal of Chemical Theory and Computation}\
  }\textbf {\bibinfo {volume} {17}},\ \bibinfo {pages} {290--301} (\bibinfo
  {year} {2021})}\BibitemShut {NoStop}%
\bibitem [{\citenamefont {Maclaurin}, \citenamefont {Duvenaud},\ and\
  \citenamefont {Adams}(2015)}]{Autograd}%
  \BibitemOpen
  \bibfield  {author} {\bibinfo {author} {\bibfnamefont {D.}~\bibnamefont
  {Maclaurin}}, \bibinfo {author} {\bibfnamefont {D.}~\bibnamefont
  {Duvenaud}},\ and\ \bibinfo {author} {\bibfnamefont {R.~P.}\ \bibnamefont
  {Adams}},\ }\bibfield  {title} {\enquote {\bibinfo {title} {Autograd:
  Effortless gradients in numpy},}\ }in\ \href@noop {} {\emph {\bibinfo
  {booktitle} {ICML 2015 AutoML workshop}}},\ Vol.\ \bibinfo {volume} {238}\
  (\bibinfo {year} {2015})\BibitemShut {NoStop}%
\bibitem [{\citenamefont {Bradbury}\ \emph {et~al.}(2018)\citenamefont
  {Bradbury}, \citenamefont {Frostig}, \citenamefont {Hawkins}, \citenamefont
  {Johnson}, \citenamefont {Leary}, \citenamefont {Maclaurin}, \citenamefont
  {Necula}, \citenamefont {Paszke}, \citenamefont {Vander{P}las}, \citenamefont
  {Wanderman-{M}ilne},\ and\ \citenamefont {Zhang}}]{Jax2018}%
  \BibitemOpen
  \bibfield  {author} {\bibinfo {author} {\bibfnamefont {J.}~\bibnamefont
  {Bradbury}}, \bibinfo {author} {\bibfnamefont {R.}~\bibnamefont {Frostig}},
  \bibinfo {author} {\bibfnamefont {P.}~\bibnamefont {Hawkins}}, \bibinfo
  {author} {\bibfnamefont {M.~J.}\ \bibnamefont {Johnson}}, \bibinfo {author}
  {\bibfnamefont {C.}~\bibnamefont {Leary}}, \bibinfo {author} {\bibfnamefont
  {D.}~\bibnamefont {Maclaurin}}, \bibinfo {author} {\bibfnamefont
  {G.}~\bibnamefont {Necula}}, \bibinfo {author} {\bibfnamefont
  {A.}~\bibnamefont {Paszke}}, \bibinfo {author} {\bibfnamefont
  {J.}~\bibnamefont {Vander{P}las}}, \bibinfo {author} {\bibfnamefont
  {S.}~\bibnamefont {Wanderman-{M}ilne}},\ and\ \bibinfo {author}
  {\bibfnamefont {Q.}~\bibnamefont {Zhang}},\ }\href
  {http://github.com/google/jax} {\enquote {\bibinfo {title} {{JAX}: composable
  transformations of {P}ython+{N}um{P}y programs},}\ } (\bibinfo {year}
  {2018}),\ \bibinfo {note} {available at
  \url{http://github.com/google/jax}}\BibitemShut {NoStop}%
\bibitem [{\citenamefont {Paszke}\ \emph {et~al.}(2019)\citenamefont {Paszke},
  \citenamefont {Gross}, \citenamefont {Massa}, \citenamefont {Lerer},
  \citenamefont {Bradbury}, \citenamefont {Chanan}, \citenamefont {Killeen},
  \citenamefont {Lin}, \citenamefont {Gimelshein}, \citenamefont {Antiga} \emph
  {et~al.}}]{Pytorch}%
  \BibitemOpen
  \bibfield  {author} {\bibinfo {author} {\bibfnamefont {A.}~\bibnamefont
  {Paszke}}, \bibinfo {author} {\bibfnamefont {S.}~\bibnamefont {Gross}},
  \bibinfo {author} {\bibfnamefont {F.}~\bibnamefont {Massa}}, \bibinfo
  {author} {\bibfnamefont {A.}~\bibnamefont {Lerer}}, \bibinfo {author}
  {\bibfnamefont {J.}~\bibnamefont {Bradbury}}, \bibinfo {author}
  {\bibfnamefont {G.}~\bibnamefont {Chanan}}, \bibinfo {author} {\bibfnamefont
  {T.}~\bibnamefont {Killeen}}, \bibinfo {author} {\bibfnamefont
  {Z.}~\bibnamefont {Lin}}, \bibinfo {author} {\bibfnamefont {N.}~\bibnamefont
  {Gimelshein}}, \bibinfo {author} {\bibfnamefont {L.}~\bibnamefont {Antiga}},
  \emph {et~al.},\ }\bibfield  {title} {\enquote {\bibinfo {title} {Pytorch: An
  imperative style, high-performance deep learning library},}\ }\href@noop {}
  {\bibfield  {journal} {\bibinfo  {journal} {Advances in neural information
  processing systems}\ }\textbf {\bibinfo {volume} {32}} (\bibinfo {year}
  {2019})}\BibitemShut {NoStop}%
\bibitem [{\citenamefont {Frostig}, \citenamefont {Johnson},\ and\
  \citenamefont {Leary}(2018)}]{Frostig2018}%
  \BibitemOpen
  \bibfield  {author} {\bibinfo {author} {\bibfnamefont {R.}~\bibnamefont
  {Frostig}}, \bibinfo {author} {\bibfnamefont {M.~J.}\ \bibnamefont
  {Johnson}},\ and\ \bibinfo {author} {\bibfnamefont {C.}~\bibnamefont
  {Leary}},\ }\bibfield  {title} {\enquote {\bibinfo {title} {Compiling machine
  learning programs via high-level tracing},}\ }\href@noop {} {\bibfield
  {journal} {\bibinfo  {journal} {Systems for Machine Learning}\ }\textbf
  {\bibinfo {volume} {4}} (\bibinfo {year} {2018})}\BibitemShut {NoStop}%
\bibitem [{\citenamefont {Reed}\ \emph {et~al.}(2022)\citenamefont {Reed},
  \citenamefont {DeVito}, \citenamefont {He}, \citenamefont {Ussery},\ and\
  \citenamefont {Ansel}}]{TorchFx}%
  \BibitemOpen
  \bibfield  {author} {\bibinfo {author} {\bibfnamefont {J.}~\bibnamefont
  {Reed}}, \bibinfo {author} {\bibfnamefont {Z.}~\bibnamefont {DeVito}},
  \bibinfo {author} {\bibfnamefont {H.}~\bibnamefont {He}}, \bibinfo {author}
  {\bibfnamefont {A.}~\bibnamefont {Ussery}},\ and\ \bibinfo {author}
  {\bibfnamefont {J.}~\bibnamefont {Ansel}},\ }\bibfield  {title} {\enquote
  {\bibinfo {title} {torch.fx: Practical program capture and transformation for
  deep learning in python},}\ }in\ \href
  {https://proceedings.mlsys.org/paper_files/paper/2022/file/7c98f9c7ab2df90911da23f9ce72ed6e-Paper.pdf}
  {\emph {\bibinfo {booktitle} {Proceedings of Machine Learning and
  Systems}}},\ Vol.~\bibinfo {volume} {4},\ \bibinfo {editor} {edited by\
  \bibinfo {editor} {\bibfnamefont {D.}~\bibnamefont {Marculescu}}, \bibinfo
  {editor} {\bibfnamefont {Y.}~\bibnamefont {Chi}},\ and\ \bibinfo {editor}
  {\bibfnamefont {C.}~\bibnamefont {Wu}}}\ (\bibinfo {year} {2022})\ pp.\
  \bibinfo {pages} {638--651}\BibitemShut {NoStop}%
\bibitem [{\citenamefont {Tamayo-Mendoza}\ \emph {et~al.}(2018)\citenamefont
  {Tamayo-Mendoza}, \citenamefont {Kreisbeck}, \citenamefont {Lindh},\ and\
  \citenamefont {Aspuru-Guzik}}]{Tamayo2018}%
  \BibitemOpen
  \bibfield  {author} {\bibinfo {author} {\bibfnamefont {T.}~\bibnamefont
  {Tamayo-Mendoza}}, \bibinfo {author} {\bibfnamefont {C.}~\bibnamefont
  {Kreisbeck}}, \bibinfo {author} {\bibfnamefont {R.}~\bibnamefont {Lindh}},\
  and\ \bibinfo {author} {\bibfnamefont {A.}~\bibnamefont {Aspuru-Guzik}},\
  }\bibfield  {title} {\enquote {\bibinfo {title} {Automatic differentiation in
  quantum chemistry with applications to fully variational {Hartree--Fock}},}\
  }\href@noop {} {\bibfield  {journal} {\bibinfo  {journal} {ACS central
  science}\ }\textbf {\bibinfo {volume} {4}},\ \bibinfo {pages} {559--566}
  (\bibinfo {year} {2018})}\BibitemShut {NoStop}%
\bibitem [{\citenamefont {Song}, \citenamefont {Mart{\'\i}nez},\ and\
  \citenamefont {Neaton}(2021)}]{Song2021}%
  \BibitemOpen
  \bibfield  {author} {\bibinfo {author} {\bibfnamefont {C.}~\bibnamefont
  {Song}}, \bibinfo {author} {\bibfnamefont {T.~J.}\ \bibnamefont
  {Mart{\'\i}nez}},\ and\ \bibinfo {author} {\bibfnamefont {J.~B.}\
  \bibnamefont {Neaton}},\ }\bibfield  {title} {\enquote {\bibinfo {title} {A
  diagrammatic approach for automatically deriving analytical gradients of
  tensor hyper-contracted electronic structure methods},}\ }\href@noop {}
  {\bibfield  {journal} {\bibinfo  {journal} {The Journal of Chemical Physics}\
  }\textbf {\bibinfo {volume} {155}},\ \bibinfo {pages} {024108} (\bibinfo
  {year} {2021})}\BibitemShut {NoStop}%
\bibitem [{\citenamefont {Abbott}\ \emph {et~al.}(2021)\citenamefont {Abbott},
  \citenamefont {Abbott}, \citenamefont {Turney},\ and\ \citenamefont
  {Schaefer~III}}]{Abbott2021}%
  \BibitemOpen
  \bibfield  {author} {\bibinfo {author} {\bibfnamefont {A.~S.}\ \bibnamefont
  {Abbott}}, \bibinfo {author} {\bibfnamefont {B.~Z.}\ \bibnamefont {Abbott}},
  \bibinfo {author} {\bibfnamefont {J.~M.}\ \bibnamefont {Turney}},\ and\
  \bibinfo {author} {\bibfnamefont {H.~F.}\ \bibnamefont {Schaefer~III}},\
  }\bibfield  {title} {\enquote {\bibinfo {title} {Arbitrary-order derivatives
  of quantum chemical methods via automatic differentiation},}\ }\href@noop {}
  {\bibfield  {journal} {\bibinfo  {journal} {The journal of physical chemistry
  letters}\ }\textbf {\bibinfo {volume} {12}},\ \bibinfo {pages} {3232--3239}
  (\bibinfo {year} {2021})}\BibitemShut {NoStop}%
\bibitem [{\citenamefont {Kasim}, \citenamefont {Lehtola},\ and\ \citenamefont
  {Vinko}(2022)}]{Kasim2022}%
  \BibitemOpen
  \bibfield  {author} {\bibinfo {author} {\bibfnamefont {M.~F.}\ \bibnamefont
  {Kasim}}, \bibinfo {author} {\bibfnamefont {S.}~\bibnamefont {Lehtola}},\
  and\ \bibinfo {author} {\bibfnamefont {S.~M.}\ \bibnamefont {Vinko}},\
  }\bibfield  {title} {\enquote {\bibinfo {title} {{DQC}: A python program
  package for differentiable quantum chemistry},}\ }\href@noop {} {\bibfield
  {journal} {\bibinfo  {journal} {The Journal of chemical physics}\ }\textbf
  {\bibinfo {volume} {156}} (\bibinfo {year} {2022})}\BibitemShut {NoStop}%
\bibitem [{\citenamefont {Zhang}\ and\ \citenamefont {Chan}(2022)}]{pyscfad}%
  \BibitemOpen
  \bibfield  {author} {\bibinfo {author} {\bibfnamefont {X.}~\bibnamefont
  {Zhang}}\ and\ \bibinfo {author} {\bibfnamefont {G.~K.}\ \bibnamefont
  {Chan}},\ }\bibfield  {title} {\enquote {\bibinfo {title} {{Differentiable
  quantum chemistry with {PySCF} for molecules and materials at the mean-field
  level and beyond}},}\ }\href {https://doi.org/10.1063/5.0118200} {\bibfield
  {journal} {\bibinfo  {journal} {The Journal of Chemical Physics}\ }\textbf
  {\bibinfo {volume} {157}},\ \bibinfo {pages} {204801} (\bibinfo {year}
  {2022})}\BibitemShut {NoStop}%
\bibitem [{\citenamefont {Vargas-Hern{\'a}ndez}\ \emph
  {et~al.}(2023)\citenamefont {Vargas-Hern{\'a}ndez}, \citenamefont {Jorner},
  \citenamefont {Pollice},\ and\ \citenamefont {Aspuru-Guzik}}]{Vargas2023}%
  \BibitemOpen
  \bibfield  {author} {\bibinfo {author} {\bibfnamefont {R.~A.}\ \bibnamefont
  {Vargas-Hern{\'a}ndez}}, \bibinfo {author} {\bibfnamefont {K.}~\bibnamefont
  {Jorner}}, \bibinfo {author} {\bibfnamefont {R.}~\bibnamefont {Pollice}},\
  and\ \bibinfo {author} {\bibfnamefont {A.}~\bibnamefont {Aspuru-Guzik}},\
  }\bibfield  {title} {\enquote {\bibinfo {title} {Inverse molecular design and
  parameter optimization with {H{\"u}ckel} theory using automatic
  differentiation},}\ }\href@noop {} {\bibfield  {journal} {\bibinfo  {journal}
  {The Journal of Chemical Physics}\ }\textbf {\bibinfo {volume} {158}},\
  \bibinfo {pages} {104801} (\bibinfo {year} {2023})}\BibitemShut {NoStop}%
\bibitem [{\citenamefont {Mahajan}\ \emph {et~al.}(2023)\citenamefont
  {Mahajan}, \citenamefont {Kurian}, \citenamefont {Lee}, \citenamefont
  {Reichman},\ and\ \citenamefont {Sharma}}]{Mahajan2023}%
  \BibitemOpen
  \bibfield  {author} {\bibinfo {author} {\bibfnamefont {A.}~\bibnamefont
  {Mahajan}}, \bibinfo {author} {\bibfnamefont {J.~S.}\ \bibnamefont {Kurian}},
  \bibinfo {author} {\bibfnamefont {J.}~\bibnamefont {Lee}}, \bibinfo {author}
  {\bibfnamefont {D.~R.}\ \bibnamefont {Reichman}},\ and\ \bibinfo {author}
  {\bibfnamefont {S.}~\bibnamefont {Sharma}},\ }\bibfield  {title} {\enquote
  {\bibinfo {title} {{Response properties in phaseless auxiliary field quantum
  {Monte Carlo}}},}\ }\href {https://doi.org/10.1063/5.0171996} {\bibfield
  {journal} {\bibinfo  {journal} {The Journal of Chemical Physics}\ }\textbf
  {\bibinfo {volume} {159}},\ \bibinfo {pages} {184101} (\bibinfo {year}
  {2023})}\BibitemShut {NoStop}%
\bibitem [{\citenamefont {M~Casares}\ \emph {et~al.}(2024)\citenamefont
  {M~Casares}, \citenamefont {Baker}, \citenamefont {Medvidovi{\'c}},
  \citenamefont {Reis},\ and\ \citenamefont {Arrazola}}]{MCasares2024}%
  \BibitemOpen
  \bibfield  {author} {\bibinfo {author} {\bibfnamefont {P.~A.}\ \bibnamefont
  {M~Casares}}, \bibinfo {author} {\bibfnamefont {J.~S.}\ \bibnamefont
  {Baker}}, \bibinfo {author} {\bibfnamefont {M.}~\bibnamefont
  {Medvidovi{\'c}}}, \bibinfo {author} {\bibfnamefont {R.~d.}\ \bibnamefont
  {Reis}},\ and\ \bibinfo {author} {\bibfnamefont {J.~M.}\ \bibnamefont
  {Arrazola}},\ }\bibfield  {title} {\enquote {\bibinfo {title} {{GradDFT}. a
  software library for machine learning enhanced density functional theory},}\
  }\href@noop {} {\bibfield  {journal} {\bibinfo  {journal} {The Journal of
  Chemical Physics}\ }\textbf {\bibinfo {volume} {160}},\ \bibinfo {pages}
  {062501} (\bibinfo {year} {2024})}\BibitemShut {NoStop}%
\bibitem [{\citenamefont {Kitaura}\ \emph {et~al.}(1999)\citenamefont
  {Kitaura}, \citenamefont {Ikeo}, \citenamefont {Asada}, \citenamefont
  {Nakano},\ and\ \citenamefont {Uebayasi}}]{FMO}%
  \BibitemOpen
  \bibfield  {author} {\bibinfo {author} {\bibfnamefont {K.}~\bibnamefont
  {Kitaura}}, \bibinfo {author} {\bibfnamefont {E.}~\bibnamefont {Ikeo}},
  \bibinfo {author} {\bibfnamefont {T.}~\bibnamefont {Asada}}, \bibinfo
  {author} {\bibfnamefont {T.}~\bibnamefont {Nakano}},\ and\ \bibinfo {author}
  {\bibfnamefont {M.}~\bibnamefont {Uebayasi}},\ }\bibfield  {title} {\enquote
  {\bibinfo {title} {Fragment molecular orbital method: an approximate
  computational method for large molecules},}\ }\href
  {https://doi.org/https://doi.org/10.1016/S0009-2614(99)00874-X} {\bibfield
  {journal} {\bibinfo  {journal} {Chemical Physics Letters}\ }\textbf {\bibinfo
  {volume} {313}},\ \bibinfo {pages} {701--706} (\bibinfo {year}
  {1999})}\BibitemShut {NoStop}%
\bibitem [{\citenamefont {Sun}\ \emph {et~al.}(2020)\citenamefont {Sun},
  \citenamefont {Zhang}, \citenamefont {Banerjee}, \citenamefont {Bao},
  \citenamefont {Barbry}, \citenamefont {Blunt}, \citenamefont {Bogdanov},
  \citenamefont {Booth}, \citenamefont {Chen}, \citenamefont {Cui},
  \citenamefont {Eriksen}, \citenamefont {Gao}, \citenamefont {Guo},
  \citenamefont {Hermann}, \citenamefont {Hermes}, \citenamefont {Koh},
  \citenamefont {Koval}, \citenamefont {Lehtola}, \citenamefont {Li},
  \citenamefont {Liu}, \citenamefont {Mardirossian}, \citenamefont {McClain},
  \citenamefont {Motta}, \citenamefont {Mussard}, \citenamefont {Pham},
  \citenamefont {Pulkin}, \citenamefont {Purwanto}, \citenamefont {Robinson},
  \citenamefont {Ronca}, \citenamefont {Sayfutyarova}, \citenamefont
  {Scheurer}, \citenamefont {Schurkus}, \citenamefont {Smith}, \citenamefont
  {Sun}, \citenamefont {Sun}, \citenamefont {Upadhyay}, \citenamefont {Wagner},
  \citenamefont {Wang}, \citenamefont {White}, \citenamefont {Whitfield},
  \citenamefont {Williamson}, \citenamefont {Wouters}, \citenamefont {Yang},
  \citenamefont {Yu}, \citenamefont {Zhu}, \citenamefont {Berkelbach},
  \citenamefont {Sharma}, \citenamefont {Sokolov},\ and\ \citenamefont
  {Chan}}]{pyscf}%
  \BibitemOpen
  \bibfield  {author} {\bibinfo {author} {\bibfnamefont {Q.}~\bibnamefont
  {Sun}}, \bibinfo {author} {\bibfnamefont {X.}~\bibnamefont {Zhang}}, \bibinfo
  {author} {\bibfnamefont {S.}~\bibnamefont {Banerjee}}, \bibinfo {author}
  {\bibfnamefont {P.}~\bibnamefont {Bao}}, \bibinfo {author} {\bibfnamefont
  {M.}~\bibnamefont {Barbry}}, \bibinfo {author} {\bibfnamefont {N.~S.}\
  \bibnamefont {Blunt}}, \bibinfo {author} {\bibfnamefont {N.~A.}\ \bibnamefont
  {Bogdanov}}, \bibinfo {author} {\bibfnamefont {G.~H.}\ \bibnamefont {Booth}},
  \bibinfo {author} {\bibfnamefont {J.}~\bibnamefont {Chen}}, \bibinfo {author}
  {\bibfnamefont {Z.}~\bibnamefont {Cui}}, \bibinfo {author} {\bibfnamefont
  {J.~J.}\ \bibnamefont {Eriksen}}, \bibinfo {author} {\bibfnamefont
  {Y.}~\bibnamefont {Gao}}, \bibinfo {author} {\bibfnamefont {S.}~\bibnamefont
  {Guo}}, \bibinfo {author} {\bibfnamefont {J.}~\bibnamefont {Hermann}},
  \bibinfo {author} {\bibfnamefont {M.~R.}\ \bibnamefont {Hermes}}, \bibinfo
  {author} {\bibfnamefont {K.}~\bibnamefont {Koh}}, \bibinfo {author}
  {\bibfnamefont {P.}~\bibnamefont {Koval}}, \bibinfo {author} {\bibfnamefont
  {S.}~\bibnamefont {Lehtola}}, \bibinfo {author} {\bibfnamefont
  {Z.}~\bibnamefont {Li}}, \bibinfo {author} {\bibfnamefont {J.}~\bibnamefont
  {Liu}}, \bibinfo {author} {\bibfnamefont {N.}~\bibnamefont {Mardirossian}},
  \bibinfo {author} {\bibfnamefont {J.~D.}\ \bibnamefont {McClain}}, \bibinfo
  {author} {\bibfnamefont {M.}~\bibnamefont {Motta}}, \bibinfo {author}
  {\bibfnamefont {B.}~\bibnamefont {Mussard}}, \bibinfo {author} {\bibfnamefont
  {H.~Q.}\ \bibnamefont {Pham}}, \bibinfo {author} {\bibfnamefont
  {A.}~\bibnamefont {Pulkin}}, \bibinfo {author} {\bibfnamefont
  {W.}~\bibnamefont {Purwanto}}, \bibinfo {author} {\bibfnamefont {P.~J.}\
  \bibnamefont {Robinson}}, \bibinfo {author} {\bibfnamefont {E.}~\bibnamefont
  {Ronca}}, \bibinfo {author} {\bibfnamefont {E.~R.}\ \bibnamefont
  {Sayfutyarova}}, \bibinfo {author} {\bibfnamefont {M.}~\bibnamefont
  {Scheurer}}, \bibinfo {author} {\bibfnamefont {H.~F.}\ \bibnamefont
  {Schurkus}}, \bibinfo {author} {\bibfnamefont {J.~E.~T.}\ \bibnamefont
  {Smith}}, \bibinfo {author} {\bibfnamefont {C.}~\bibnamefont {Sun}}, \bibinfo
  {author} {\bibfnamefont {S.}~\bibnamefont {Sun}}, \bibinfo {author}
  {\bibfnamefont {S.}~\bibnamefont {Upadhyay}}, \bibinfo {author}
  {\bibfnamefont {L.~K.}\ \bibnamefont {Wagner}}, \bibinfo {author}
  {\bibfnamefont {X.}~\bibnamefont {Wang}}, \bibinfo {author} {\bibfnamefont
  {A.}~\bibnamefont {White}}, \bibinfo {author} {\bibfnamefont {J.~D.}\
  \bibnamefont {Whitfield}}, \bibinfo {author} {\bibfnamefont {M.~J.}\
  \bibnamefont {Williamson}}, \bibinfo {author} {\bibfnamefont
  {S.}~\bibnamefont {Wouters}}, \bibinfo {author} {\bibfnamefont
  {J.}~\bibnamefont {Yang}}, \bibinfo {author} {\bibfnamefont {J.~M.}\
  \bibnamefont {Yu}}, \bibinfo {author} {\bibfnamefont {T.}~\bibnamefont
  {Zhu}}, \bibinfo {author} {\bibfnamefont {T.~C.}\ \bibnamefont {Berkelbach}},
  \bibinfo {author} {\bibfnamefont {S.}~\bibnamefont {Sharma}}, \bibinfo
  {author} {\bibfnamefont {A.~Y.}\ \bibnamefont {Sokolov}},\ and\ \bibinfo
  {author} {\bibfnamefont {G.~K.-L.}\ \bibnamefont {Chan}},\ }\bibfield
  {title} {\enquote {\bibinfo {title} {{Recent developments in the {PySCF}
  program package}},}\ }\href {https://doi.org/10.1063/5.0006074} {\bibfield
  {journal} {\bibinfo  {journal} {The Journal of Chemical Physics}\ }\textbf
  {\bibinfo {volume} {153}},\ \bibinfo {pages} {024109} (\bibinfo {year}
  {2020})}\BibitemShut {NoStop}%
\bibitem [{\citenamefont {Dalc{\'\i}n}, \citenamefont {Paz},\ and\
  \citenamefont {Storti}(2005)}]{mpi4py}%
  \BibitemOpen
  \bibfield  {author} {\bibinfo {author} {\bibfnamefont {L.}~\bibnamefont
  {Dalc{\'\i}n}}, \bibinfo {author} {\bibfnamefont {R.}~\bibnamefont {Paz}},\
  and\ \bibinfo {author} {\bibfnamefont {M.}~\bibnamefont {Storti}},\
  }\bibfield  {title} {\enquote {\bibinfo {title} {{MPI} for {Python}},}\
  }\href@noop {} {\bibfield  {journal} {\bibinfo  {journal} {Journal of
  Parallel and Distributed Computing}\ }\textbf {\bibinfo {volume} {65}},\
  \bibinfo {pages} {1108--1115} (\bibinfo {year} {2005})}\BibitemShut {NoStop}%
\bibitem [{\citenamefont {Scheurer}\ and\ \citenamefont
  {Schwarz}(2000{\natexlab{a}})}]{Scheurer2000a}%
  \BibitemOpen
  \bibfield  {author} {\bibinfo {author} {\bibfnamefont {P.}~\bibnamefont
  {Scheurer}}\ and\ \bibinfo {author} {\bibfnamefont {W.}~\bibnamefont
  {Schwarz}},\ }\bibfield  {title} {\enquote {\bibinfo {title} {Externally
  localized molecular orbitals: A numerical investigation of localization
  degeneracy},}\ }\href@noop {} {\bibfield  {journal} {\bibinfo  {journal}
  {International Journal of Quantum Chemistry}\ }\textbf {\bibinfo {volume}
  {76}},\ \bibinfo {pages} {420--427} (\bibinfo {year}
  {2000}{\natexlab{a}})}\BibitemShut {NoStop}%
\bibitem [{\citenamefont {Scheurer}\ and\ \citenamefont
  {Schwarz}(2000{\natexlab{b}})}]{Scheurer2000b}%
  \BibitemOpen
  \bibfield  {author} {\bibinfo {author} {\bibfnamefont {P.}~\bibnamefont
  {Scheurer}}\ and\ \bibinfo {author} {\bibfnamefont {W.}~\bibnamefont
  {Schwarz}},\ }\bibfield  {title} {\enquote {\bibinfo {title} {Continuous
  degeneracy of sets of localized orbitals},}\ }\href@noop {} {\bibfield
  {journal} {\bibinfo  {journal} {International Journal of Quantum Chemistry}\
  }\textbf {\bibinfo {volume} {76}},\ \bibinfo {pages} {428--433} (\bibinfo
  {year} {2000}{\natexlab{b}})}\BibitemShut {NoStop}%
\bibitem [{\citenamefont {Whitten}(1973)}]{Whitten1973}%
  \BibitemOpen
  \bibfield  {author} {\bibinfo {author} {\bibfnamefont {J.~L.}\ \bibnamefont
  {Whitten}},\ }\bibfield  {title} {\enquote {\bibinfo {title} {Coulombic
  potential energy integrals and approximations},}\ }\href@noop {} {\bibfield
  {journal} {\bibinfo  {journal} {The Journal of Chemical Physics}\ }\textbf
  {\bibinfo {volume} {58}},\ \bibinfo {pages} {4496--4501} (\bibinfo {year}
  {1973})}\BibitemShut {NoStop}%
\bibitem [{\citenamefont {Dunlap}, \citenamefont {Connolly},\ and\
  \citenamefont {Sabin}(1979)}]{Dunlap1979}%
  \BibitemOpen
  \bibfield  {author} {\bibinfo {author} {\bibfnamefont {B.~I.}\ \bibnamefont
  {Dunlap}}, \bibinfo {author} {\bibfnamefont {J.}~\bibnamefont {Connolly}},\
  and\ \bibinfo {author} {\bibfnamefont {J.}~\bibnamefont {Sabin}},\ }\bibfield
   {title} {\enquote {\bibinfo {title} {On some approximations in applications
  of {X$\alpha$} theory},}\ }\href@noop {} {\bibfield  {journal} {\bibinfo
  {journal} {The Journal of Chemical Physics}\ }\textbf {\bibinfo {volume}
  {71}},\ \bibinfo {pages} {3396--3402} (\bibinfo {year} {1979})}\BibitemShut
  {NoStop}%
\bibitem [{\citenamefont {Pipek}\ and\ \citenamefont
  {Mezey}(1989)}]{pipek1989fast}%
  \BibitemOpen
  \bibfield  {author} {\bibinfo {author} {\bibfnamefont {J.}~\bibnamefont
  {Pipek}}\ and\ \bibinfo {author} {\bibfnamefont {P.~G.}\ \bibnamefont
  {Mezey}},\ }\bibfield  {title} {\enquote {\bibinfo {title} {A fast intrinsic
  localization procedure applicable for ab initio and semiempirical linear
  combination of atomic orbital wave functions},}\ }\href@noop {} {\bibfield
  {journal} {\bibinfo  {journal} {The Journal of Chemical Physics}\ }\textbf
  {\bibinfo {volume} {90}},\ \bibinfo {pages} {4916--4926} (\bibinfo {year}
  {1989})}\BibitemShut {NoStop}%
\bibitem [{\citenamefont {Baker}(1993)}]{Baker}%
  \BibitemOpen
  \bibfield  {author} {\bibinfo {author} {\bibfnamefont {J.}~\bibnamefont
  {Baker}},\ }\bibfield  {title} {\enquote {\bibinfo {title} {Techniques for
  geometry optimization: A comparison of cartesian and natural internal
  coordinates},}\ }\href
  {https://doi.org/https://doi.org/10.1002/jcc.540140910} {\bibfield  {journal}
  {\bibinfo  {journal} {Journal of Computational Chemistry}\ }\textbf {\bibinfo
  {volume} {14}},\ \bibinfo {pages} {1085--1100} (\bibinfo {year}
  {1993})}\BibitemShut {NoStop}%
\bibitem [{\citenamefont {Knizia}(2013)}]{IAO}%
  \BibitemOpen
  \bibfield  {author} {\bibinfo {author} {\bibfnamefont {G.}~\bibnamefont
  {Knizia}},\ }\bibfield  {title} {\enquote {\bibinfo {title} {Intrinsic atomic
  orbitals: An unbiased bridge between quantum theory and chemical concepts},}\
  }\href@noop {} {\bibfield  {journal} {\bibinfo  {journal} {Journal of
  chemical theory and computation}\ }\textbf {\bibinfo {volume} {9}},\ \bibinfo
  {pages} {4834--4843} (\bibinfo {year} {2013})}\BibitemShut {NoStop}%
\bibitem [{\citenamefont {Dunning~Jr}(1989)}]{ccpvtz}%
  \BibitemOpen
  \bibfield  {author} {\bibinfo {author} {\bibfnamefont {T.~H.}\ \bibnamefont
  {Dunning~Jr}},\ }\bibfield  {title} {\enquote {\bibinfo {title} {Gaussian
  basis sets for use in correlated molecular calculations. {I}. the atoms boron
  through neon and hydrogen},}\ }\href@noop {} {\bibfield  {journal} {\bibinfo
  {journal} {The Journal of chemical physics}\ }\textbf {\bibinfo {volume}
  {90}},\ \bibinfo {pages} {1007--1023} (\bibinfo {year} {1989})}\BibitemShut
  {NoStop}%
\bibitem [{\citenamefont {Wang}\ and\ \citenamefont {Song}(2016)}]{geometric}%
  \BibitemOpen
  \bibfield  {author} {\bibinfo {author} {\bibfnamefont {L.-P.}\ \bibnamefont
  {Wang}}\ and\ \bibinfo {author} {\bibfnamefont {C.}~\bibnamefont {Song}},\
  }\bibfield  {title} {\enquote {\bibinfo {title} {Geometry optimization made
  simple with translation and rotation coordinates},}\ }\href@noop {}
  {\bibfield  {journal} {\bibinfo  {journal} {The Journal of chemical physics}\
  }\textbf {\bibinfo {volume} {144}} (\bibinfo {year} {2016})}\BibitemShut
  {NoStop}%
\bibitem [{\citenamefont {Shomura}\ \emph {et~al.}(2017)\citenamefont
  {Shomura}, \citenamefont {Taketa}, \citenamefont {Nakashima}, \citenamefont
  {Tai}, \citenamefont {Nakagawa}, \citenamefont {Ikeda}, \citenamefont
  {Ishii}, \citenamefont {Igarashi}, \citenamefont {Nishihara}, \citenamefont
  {Yoon}, \citenamefont {Ogo}, \citenamefont {Hirota},\ and\ \citenamefont
  {Higuchi}}]{Shomura2017}%
  \BibitemOpen
  \bibfield  {author} {\bibinfo {author} {\bibfnamefont {Y.}~\bibnamefont
  {Shomura}}, \bibinfo {author} {\bibfnamefont {M.}~\bibnamefont {Taketa}},
  \bibinfo {author} {\bibfnamefont {H.}~\bibnamefont {Nakashima}}, \bibinfo
  {author} {\bibfnamefont {H.}~\bibnamefont {Tai}}, \bibinfo {author}
  {\bibfnamefont {H.}~\bibnamefont {Nakagawa}}, \bibinfo {author}
  {\bibfnamefont {Y.}~\bibnamefont {Ikeda}}, \bibinfo {author} {\bibfnamefont
  {M.}~\bibnamefont {Ishii}}, \bibinfo {author} {\bibfnamefont
  {Y.}~\bibnamefont {Igarashi}}, \bibinfo {author} {\bibfnamefont
  {H.}~\bibnamefont {Nishihara}}, \bibinfo {author} {\bibfnamefont {K.-S.}\
  \bibnamefont {Yoon}}, \bibinfo {author} {\bibfnamefont {S.}~\bibnamefont
  {Ogo}}, \bibinfo {author} {\bibfnamefont {S.}~\bibnamefont {Hirota}},\ and\
  \bibinfo {author} {\bibfnamefont {Y.}~\bibnamefont {Higuchi}},\ }\bibfield
  {title} {\enquote {\bibinfo {title} {Structural basis of the redox switches
  in the {NAD$^+$}-reducing soluble {[NiFe]}-hydrogenase},}\ }\href
  {https://doi.org/10.1126/science.aan4497} {\bibfield  {journal} {\bibinfo
  {journal} {Science}\ }\textbf {\bibinfo {volume} {357}},\ \bibinfo {pages}
  {928--932} (\bibinfo {year} {2017})}\BibitemShut {NoStop}%
\bibitem [{\citenamefont {Preissler}\ \emph {et~al.}(2018)\citenamefont
  {Preissler}, \citenamefont {Wahlefeld}, \citenamefont {Lorent}, \citenamefont
  {Teutloff}, \citenamefont {Horch}, \citenamefont {Lauterbach}, \citenamefont
  {Cramer}, \citenamefont {Zebger},\ and\ \citenamefont
  {Lenz}}]{Preissler2018}%
  \BibitemOpen
  \bibfield  {author} {\bibinfo {author} {\bibfnamefont {J.}~\bibnamefont
  {Preissler}}, \bibinfo {author} {\bibfnamefont {S.}~\bibnamefont
  {Wahlefeld}}, \bibinfo {author} {\bibfnamefont {C.}~\bibnamefont {Lorent}},
  \bibinfo {author} {\bibfnamefont {C.}~\bibnamefont {Teutloff}}, \bibinfo
  {author} {\bibfnamefont {M.}~\bibnamefont {Horch}}, \bibinfo {author}
  {\bibfnamefont {L.}~\bibnamefont {Lauterbach}}, \bibinfo {author}
  {\bibfnamefont {S.~P.}\ \bibnamefont {Cramer}}, \bibinfo {author}
  {\bibfnamefont {I.}~\bibnamefont {Zebger}},\ and\ \bibinfo {author}
  {\bibfnamefont {O.}~\bibnamefont {Lenz}},\ }\bibfield  {title} {\enquote
  {\bibinfo {title} {Enzymatic and spectroscopic properties of a thermostable
  {[NiFe]}‑hydrogenase performing {H$_2$}-driven {NAD$^+$}-reduction in the
  presence of {O$_2$}},}\ }\href
  {https://doi.org/https://doi.org/10.1016/j.bbabio.2017.09.006} {\bibfield
  {journal} {\bibinfo  {journal} {Biochimica et Biophysica Acta (BBA) -
  Bioenergetics}\ }\textbf {\bibinfo {volume} {1859}},\ \bibinfo {pages}
  {8--18} (\bibinfo {year} {2018})}\BibitemShut {NoStop}%
\bibitem [{\citenamefont {Kulka-Peschke}\ \emph {et~al.}(2022)\citenamefont
  {Kulka-Peschke}, \citenamefont {Schulz}, \citenamefont {Lorent},
  \citenamefont {Rippers}, \citenamefont {Wahlefeld}, \citenamefont
  {Preissler}, \citenamefont {Schulz}, \citenamefont {Wiemann}, \citenamefont
  {Bernitzky}, \citenamefont {Karafoulidi-Retsou}, \citenamefont {Wrathall},
  \citenamefont {Procacci}, \citenamefont {Matsuura}, \citenamefont {Greetham},
  \citenamefont {Teutloff}, \citenamefont {Lauterbach}, \citenamefont
  {Higuchi}, \citenamefont {Ishii}, \citenamefont {Hunt}, \citenamefont {Lenz},
  \citenamefont {Zebger},\ and\ \citenamefont {Horch}}]{Kulka-Peschke2022}%
  \BibitemOpen
  \bibfield  {author} {\bibinfo {author} {\bibfnamefont {C.~J.}\ \bibnamefont
  {Kulka-Peschke}}, \bibinfo {author} {\bibfnamefont {A.-C.}\ \bibnamefont
  {Schulz}}, \bibinfo {author} {\bibfnamefont {C.}~\bibnamefont {Lorent}},
  \bibinfo {author} {\bibfnamefont {Y.}~\bibnamefont {Rippers}}, \bibinfo
  {author} {\bibfnamefont {S.}~\bibnamefont {Wahlefeld}}, \bibinfo {author}
  {\bibfnamefont {J.}~\bibnamefont {Preissler}}, \bibinfo {author}
  {\bibfnamefont {C.}~\bibnamefont {Schulz}}, \bibinfo {author} {\bibfnamefont
  {C.}~\bibnamefont {Wiemann}}, \bibinfo {author} {\bibfnamefont {C.~C.~M.}\
  \bibnamefont {Bernitzky}}, \bibinfo {author} {\bibfnamefont {C.}~\bibnamefont
  {Karafoulidi-Retsou}}, \bibinfo {author} {\bibfnamefont {S.~L.~D.}\
  \bibnamefont {Wrathall}}, \bibinfo {author} {\bibfnamefont {B.}~\bibnamefont
  {Procacci}}, \bibinfo {author} {\bibfnamefont {H.}~\bibnamefont {Matsuura}},
  \bibinfo {author} {\bibfnamefont {G.~M.}\ \bibnamefont {Greetham}}, \bibinfo
  {author} {\bibfnamefont {C.}~\bibnamefont {Teutloff}}, \bibinfo {author}
  {\bibfnamefont {L.}~\bibnamefont {Lauterbach}}, \bibinfo {author}
  {\bibfnamefont {Y.}~\bibnamefont {Higuchi}}, \bibinfo {author} {\bibfnamefont
  {M.}~\bibnamefont {Ishii}}, \bibinfo {author} {\bibfnamefont {N.~T.}\
  \bibnamefont {Hunt}}, \bibinfo {author} {\bibfnamefont {O.}~\bibnamefont
  {Lenz}}, \bibinfo {author} {\bibfnamefont {I.}~\bibnamefont {Zebger}},\ and\
  \bibinfo {author} {\bibfnamefont {M.}~\bibnamefont {Horch}},\ }\bibfield
  {title} {\enquote {\bibinfo {title} {Reversible glutamate coordination to
  high-valent nickel protects the active site of a {[NiFe]} hydrogenase from
  oxygen},}\ }\href {https://doi.org/10.1021/jacs.2c06400} {\bibfield
  {journal} {\bibinfo  {journal} {Journal of the American Chemical Society}\
  }\textbf {\bibinfo {volume} {144}},\ \bibinfo {pages} {17022--17032}
  (\bibinfo {year} {2022})}\BibitemShut {NoStop}%
\bibitem [{\citenamefont {Kumar}\ and\ \citenamefont
  {Stein}(2023)}]{Kumar2023}%
  \BibitemOpen
  \bibfield  {author} {\bibinfo {author} {\bibfnamefont {R.}~\bibnamefont
  {Kumar}}\ and\ \bibinfo {author} {\bibfnamefont {M.}~\bibnamefont {Stein}},\
  }\bibfield  {title} {\enquote {\bibinfo {title} {The fully oxidized state of
  the glutamate coordinated {O$_2$}-tolerant [nife]-hydrogenase shows a
  {Ni(III)/Fe(III)} open-shell singlet ground state},}\ }\href
  {https://doi.org/10.1021/jacs.3c02438} {\bibfield  {journal} {\bibinfo
  {journal} {Journal of the American Chemical Society}\ }\textbf {\bibinfo
  {volume} {145}},\ \bibinfo {pages} {10954--10959} (\bibinfo {year}
  {2023})}\BibitemShut {NoStop}%
\bibitem [{\citenamefont {Staroverov}\ \emph {et~al.}(2003)\citenamefont
  {Staroverov}, \citenamefont {Scuseria}, \citenamefont {Tao},\ and\
  \citenamefont {Perdew}}]{TPSSH}%
  \BibitemOpen
  \bibfield  {author} {\bibinfo {author} {\bibfnamefont {V.~N.}\ \bibnamefont
  {Staroverov}}, \bibinfo {author} {\bibfnamefont {G.~E.}\ \bibnamefont
  {Scuseria}}, \bibinfo {author} {\bibfnamefont {J.}~\bibnamefont {Tao}},\ and\
  \bibinfo {author} {\bibfnamefont {J.~P.}\ \bibnamefont {Perdew}},\ }\bibfield
   {title} {\enquote {\bibinfo {title} {Comparative assessment of a new
  nonempirical density functional: Molecules and hydrogen-bonded complexes},}\
  }\href@noop {} {\bibfield  {journal} {\bibinfo  {journal} {The Journal of
  chemical physics}\ }\textbf {\bibinfo {volume} {119}},\ \bibinfo {pages}
  {12129--12137} (\bibinfo {year} {2003})}\BibitemShut {NoStop}%
\bibitem [{\citenamefont {Weigend}\ and\ \citenamefont
  {Ahlrichs}(2005)}]{def2TZVP}%
  \BibitemOpen
  \bibfield  {author} {\bibinfo {author} {\bibfnamefont {F.}~\bibnamefont
  {Weigend}}\ and\ \bibinfo {author} {\bibfnamefont {R.}~\bibnamefont
  {Ahlrichs}},\ }\bibfield  {title} {\enquote {\bibinfo {title} {Balanced basis
  sets of split valence, triple zeta valence and quadruple zeta valence quality
  for {H} to {Rn}: Design and assessment of accuracy},}\ }\href@noop {}
  {\bibfield  {journal} {\bibinfo  {journal} {Physical Chemistry Chemical
  Physics}\ }\textbf {\bibinfo {volume} {7}},\ \bibinfo {pages} {3297--3305}
  (\bibinfo {year} {2005})}\BibitemShut {NoStop}%
\bibitem [{\citenamefont {Hanwell}\ \emph {et~al.}(2012)\citenamefont
  {Hanwell}, \citenamefont {Curtis}, \citenamefont {Lonie}, \citenamefont
  {Vandermeersch}, \citenamefont {Zurek},\ and\ \citenamefont
  {Hutchison}}]{hanwell2012avogadro}%
  \BibitemOpen
  \bibfield  {author} {\bibinfo {author} {\bibfnamefont {M.~D.}\ \bibnamefont
  {Hanwell}}, \bibinfo {author} {\bibfnamefont {D.~E.}\ \bibnamefont {Curtis}},
  \bibinfo {author} {\bibfnamefont {D.~C.}\ \bibnamefont {Lonie}}, \bibinfo
  {author} {\bibfnamefont {T.}~\bibnamefont {Vandermeersch}}, \bibinfo {author}
  {\bibfnamefont {E.}~\bibnamefont {Zurek}},\ and\ \bibinfo {author}
  {\bibfnamefont {G.~R.}\ \bibnamefont {Hutchison}},\ }\bibfield  {title}
  {\enquote {\bibinfo {title} {Avogadro: an advanced semantic chemical editor,
  visualization, and analysis platform},}\ }\href@noop {} {\bibfield  {journal}
  {\bibinfo  {journal} {Journal of cheminformatics}\ }\textbf {\bibinfo
  {volume} {4}},\ \bibinfo {pages} {1--17} (\bibinfo {year}
  {2012})}\BibitemShut {NoStop}%
\bibitem [{\citenamefont {Chai}\ and\ \citenamefont
  {Head-Gordon}(2008)}]{wB97X}%
  \BibitemOpen
  \bibfield  {author} {\bibinfo {author} {\bibfnamefont {J.-D.}\ \bibnamefont
  {Chai}}\ and\ \bibinfo {author} {\bibfnamefont {M.}~\bibnamefont
  {Head-Gordon}},\ }\bibfield  {title} {\enquote {\bibinfo {title} {Systematic
  optimization of long-range corrected hybrid density functionals},}\
  }\href@noop {} {\bibfield  {journal} {\bibinfo  {journal} {The Journal of
  chemical physics}\ }\textbf {\bibinfo {volume} {128}},\ \bibinfo {pages}
  {084106} (\bibinfo {year} {2008})}\BibitemShut {NoStop}%
\bibitem [{\citenamefont {Li}\ and\ \citenamefont {Voth}(2022)}]{li2022using}%
  \BibitemOpen
  \bibfield  {author} {\bibinfo {author} {\bibfnamefont {C.}~\bibnamefont
  {Li}}\ and\ \bibinfo {author} {\bibfnamefont {G.~A.}\ \bibnamefont {Voth}},\
  }\bibfield  {title} {\enquote {\bibinfo {title} {Using machine learning to
  greatly accelerate path integral ab initio molecular dynamics},}\ }\href@noop
  {} {\bibfield  {journal} {\bibinfo  {journal} {Journal of Chemical Theory and
  Computation}\ }\textbf {\bibinfo {volume} {18}},\ \bibinfo {pages} {599--604}
  (\bibinfo {year} {2022})}\BibitemShut {NoStop}%
\bibitem [{\citenamefont {Musaelian}\ \emph {et~al.}(2023)\citenamefont
  {Musaelian}, \citenamefont {Batzner}, \citenamefont {Johansson},
  \citenamefont {Sun}, \citenamefont {Owen}, \citenamefont {Kornbluth},\ and\
  \citenamefont {Kozinsky}}]{musaelian2023learning}%
  \BibitemOpen
  \bibfield  {author} {\bibinfo {author} {\bibfnamefont {A.}~\bibnamefont
  {Musaelian}}, \bibinfo {author} {\bibfnamefont {S.}~\bibnamefont {Batzner}},
  \bibinfo {author} {\bibfnamefont {A.}~\bibnamefont {Johansson}}, \bibinfo
  {author} {\bibfnamefont {L.}~\bibnamefont {Sun}}, \bibinfo {author}
  {\bibfnamefont {C.~J.}\ \bibnamefont {Owen}}, \bibinfo {author}
  {\bibfnamefont {M.}~\bibnamefont {Kornbluth}},\ and\ \bibinfo {author}
  {\bibfnamefont {B.}~\bibnamefont {Kozinsky}},\ }\bibfield  {title} {\enquote
  {\bibinfo {title} {Learning local equivariant representations for large-scale
  atomistic dynamics},}\ }\href@noop {} {\bibfield  {journal} {\bibinfo
  {journal} {Nature Communications}\ }\textbf {\bibinfo {volume} {14}},\
  \bibinfo {pages} {579} (\bibinfo {year} {2023})}\BibitemShut {NoStop}%
\bibitem [{\citenamefont {Heine}\ \emph {et~al.}(2013)\citenamefont {Heine},
  \citenamefont {Fagiani}, \citenamefont {Rossi}, \citenamefont {Wende},
  \citenamefont {Berden}, \citenamefont {Blum},\ and\ \citenamefont
  {Asmis}}]{heine2013isomer}%
  \BibitemOpen
  \bibfield  {author} {\bibinfo {author} {\bibfnamefont {N.}~\bibnamefont
  {Heine}}, \bibinfo {author} {\bibfnamefont {M.~R.}\ \bibnamefont {Fagiani}},
  \bibinfo {author} {\bibfnamefont {M.}~\bibnamefont {Rossi}}, \bibinfo
  {author} {\bibfnamefont {T.}~\bibnamefont {Wende}}, \bibinfo {author}
  {\bibfnamefont {G.}~\bibnamefont {Berden}}, \bibinfo {author} {\bibfnamefont
  {V.}~\bibnamefont {Blum}},\ and\ \bibinfo {author} {\bibfnamefont {K.~R.}\
  \bibnamefont {Asmis}},\ }\bibfield  {title} {\enquote {\bibinfo {title}
  {Isomer-selective detection of hydrogen-bond vibrations in the protonated
  water hexamer},}\ }\href@noop {} {\bibfield  {journal} {\bibinfo  {journal}
  {Journal of the American Chemical Society}\ }\textbf {\bibinfo {volume}
  {135}},\ \bibinfo {pages} {8266--8273} (\bibinfo {year} {2013})}\BibitemShut
  {NoStop}%
\bibitem [{\citenamefont {Rossi}, \citenamefont {Ceriotti},\ and\ \citenamefont
  {Manolopoulos}(2014)}]{rossi2014remove}%
  \BibitemOpen
  \bibfield  {author} {\bibinfo {author} {\bibfnamefont {M.}~\bibnamefont
  {Rossi}}, \bibinfo {author} {\bibfnamefont {M.}~\bibnamefont {Ceriotti}},\
  and\ \bibinfo {author} {\bibfnamefont {D.~E.}\ \bibnamefont {Manolopoulos}},\
  }\bibfield  {title} {\enquote {\bibinfo {title} {How to remove the spurious
  resonances from ring polymer molecular dynamics},}\ }\href@noop {} {\bibfield
   {journal} {\bibinfo  {journal} {The Journal of chemical physics}\ }\textbf
  {\bibinfo {volume} {140}},\ \bibinfo {pages} {234116} (\bibinfo {year}
  {2014})}\BibitemShut {NoStop}%
\bibitem [{\citenamefont {Paesani}\ and\ \citenamefont
  {Voth}(2010)}]{paesani2010quantitative}%
  \BibitemOpen
  \bibfield  {author} {\bibinfo {author} {\bibfnamefont {F.}~\bibnamefont
  {Paesani}}\ and\ \bibinfo {author} {\bibfnamefont {G.~A.}\ \bibnamefont
  {Voth}},\ }\bibfield  {title} {\enquote {\bibinfo {title} {A quantitative
  assessment of the accuracy of centroid molecular dynamics for the calculation
  of the infrared spectrum of liquid water},}\ }\href@noop {} {\bibfield
  {journal} {\bibinfo  {journal} {The Journal of chemical physics}\ }\textbf
  {\bibinfo {volume} {132}},\ \bibinfo {pages} {014105} (\bibinfo {year}
  {2010})}\BibitemShut {NoStop}%
\bibitem [{\citenamefont {Kjaergaard}\ \emph {et~al.}(2008)\citenamefont
  {Kjaergaard}, \citenamefont {Garden}, \citenamefont {Chaban}, \citenamefont
  {Gerber}, \citenamefont {Matthews},\ and\ \citenamefont
  {Stanton}}]{kjaergaard2008calculation}%
  \BibitemOpen
  \bibfield  {author} {\bibinfo {author} {\bibfnamefont {H.~G.}\ \bibnamefont
  {Kjaergaard}}, \bibinfo {author} {\bibfnamefont {A.~L.}\ \bibnamefont
  {Garden}}, \bibinfo {author} {\bibfnamefont {G.~M.}\ \bibnamefont {Chaban}},
  \bibinfo {author} {\bibfnamefont {R.~B.}\ \bibnamefont {Gerber}}, \bibinfo
  {author} {\bibfnamefont {D.~A.}\ \bibnamefont {Matthews}},\ and\ \bibinfo
  {author} {\bibfnamefont {J.~F.}\ \bibnamefont {Stanton}},\ }\bibfield
  {title} {\enquote {\bibinfo {title} {Calculation of vibrational transition
  frequencies and intensities in water dimer: Comparison of different
  vibrational approaches},}\ }\href@noop {} {\bibfield  {journal} {\bibinfo
  {journal} {The Journal of Physical Chemistry A}\ }\textbf {\bibinfo {volume}
  {112}},\ \bibinfo {pages} {4324--4335} (\bibinfo {year} {2008})}\BibitemShut
  {NoStop}%
\bibitem [{\citenamefont {Yu}\ and\ \citenamefont
  {Bowman}(2019)}]{yu2019classical}%
  \BibitemOpen
  \bibfield  {author} {\bibinfo {author} {\bibfnamefont {Q.}~\bibnamefont
  {Yu}}\ and\ \bibinfo {author} {\bibfnamefont {J.~M.}\ \bibnamefont
  {Bowman}},\ }\bibfield  {title} {\enquote {\bibinfo {title} {Classical,
  thermostated ring polymer, and quantum {VSCF/VCI} calculations of {IR}
  spectra of {$\text{H}_7\text{O}_3^+$} and {$\text{H}_9\text{O}_4^+$}
  ({Eigen}) and comparison with experiment},}\ }\href@noop {} {\bibfield
  {journal} {\bibinfo  {journal} {The Journal of Physical Chemistry A}\
  }\textbf {\bibinfo {volume} {123}},\ \bibinfo {pages} {1399--1409} (\bibinfo
  {year} {2019})}\BibitemShut {NoStop}%
\bibitem [{\citenamefont {Carpenter}\ \emph {et~al.}(2020)\citenamefont
  {Carpenter}, \citenamefont {Yu}, \citenamefont {Hack}, \citenamefont
  {Dereka}, \citenamefont {Bowman},\ and\ \citenamefont
  {Tokmakoff}}]{carpenter2020decoding}%
  \BibitemOpen
  \bibfield  {author} {\bibinfo {author} {\bibfnamefont {W.~B.}\ \bibnamefont
  {Carpenter}}, \bibinfo {author} {\bibfnamefont {Q.}~\bibnamefont {Yu}},
  \bibinfo {author} {\bibfnamefont {J.~H.}\ \bibnamefont {Hack}}, \bibinfo
  {author} {\bibfnamefont {B.}~\bibnamefont {Dereka}}, \bibinfo {author}
  {\bibfnamefont {J.~M.}\ \bibnamefont {Bowman}},\ and\ \bibinfo {author}
  {\bibfnamefont {A.}~\bibnamefont {Tokmakoff}},\ }\bibfield  {title} {\enquote
  {\bibinfo {title} {Decoding the {2D} {IR} spectrum of the aqueous proton with
  high-level {VSCF/VCI} calculations},}\ }\href@noop {} {\bibfield  {journal}
  {\bibinfo  {journal} {The Journal of Chemical Physics}\ }\textbf {\bibinfo
  {volume} {153}},\ \bibinfo {pages} {124506} (\bibinfo {year}
  {2020})}\BibitemShut {NoStop}%
\bibitem [{\citenamefont {Christiansen}, \citenamefont {J{\o}rgensen},\ and\
  \citenamefont {H{\"a}ttig}(1998)}]{Christiansen1998}%
  \BibitemOpen
  \bibfield  {author} {\bibinfo {author} {\bibfnamefont {O.}~\bibnamefont
  {Christiansen}}, \bibinfo {author} {\bibfnamefont {P.}~\bibnamefont
  {J{\o}rgensen}},\ and\ \bibinfo {author} {\bibfnamefont {C.}~\bibnamefont
  {H{\"a}ttig}},\ }\bibfield  {title} {\enquote {\bibinfo {title} {Response
  functions from fourier component variational perturbation theory applied to a
  time-averaged quasienergy},}\ }\href@noop {} {\bibfield  {journal} {\bibinfo
  {journal} {International Journal of Quantum Chemistry}\ }\textbf {\bibinfo
  {volume} {68}},\ \bibinfo {pages} {1--52} (\bibinfo {year}
  {1998})}\BibitemShut {NoStop}%
\end{thebibliography}%


\begin{thebibliography}{10}%
\makeatletter
\providecommand \@ifxundefined [1]{%
 \@ifx{#1\undefined}
}%
\providecommand \@ifnum [1]{%
 \ifnum #1\expandafter \@firstoftwo
 \else \expandafter \@secondoftwo
 \fi
}%
\providecommand \@ifx [1]{%
 \ifx #1\expandafter \@firstoftwo
 \else \expandafter \@secondoftwo
 \fi
}%
\providecommand \natexlab [1]{#1}%
\providecommand \enquote  [1]{``#1''}%
\providecommand \bibnamefont  [1]{#1}%
\providecommand \bibfnamefont [1]{#1}%
\providecommand \citenamefont [1]{#1}%
\providecommand \href@noop [0]{\@secondoftwo}%
\providecommand \href [0]{\begingroup \@sanitize@url \@href}%
\providecommand \@href[1]{\@@startlink{#1}\@@href}%
\providecommand \@@href[1]{\endgroup#1\@@endlink}%
\providecommand \@sanitize@url [0]{\catcode `\\12\catcode `\$12\catcode
  `\&12\catcode `\#12\catcode `\^12\catcode `\_12\catcode `\%12\relax}%
\providecommand \@@startlink[1]{}%
\providecommand \@@endlink[0]{}%
\providecommand \url  [0]{\begingroup\@sanitize@url \@url }%
\providecommand \@url [1]{\endgroup\@href {#1}{\urlprefix }}%
\providecommand \urlprefix  [0]{URL }%
\providecommand \Eprint [0]{\href }%
\providecommand \doibase [0]{https://doi.org/}%
\providecommand \selectlanguage [0]{\@gobble}%
\providecommand \bibinfo  [0]{\@secondoftwo}%
\providecommand \bibfield  [0]{\@secondoftwo}%
\providecommand \translation [1]{[#1]}%
\providecommand \BibitemOpen [0]{}%
\providecommand \bibitemStop [0]{}%
\providecommand \bibitemNoStop [0]{.\EOS\space}%
\providecommand \EOS [0]{\spacefactor3000\relax}%
\providecommand \BibitemShut  [1]{\csname bibitem#1\endcsname}%
\let\auto@bib@innerbib\@empty
\bibitem [{\citenamefont {Watts}, \citenamefont {Gauss},\ and\ \citenamefont
  {Bartlett}(1993)}]{Watts1993}%
  \BibitemOpen
  \bibfield  {author} {\bibinfo {author} {\bibfnamefont {J.~D.}\ \bibnamefont
  {Watts}}, \bibinfo {author} {\bibfnamefont {J.}~\bibnamefont {Gauss}},\ and\
  \bibinfo {author} {\bibfnamefont {R.~J.}\ \bibnamefont {Bartlett}},\
  }\bibfield  {title} {\enquote {\bibinfo {title} {{Coupled‐cluster methods
  with noniterative triple excitations for restricted open‐shell
  {Hartree–Fock} and other general single determinant reference functions.
  Energies and analytical gradients}},}\ }\href
  {https://doi.org/10.1063/1.464480} {\bibfield  {journal} {\bibinfo  {journal}
  {The Journal of Chemical Physics}\ }\textbf {\bibinfo {volume} {98}},\
  \bibinfo {pages} {8718--8733} (\bibinfo {year} {1993})}\BibitemShut {NoStop}%
\bibitem [{\citenamefont {Scuseria}(1991)}]{Scuseria1991}%
  \BibitemOpen
  \bibfield  {author} {\bibinfo {author} {\bibfnamefont {G.~E.}\ \bibnamefont
  {Scuseria}},\ }\bibfield  {title} {\enquote {\bibinfo {title} {{Analytic
  evaluation of energy gradients for the singles and doubles coupled cluster
  method including perturbative triple excitations: Theory and applications to
  {FOOF} and {Cr$_2$}}},}\ }\href {https://doi.org/10.1063/1.460359} {\bibfield
   {journal} {\bibinfo  {journal} {The Journal of Chemical Physics}\ }\textbf
  {\bibinfo {volume} {94}},\ \bibinfo {pages} {442--447} (\bibinfo {year}
  {1991})}\BibitemShut {NoStop}%
\bibitem [{\citenamefont {Pinski}\ and\ \citenamefont
  {Neese}(2019)}]{Pinski2019}%
  \BibitemOpen
  \bibfield  {author} {\bibinfo {author} {\bibfnamefont {P.}~\bibnamefont
  {Pinski}}\ and\ \bibinfo {author} {\bibfnamefont {F.}~\bibnamefont {Neese}},\
  }\bibfield  {title} {\enquote {\bibinfo {title} {Analytical gradient for the
  domain-based local pair natural orbital second order {M{\o}ller-Plesset}
  perturbation theory method ({DLPNO-MP2})},}\ }\href@noop {} {\bibfield
  {journal} {\bibinfo  {journal} {The Journal of Chemical Physics}\ }\textbf
  {\bibinfo {volume} {150}},\ \bibinfo {pages} {164102} (\bibinfo {year}
  {2019})}\BibitemShut {NoStop}%
\bibitem [{\citenamefont {England}(1971)}]{England1971}%
  \BibitemOpen
  \bibfield  {author} {\bibinfo {author} {\bibfnamefont {W.}~\bibnamefont
  {England}},\ }\bibfield  {title} {\enquote {\bibinfo {title} {Continuous
  degeneracy and energy-localization of molecular orbitals},}\ }\href@noop {}
  {\bibfield  {journal} {\bibinfo  {journal} {International Journal of Quantum
  Chemistry}\ }\textbf {\bibinfo {volume} {5}},\ \bibinfo {pages} {683--697}
  (\bibinfo {year} {1971})}\BibitemShut {NoStop}%
\bibitem [{\citenamefont {Scheurer}\ and\ \citenamefont
  {Schwarz}(2000)}]{Scheurer2000}%
  \BibitemOpen
  \bibfield  {author} {\bibinfo {author} {\bibfnamefont {P.}~\bibnamefont
  {Scheurer}}\ and\ \bibinfo {author} {\bibfnamefont {W.~H.~E.}\ \bibnamefont
  {Schwarz}},\ }\bibfield  {title} {\enquote {\bibinfo {title} {Continuous
  degeneracy of sets of localized orbitals},}\ }\href@noop {} {\bibfield
  {journal} {\bibinfo  {journal} {International Journal of Quantum Chemistry}\
  }\textbf {\bibinfo {volume} {76}},\ \bibinfo {pages} {428--433} (\bibinfo
  {year} {2000})}\BibitemShut {NoStop}%
\bibitem [{\citenamefont {Bradbury}\ \emph {et~al.}(2018)\citenamefont
  {Bradbury}, \citenamefont {Frostig}, \citenamefont {Hawkins}, \citenamefont
  {Johnson}, \citenamefont {Leary}, \citenamefont {Maclaurin}, \citenamefont
  {Necula}, \citenamefont {Paszke}, \citenamefont {Vander{P}las}, \citenamefont
  {Wanderman-{M}ilne},\ and\ \citenamefont {Zhang}}]{Jax2018}%
  \BibitemOpen
  \bibfield  {author} {\bibinfo {author} {\bibfnamefont {J.}~\bibnamefont
  {Bradbury}}, \bibinfo {author} {\bibfnamefont {R.}~\bibnamefont {Frostig}},
  \bibinfo {author} {\bibfnamefont {P.}~\bibnamefont {Hawkins}}, \bibinfo
  {author} {\bibfnamefont {M.~J.}\ \bibnamefont {Johnson}}, \bibinfo {author}
  {\bibfnamefont {C.}~\bibnamefont {Leary}}, \bibinfo {author} {\bibfnamefont
  {D.}~\bibnamefont {Maclaurin}}, \bibinfo {author} {\bibfnamefont
  {G.}~\bibnamefont {Necula}}, \bibinfo {author} {\bibfnamefont
  {A.}~\bibnamefont {Paszke}}, \bibinfo {author} {\bibfnamefont
  {J.}~\bibnamefont {Vander{P}las}}, \bibinfo {author} {\bibfnamefont
  {S.}~\bibnamefont {Wanderman-{M}ilne}},\ and\ \bibinfo {author}
  {\bibfnamefont {Q.}~\bibnamefont {Zhang}},\ }\href
  {http://github.com/google/jax} {\enquote {\bibinfo {title} {{JAX}: composable
  transformations of {P}ython+{N}um{P}y programs},}\ } (\bibinfo {year}
  {2018}),\ \bibinfo {note} {available at
  \url{http://github.com/google/jax}}\BibitemShut {NoStop}%
\bibitem [{\citenamefont {Drosou}, \citenamefont {Mitsopoulou},\ and\
  \citenamefont {Pantazis}(2022)}]{Drosou2022}%
  \BibitemOpen
  \bibfield  {author} {\bibinfo {author} {\bibfnamefont {M.}~\bibnamefont
  {Drosou}}, \bibinfo {author} {\bibfnamefont {C.~A.}\ \bibnamefont
  {Mitsopoulou}},\ and\ \bibinfo {author} {\bibfnamefont {D.~A.}\ \bibnamefont
  {Pantazis}},\ }\bibfield  {title} {\enquote {\bibinfo {title} {Reconciling
  local coupled cluster with multireference approaches for transition metal
  spin-state energetics},}\ }\href {https://doi.org/10.1021/acs.jctc.2c00265}
  {\bibfield  {journal} {\bibinfo  {journal} {Journal of Chemical Theory and
  Computation}\ }\textbf {\bibinfo {volume} {18}},\ \bibinfo {pages}
  {3538--3548} (\bibinfo {year} {2022})}\BibitemShut {NoStop}%
\bibitem [{\citenamefont {Mayer}(1984)}]{Mayer1984}%
  \BibitemOpen
  \bibfield  {author} {\bibinfo {author} {\bibfnamefont {I.}~\bibnamefont
  {Mayer}},\ }\bibfield  {title} {\enquote {\bibinfo {title} {Bond order and
  valence: Relations to {Mulliken}'s population analysis},}\ }\href
  {https://doi.org/https://doi.org/10.1002/qua.560260111} {\bibfield  {journal}
  {\bibinfo  {journal} {International Journal of Quantum Chemistry}\ }\textbf
  {\bibinfo {volume} {26}},\ \bibinfo {pages} {151--154} (\bibinfo {year}
  {1984})}\BibitemShut {NoStop}%
\bibitem [{\citenamefont {Petrilli}\ \emph {et~al.}(1998)\citenamefont
  {Petrilli}, \citenamefont {Bl\"ochl}, \citenamefont {Blaha},\ and\
  \citenamefont {Schwarz}}]{Petrilli1998}%
  \BibitemOpen
  \bibfield  {author} {\bibinfo {author} {\bibfnamefont {H.~M.}\ \bibnamefont
  {Petrilli}}, \bibinfo {author} {\bibfnamefont {P.~E.}\ \bibnamefont
  {Bl\"ochl}}, \bibinfo {author} {\bibfnamefont {P.}~\bibnamefont {Blaha}},\
  and\ \bibinfo {author} {\bibfnamefont {K.}~\bibnamefont {Schwarz}},\
  }\bibfield  {title} {\enquote {\bibinfo {title} {Electric-field-gradient
  calculations using the projector augmented wave method},}\ }\href
  {https://doi.org/10.1103/PhysRevB.57.14690} {\bibfield  {journal} {\bibinfo
  {journal} {Phys. Rev. B}\ }\textbf {\bibinfo {volume} {57}},\ \bibinfo
  {pages} {14690--14697} (\bibinfo {year} {1998})}\BibitemShut {NoStop}%
\bibitem [{\citenamefont {Stone}(2016)}]{Stone2016}%
  \BibitemOpen
  \bibfield  {author} {\bibinfo {author} {\bibfnamefont {N.}~\bibnamefont
  {Stone}},\ }\bibfield  {title} {\enquote {\bibinfo {title} {Table of nuclear
  electric quadrupole moments},}\ }\href
  {https://doi.org/https://doi.org/10.1016/j.adt.2015.12.002} {\bibfield
  {journal} {\bibinfo  {journal} {Atomic Data and Nuclear Data Tables}\
  }\textbf {\bibinfo {volume} {111-112}},\ \bibinfo {pages} {1--28} (\bibinfo
  {year} {2016})}\BibitemShut {NoStop}%
\end{thebibliography}%
\end{document}


\title{Supporting Information for: ``Performant Automatic Differentiation of Local Coupled Cluster Theories: Response Properties and Ab Initio Molecular Dynamics''}

\author{Xing Zhang}
\author{Chenghan Li}
\affiliation
{Division of Chemistry and Chemical Engineering,
California Institute of Technology, Pasadena, CA 91125, USA}
\author{Hong-Zhou Ye}
\author{Timothy C. Berkelbach}
\affiliation
{Department of Chemistry, Columbia University, New York, NY 10027, USA}
\author{Garnet Kin-Lic Chan}
\affiliation
{Division of Chemistry and Chemical Engineering,
California Institute of Technology, Pasadena, CA 91125, USA}
\email{gkc1000@gmail.com}

\maketitle

\section{Implementation Details}
In this section, we introduce additional details about the implementation
of our differentiable LNO-CC method.

\subsection{Energy Expressions for LNO-CCSD(T)}
\label{sec:cc_energy}
The local CCSD correlation energy on fragment $\Omega$ reads
\begin{equation}
  E^{\Omega}_{\text{CCSD}} = \sum_{I \in \Omega} \sum _{\tilde{m}\tilde{n} \in \Omega} U^{\Omega \top}_{I\tilde{m}} \left(\sum_{\tilde{a} \in \Omega}
  F_{\tilde{m}\tilde{a}} t_{\tilde{n}}^{\tilde{a}} +
  \frac{1}{4} \sum_{\tilde{j}\tilde{a}\tilde{b} \in \Omega} \langle \tilde{a}\tilde{b} || \tilde{m}\tilde{j} \rangle  \tau_{\tilde{n}\tilde{j}}^{\tilde{a}\tilde{b}} \right) U^{\Omega}_{\tilde{n}I} \;,
  \label{eq:ccsd}
\end{equation}
where $\mathbf{F}$ is the Fock matrix, $\mathbf{t}$ represents the CC
amplitudes, and
\begin{equation}
    \tau_{\tilde{i}\tilde{j}}^{\tilde{a}\tilde{b}}
    = t_{\tilde{i}\tilde{j}}^{\tilde{a}\tilde{b}}
    + t_{\tilde{i}}^{\tilde{a}} t_{\tilde{j}}^{\tilde{b}}
    - t_{\tilde{j}}^{\tilde{a}} t_{\tilde{i}}^{\tilde{b}} \;.
\end{equation}

The contribution of triple excitations
to the LNO-CCSD(T) correlation energy on fragment
$\Omega$ may be written as
\begin{equation}
  E^{\Omega}_{\text{(T)}} = \sum_{I \in \Omega} \sum _{\tilde{m}\tilde{n} \in \Omega} U^{\Omega \top}_{I\tilde{m}} \left(\frac{1}{36} \sum_{\tilde{j}\tilde{k} \in \Omega}
  \sum_{\tilde{a}\tilde{b}\tilde{c} \in \Omega}
  t(c)_{\tilde{m}\tilde{j}\tilde{k}}^{\tilde{a}\tilde{b}\tilde{c}}
  D_{\tilde{m}\tilde{j}\tilde{k}}^{\tilde{a}\tilde{b}\tilde{c}}
  \left(t(c)_{\tilde{n}\tilde{j}\tilde{k}}^{\tilde{a}\tilde{b}\tilde{c}}
  +t(d)_{\tilde{n}\tilde{j}\tilde{k}}^{\tilde{a}\tilde{b}\tilde{c}}
  \right)\right) U^{\Omega}_{\tilde{n}I} \;,
  \label{eq:et}
\end{equation}
where the definitions of $\mathbf{t}(c)$,
$\mathbf{t}(d)$, and $\mathbf{D}$ can be read from Ref.~\citenum{Watts1993}.
In practice, the implementation of Eq.~\ref{eq:et}
takes into account both spin symmetry,
and permutation symmetries of the electron integrals and the
$\mathbf{t}$ amplitudes,
leading to the formulation for closed-shell systems as
\begin{equation}
  E^{\Omega}_{\text{(T)}} = \sum_{I \in \Omega} \sum _{\tilde{m}\tilde{n} \in \Omega} U^{\Omega \top}_{I\tilde{m}} \left(\frac{2}{3}
  \sum_{\tilde{j}\tilde{k} \in \Omega}
  \sum_{\tilde{a} \in \Omega}
  \sum_{\tilde{b} \geqslant \tilde{c} \in \Omega}
  W_{\tilde{m}\tilde{j}\tilde{k}}^{\tilde{a}\tilde{b}\tilde{c}}
  D_{\tilde{m}\tilde{j}\tilde{k}}^{\tilde{a}\tilde{b}\tilde{c}}
  R\left[W_{\tilde{n}\tilde{j}\tilde{k}}^{\tilde{a}\tilde{b}\tilde{c}}
  +V_{\tilde{n}\tilde{j}\tilde{k}}^{\tilde{a}\tilde{b}\tilde{c}}
  \right]\right) U^{\Omega}_{\tilde{n}I} \;,
  \label{eq:et_R}
\end{equation}
where $\mathbf{W}$, $\mathbf{V}$, and $R$ are defined in
Ref.~\citenum{Scuseria1991}, and the orbital indices now
refer to spatial orbitals.
Similar expression can be derived for unrestricted open-shell calculations,
and will be introduced in a separate work.

\subsection{Differentiating Orbital Localization}
\label{sec:local_orb_autodiff}
For orbital localization methods
that rely on solving specific optimization problems
(e.g., Boys and Pipek-Mezey localization),
derivatives of the localized orbitals can be computed by implicitly differentiating the optimality conditions.
Suppose the local orbitals are
parameterized by the canonical molecular orbital (MO) coefficients $\mathbf{c}$,
and an anti-Hermitian matrix $\mathbf{x}$,
\begin{equation}
    C_{\mu I} = \sum_i c_{\mu i} (e^{\mathbf{x}})_{iI} \;,
\end{equation}
then the localization procedure is carried out
by finding the extremum of some objective function
$f(\mathbf{x}(\theta), \theta)$, where $\theta$ represents the variables for the perturbations,
such as molecular geometries or external electric fields.
The solution (denoted as $\mathbf{x}^{\star}$) of such optimization problems
satisfies the optimality condition,
\begin{equation}
    \mathbf{g}(\mathbf{x}^{\star}(\theta), \theta) \equiv
    \frac{\partial f}{\partial \mathbf{x}} \bigg\rvert_{\mathbf{x}=\mathbf{x}^{\star}} = \mathbf{0} \;.
    \label{eq:local_optcond}
\end{equation}
Differentiating Eq.~\ref{eq:local_optcond} with respect to $\theta$
gives the derivative of $\mathbf{x}^{\star}$,
\begin{equation}
    \frac{\partial \mathbf{x}^{\star}}{\partial \theta}=
    -\left(\frac{\partial^2 f}{\partial \mathbf{x}^2} \bigg \rvert_{\mathbf{x}=\mathbf{x}^{\star}}\right)^{-1} \frac{\partial \mathbf{g}}{\partial \theta} \;.
    \label{eq:local_deriv}
\end{equation}
During gradient backpropagation,
the actual quantity computed is the so-called vector-Jacobian product (VJP),
\begin{equation}
    \mathbf{v}^{\top} \frac{\partial \mathbf{x}^{\star}}{\partial \theta} = -\mathbf{z}^{\top} \frac{\partial \mathbf{g}}{\partial \theta} \;,
    \label{eq:local_vjp}
\end{equation}
where $\mathbf{v}$ is a vector in the cotangent space
of $\mathbf{x}^{\star}$ at $\theta$,
and $\mathbf{z}$ is the solution of the following equations,
\begin{equation}
    \left(\frac{\partial^2 f}{\partial \mathbf{x}^2} \bigg \rvert_{\mathbf{x}=\mathbf{x}^{\star}}\right)^{\top} \mathbf{z} = \mathbf{v} \;.
    \label{eq:local_zvec}
\end{equation}
Note that in Eqs.~\ref{eq:local_deriv} and \ref{eq:local_zvec},
the orbital Hessian,
\begin{equation}
    \mathbf{H}^{\text{Loc}} \equiv \frac{\partial^2 f}{\partial \mathbf{x}^2} \bigg\rvert_{\mathbf{x}=\mathbf{x}^{\star}}\;,
\end{equation}
may be rank deficient either accidentally or as required by symmetry,
which can lead to singular derivatives of $\mathbf{x}^{\star}$.
In practice, $\mathbf{H}^{\text{Loc}}$ in
Eq.~\ref{eq:local_zvec} is replaced (following Ref.~\citenum{Pinski2019}) by
\begin{equation}
    \tilde{\mathbf{H}}^{\text{Loc}} =
    (\mathbf{1}-\mathbf{u}\mathbf{u}^{\top})
    \mathbf{H}^{\text{Loc}}
    (\mathbf{1}-\mathbf{u}\mathbf{u}^{\top}) +
    \mathbf{u}\mathbf{u}^{\top} \;,
    \label{eq:local_Hloc}
\end{equation}
where the columns of $\mathbf{u}$ are the eigenvectors of $\mathbf{H}^{\text{Loc}}$ that span its null space.
As the result, Eq.~\ref{eq:local_zvec} is transformed to a
non-singular problem, and its solution $\mathbf{z}$ is further
projected onto the range space of $\mathbf{H}^{\text{Loc}}$ as
\begin{equation}
    \tilde{\mathbf{z}} = (\mathbf{1}-\mathbf{u}\mathbf{u}^{\top}) \mathbf{z} \;.
    \label{eq:local_ztilde}
\end{equation}
$\tilde{\mathbf{z}}$ then substitutes $\mathbf{z}$ in Eq.~\ref{eq:local_vjp}
for computing the VJP.

When $\mathbf{H}^{\text{Loc}}$ has full rank,
$\mathbf{x}^{\star}$ is uniquely determined and
its derivative is well-defined.
The procedure above is valid in this case as $\mathbf{u}$ is an empty matrix.
If $\mathbf{H}^{\text{Loc}}$ is rank deficient,
the derivative of $\mathbf{x}^{\star}$ becomes ill-defined in general, as can be seen from Eq.~\ref{eq:local_deriv}.
However, the energy gradient can still be well-defined
if $\mathbf{v}$ is orthogonal to $\mathbf{u}$, so that $\mathbf{z}$ in Eq.~\ref{eq:local_zvec} has unique solutions that
are confined in the range space of $\mathbf{H}^{\text{Loc}}$.
Such a situation occurs when the energy is invariant
with respect to the orbital rotations that follow $\mathbf{u}$.
For systems with symmetries that admit continuously degenerate
localized orbitals,\cite{England1971,Scheurer2000}
the energy invariance condition may be fulfilled if
the energy formulation is symmetry adapted.
Otherwise, $\mathbf{v}$ may overlap with $\mathbf{u}$,
and Eq.~\ref{eq:local_zvec} has no solution, which makes the
energy gradient ill-defined.
The procedure outlined in Eqs.~\ref{eq:local_Hloc}
and \ref{eq:local_ztilde} hence can only give approximate
energy gradients compare to the exact (symmetry-adapted) results
in this scenario.

Finally, it is worth mentioning that within the automatic differentiation (AD) framework of \textsc{PySCFAD},
only the objective function $f$ requires explicit implementation,
while all differentiation specified in Eqs.~\ref{eq:local_optcond} and
\ref{eq:local_deriv} is automatically computed.
This makes our approach generalizable to many other
orbital localization methods with little effort.

\subsection{Performance Optimization} \label{sec:opt}
Numerous optimizations have been implemented within the \textsc{PySCFAD} package to enhance performance, thereby facilitating calculations of larger sizes in this study.
The optimizations are specifically applied to the reverse-mode AD,
which is efficient when the number of objectives is less than the number of variables.
These include:
\begin{itemize}
\item Permutation (up to four-fold) symmetries of electron integrals,
including the commonly used two-, three-, and four-center integrals,
have been considered during the gradient backpropagation,
for which the VJPs are implemented in efficient multithreaded C code.
\item The VJPs for the gradient backpropagation of Coulomb/exchange matrix evaluations in the mean-field calculation
have been reimplemented in C code, leveraging efficient BLAS3
(the level 3 basic linear algebra subprograms) libraries.
\item Permutation symmetries of the CC amplitudes have been considered in both energy and gradient calculations.
The triple excitation correction in the CCSD(T) method is computed on-the-fly
in both forward and backward passes, without storing any intermediate quantities except for the singles and doubles amplitudes and the two-electron integrals required for the CCSD calculation. The biggest tensor stored in memory corresponds to the integral of the type
$\langle ov|vv \rangle$, where $o$ and $v$ denote occupied and virtual molecular orbitals,
respectively.
\item The integral transformations employ the efficient implementation of \textsc{PySCF},
and the corresponding VJPs for the gradient backpropagation are also implemented
in C code, involving mainly BLAS3 routines.
\item Other straightforward but memory intensive calculations,
such as computation of the canonical MP2 density matrix,
are repeated during the gradient backpropagation rather than by keeping the intermediate quantities in memory. This is achieved simply by calling the \texttt{jax.checkpoint} function.
\end{itemize}

\section{Additional Results}
In this section, additional data are presented to support the findings of the main text.

\subsection{Performance of \textsc{PySCFAD}}
\begin{figure*}
\includegraphics{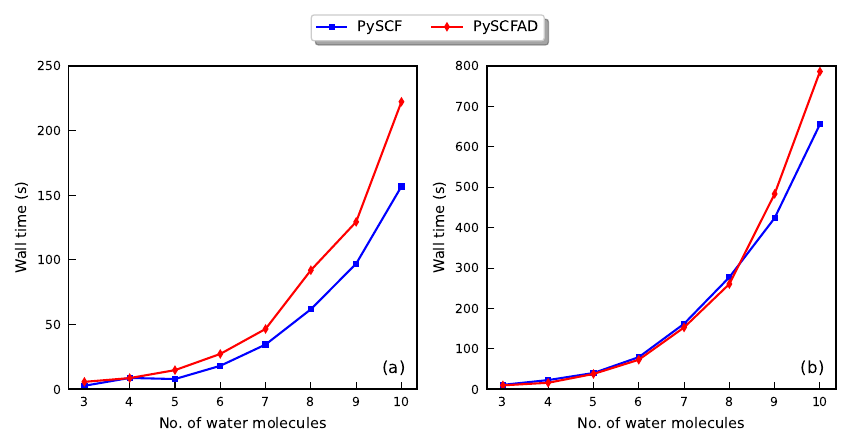}
\centering
\caption{Wall times for (a) MP2 energy and nuclear gradient,
and (b) CCSD energy, computed with \textsc{PySCF} and
\textsc{PySCFAD} for water clusters using the cc-pVDZ basis set.
The calculations were performed on one computer node with
\revision{12/}24 Intel Xeon Broadwell (E5-2697v4) \revision{CPU cores/threads}.}
\label{fig:timing}
\end{figure*}

The performance of the newly optimized \textsc{PySCFAD} package is evaluated
through medium-sized calculations, involving a few hundred basis functions.
These dimensions are typical for local fragment calculations of the LNO-CC method.
In Fig.~\ref{fig:timing}, we compare the efficiency of \textsc{PySCFAD}
and its parent software \textsc{PySCF}, specifically in performing energy and gradient calculations at the post-HF levels.
We plot the wall times required for computing the MP2 (without density fitting) energies and nuclear gradients in Fig.~\ref{fig:timing}(a),
and the CCSD energies in Fig.~\ref{fig:timing}(b), for a series of water clusters.
Comparable performance of the two packages is observed,
albeit with \textsc{PySCFAD} slightly falling behind \textsc{PySCF}.
This difference can primarily be attributed to the additional time for
program tracing and just-in-time (JIT) compilation required by \textsc{Jax},\cite{Jax2018}
which is the AD tool employed in \textsc{PySCFAD}.
Nevertheless, the advantage of methods implemented in
\textsc{PySCFAD} being differentiable significantly mitigates the small performance gap compared to \textsc{PySCF}. It is worth noting that the implementation of analytic nuclear gradients for most post-HF methods with density fitting is still lacking in \textsc{PySCF}.

\subsection{Additional Results for the Baker Test Set}
\label{sec:supp_baker}
\begin{figure*}
\includegraphics{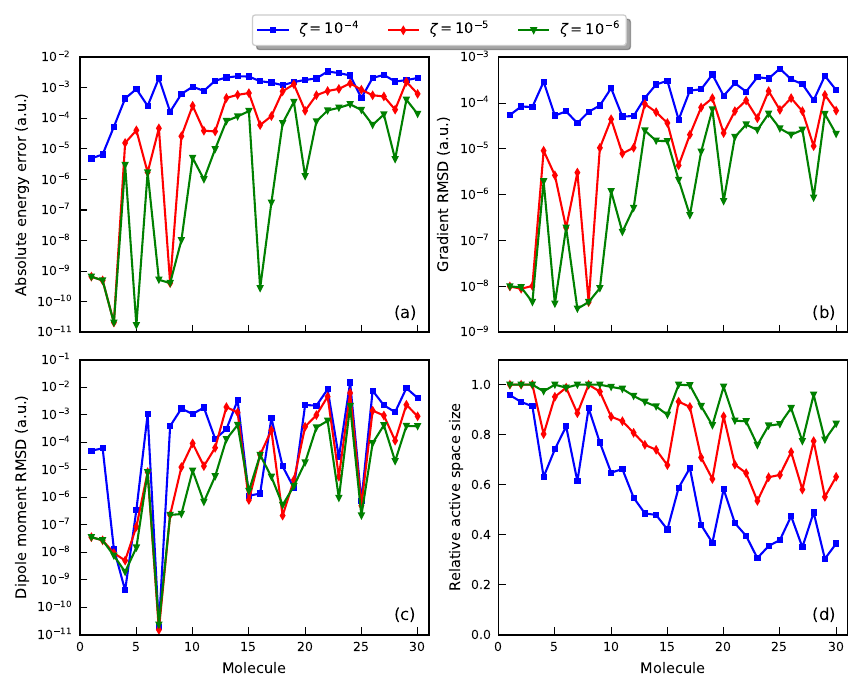}
\centering
\caption{Absolute energy error, nuclear gradient RMSD, dipole moment RMSD, and relative active space size computed at the
LNO-CCSD(T)/IAO/cc-pVDZ level for the Baker test set.
The reference data was obtained at the canonical CCSD(T) level with the same basis set.}
\label{fig:grad_iao}
\end{figure*}

In Fig.~\ref{fig:grad_iao} and Table~\ref{tab:geom_opt_iao},
we present results of nuclear gradient, dipole moment, and
geometry optimization calculations for the Baker test set
using the LNO-CCSD(T) method with IAO local orbitals.
Typically, for the same LNO cutoff value,
less compact local active spaces are obtained when
employing IAOs compared with Pipek-Mezey (PM) local orbitals.
However, due to the atomic orbital nature inherent in IAOs,
their application in atom-based fragmentation approaches,
such as the multi-orbital scheme used here,
tends to give results that preserve the point-group symmetry.
[See benzene (molecule 7 in Fig.~\ref{fig:grad_iao}) as an example.]
The geometry optimizations show slightly higher accuracy when
utilizing IAOs in comparison to PM local orbitals, which is
primarily attributed to the incorporation of larger local correlation domains.

\begin{table*}
  \centering
  \caption{Mean absolute error (MAE)
and maximum error (max) in bond lengths,
angles, dihedral angles and out-of-plane angles of
the geometries optimized at the LNO-CCSD(T)/IAO/cc-pVDZ level
for the Baker test set. The reference geometries were
optimized at the canonical CCSD(T) level with the same basis set.}
  \begin{tabular*}{1.0\textwidth}{@{\extracolsep{\fill}}llcccc}
    \hline\hline
    $\zeta$   &     & bond length (\AA) & angle (\degree) & dihedral angle (\degree) & out-of-plane angle (\degree) \\
    \hline
    $1 \times 10^{-4}$ & MAE & 0.0003          & 0.026 & 0.059         & 0.008 \\
              & max & 0.0031          & 0.138 & 1.459         & 0.113 \\
    $1 \times 10^{-5}$ & MAE & 0.0001          & 0.007 & 0.015         & 0.002 \\
              & max & 0.0006          & 0.073 & 0.295         & 0.079 \\
    \hline
  \end{tabular*}
  \label{tab:geom_opt_iao}
\end{table*}

\begin{figure*}
\includegraphics{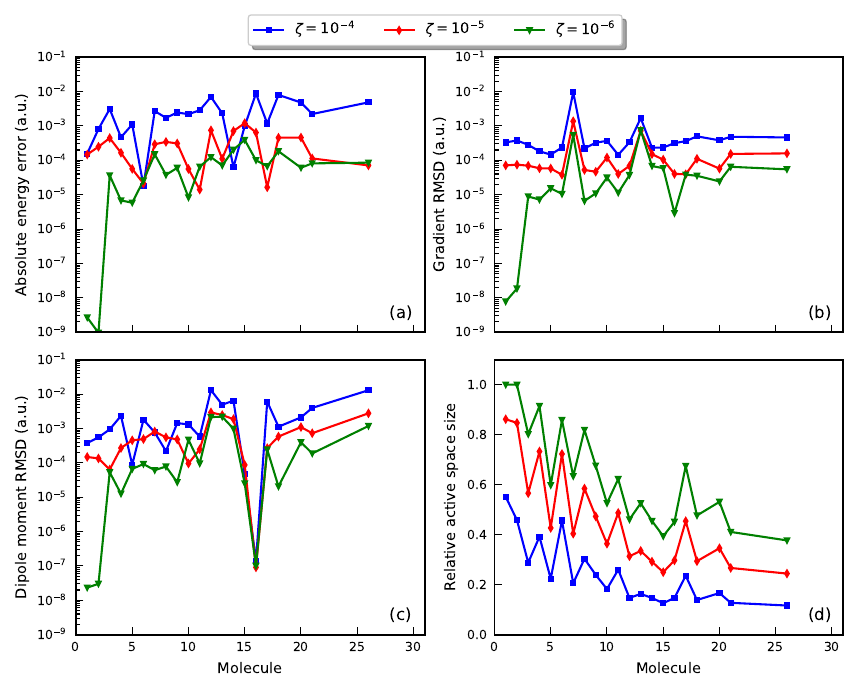}
\centering
\caption{\revision{Absolute energy error, nuclear gradient RMSD, dipole moment RMSD, and relative active space size computed at the
LNO-CCSD(T)/PM/cc-pVTZ level for a subset of the Baker test set.
The reference data were obtained at the CCSD(T)/cc-pVTZ level,
and the geometries were optimized at the MP2/cc-pVDZ level.}}
\label{fig:grad_pm_TZ}
\end{figure*}

\begin{figure}
\includegraphics{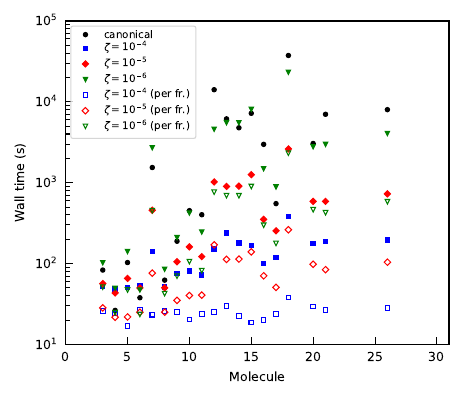}
\centering
\caption{\revision{Total wall time of computing energy, nuclear gradient, and dipole moment for a subset of the Baker test set.
Comparing the results by LNO-CCSD(T)/PM (blue, red, and green) and canonical CCSD(T) (black) methods with the cc-pVTZ basis set.
The average wall time per fragment is plotted as unfilled markers.
The calculations were performed on one computer node with
14/28 Intel Xeon Broadwell (E5-2680v4) CPU cores/threads.}}
\label{fig:time_pm_TZ}
\end{figure}

\revision{
In Figs.~\ref{fig:grad_pm_TZ} and \ref{fig:time_pm_TZ},
we present the errors for energy, nuclear gradient, and dipole moment
computed with the LNO-CCSD(T) method,
as well as the corresponding computational time
compared to that of the canonical CCSD(T) calculations
for a subset of the Baker test set using the cc-pVTZ basis set.
No noticeable increase in errors is observed
compared to the results obtained with the cc-pVDZ basis set.
Additionally, Fig.~\ref{fig:time_pm_TZ} shows that,
except for very small molecules, LNO-CCSD(T) is already one to two orders of
magnitude more efficient than canonical CCSD(T)
even without parallelization over the fragments.
This efficiency is observed for loose ($\zeta=10^{-4}$) to
moderate ($\zeta=10^{-5}$) LNO cutoff values, offering an accuracy of
approximately $10^{-3}$ and $10^{-4}$ a.u. in nuclear gradients, respectively.
On the other hand, using tighter cutoffs can make each fragment calculation
more time-consuming due to the increased size of the local correlation domain.
As multiple fragments need to be computed, the overall computational time might
surpass that of the canonical CCSD(T) calculation.
Nevertheless, further speedups can be achieved by parallelizing over the fragments,
which is common practice for fragment-based methods.
This is illustrated by the wall time per fragment in Fig.~\ref{fig:time_pm_TZ}
(see e.g., the blue unfilled markers),
which indicates an almost constant computational cost across a range of molecules with varying sizes.
}

\subsection{Additional Results for the \textit{Ht}-SH Active Site Model}
\label{sec:HtSH}
\subsubsection{Energy Convergence}
\begin{figure}[h]
\includegraphics{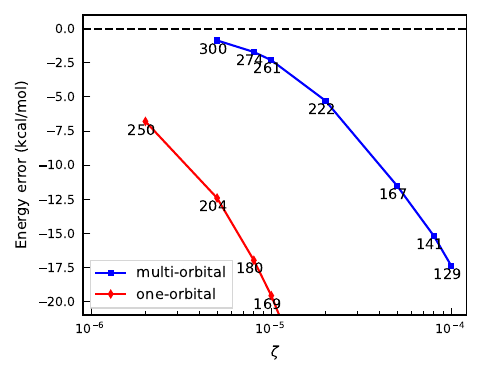}
\centering
\caption{
  Energy errors (in kcal/mol) versus LNO cutoff values ($\zeta$)
  computed for the \textit{Ht}-SH active site model system
  at the LNO-CCSD/IAO/def2-SVP level,
  using the multi-orbital scheme (blue) and
  the one-orbital scheme (red), respectively.
  The reference energy was computed at the canonical CCSD level with the same basis set.
  The largest fragment size in each calculation is labeled on the corresponding data point. The complete Hilbert space contains 549 orbitals.
}
\label{fig:NiFe_ccsd_svp}
\end{figure}

We investigate the energy convergence patterns of the LNO-CC methods
with respect to the LNO cutoff values ($\zeta$) for the \textit{Ht}-SH active site model system.
In Fig.~\ref{fig:NiFe_ccsd_svp}, we compare the multi-orbital scheme with the one-orbital scheme at the LNO-CCSD/IAO/def2-SVP level.
It is clear that the energy converges much faster with the multi-orbital scheme,
reaching one-millihartree accuracy at $\zeta=5\times10^{-6}$,
whereas large energy errors persist with the one-orbital scheme,
even as $\zeta$ decreases to $2\times10^{-6}$.
While it is true that the one-orbital scheme generally produces much smaller
fragments (in terms of the local active space size), the overall computational cost may not necessarily be lower than that of the multi-orbital scheme, owing to the increased number of fragments.
Indeed, for the data points depicted in the most left side of Fig.~\ref{fig:NiFe_ccsd_svp},
the largest fragments given by the two schemes are approximately the same size
($\sim50\%$ of the complete Hilbert space size),
rendering the one-orbital scheme nearly three times more expensive than the
multi-orbital scheme.
As such, we opt to proceed with the multi-orbital scheme for all subsequent calculations.

\begin{figure}[h]
\includegraphics{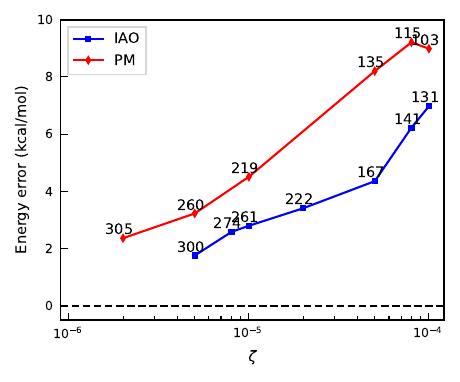}
\centering
\caption{
  Energy errors (in kcal/mol) versus LNO cutoff values ($\zeta$)
  computed for the \textit{Ht}-SH active site model system
  at the LNO-CCSD(T)/def2-SVP level with IAO (blue)
  and PM (red) local orbitals, respectively.
  The reference energy was estimated
  based on a polynomial fit to the differences between
  the energies calculated at the LNO-CCSD(T) level and the LNO-CCSD level.
  The largest fragment size in each calculation is labeled on the corresponding data point. The complete Hilbert space contains 549 orbitals.
}
\label{fig:NiFe_ccsdt_svp}
\end{figure}

\begin{figure}[h]
\includegraphics{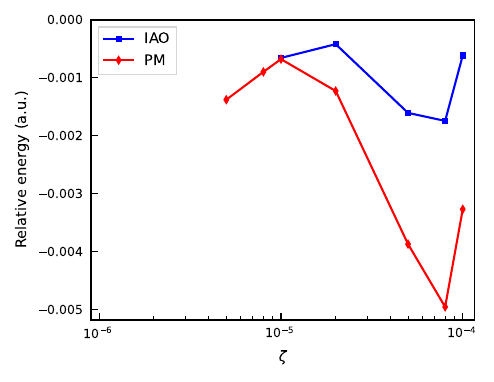}
\centering
\caption{
  Relative energies (in hartree) versus LNO cutoff values ($\zeta$)
  computed for the \textit{Ht}-SH active site model system
  at the LNO-CCSD(T)/def2-TZVP level with IAO (blue)
  and PM (red) local orbitals, respectively.
  The zero energy is arbitrarily set at -5283.793 hartree.
}
\label{fig:NiFe_ccsdt_tzvp}
\end{figure}

Next, we examine the influence of employing different orbital localization approaches.
In Figs.~\ref{fig:NiFe_ccsdt_svp} and \ref{fig:NiFe_ccsdt_tzvp},
energies computed using the PM local orbitals and the IAOs are compared
at the LNO-CCSD(T) level with def2-SVP and def2-TZVP basis sets, respectively.
When the local active space is not capable of capturing significant
electron correlations, it can result in oscillatory energy convergence patterns
against $\zeta$, as can be seen in
Fig.~\ref{fig:NiFe_ccsdt_tzvp} for both orbital localization methods,
and in Figs.~\ref{fig:NiFe_ccsdt_svp} for the PM localization.
Usually, for the same $\zeta$ value, employing PM local orbitals yields smaller fragments
compared to utilizing IAOs. Nevertheless, even after accounting for the fragment size,
the energy convergence exhibits a slightly slower rate when using PM orbitals (see Fig.~\ref{fig:NiFe_ccsdt_svp}).

Lastly, we observe a slow convergence of the LNO-CCSD(T) energy towards
the reference canonical CCSD(T) energy for the \textit{Ht}-SH active site model studied here.
As shown in Figs.~\ref{fig:NiFe_ccsd_svp} and \ref{fig:NiFe_ccsdt_svp},
it is evident that fragments of approximately half the size of the complete Hilbert space
are necessary to achieve one-millihartree accuracy.
This implies that the local correlation domains constructed using the current LNO
truncation approach might not be sufficiently compact when treating transition metal complexes. A recent study by Drosou \textit{et al.}\cite{Drosou2022} suggests employing
the iterative approach rather than the semicanonical approach for computing the perturbative triple excitation correction, in addition to extrapolating the local correlation energies to reach the complete active space limit. These will be explored in a separate work.

Considering the energy convergence tests above,
we decided to use the IAOs along with an LNO cutoff of $\zeta=2\times10^{-5}$
for the subsequent property calculations with our LNO-CCSD(T) method.
The selected value of $\zeta$ corresponds to the turning point depicted in
Fig.~\ref{fig:NiFe_ccsdt_tzvp}, marking the onset of energy convergence.
It should be regarded as the biggest $\zeta$ value
capable of giving qualitative results compared to the canonical CCSD(T) reference.

\subsubsection{Bond Order}
The Mayer bond order\cite{Mayer1984} is defined as
\begin{equation}
b_{\text{AB}} = \sum_{\mu \in \text{A}}\sum_{\nu \in \text{B}}
(\mathbf{D}\mathbf{S})_{\mu\nu} (\mathbf{D}\mathbf{S})_{\nu\mu} \;,
\label{eq:bond_order}
\end{equation}
where A and B label the atoms or the shells of atomic orbitals,
$\mathbf{D}$ is the one-electron reduced density matrix,
and $\mathbf{S}$ is the atomic orbital overlap integral matrix.
In Table\revision{s}~\ref{tab:NiFe_bond_order1_full} \revision{and \ref{tab:NiFe_bond_order}}, additional results of bond orders
for the atoms coordinated to the two metal centers of
the \textit{Ht}-SH active site model are presented.

\begin{table}[h]
  \centering
  \caption{
  \revision{Orbital resolved bond orders for the \textit{Ht}-SH active site model system,
  computed at the DFT/TPSSh and LNO-CCSD(T)/IAO
  ($\zeta=2\times 10^{-5}$) levels
  with the def2-TZVP basis set.}}
  \begin{tabular*}{0.48\textwidth}{@{\extracolsep{\fill}}lcc}
    \hline\hline
    Bond & TPSSh & LNO-CCSD(T) \\
    \hline
    Fe(3d)--C$^1$(2p)N & 0.24 & 0.17 \\
    Fe(4s)--C$^1$(2p)N & 0.03 & 0.02 \\
    Fe(3d)--C$^2$(2p)N & 0.24 & 0.16 \\
    Fe(4s)--C$^2$(2p)N & 0.03 & 0.02 \\
    Fe(3d)--C(2p)O & 0.57 & 0.87 \\
    Fe(4s)--C(2p)O & 0.01 & 0.00 \\
    Fe(3d)--S$^1$(3p)Cys & 0.21 & 0.09 \\
    Fe(4s)--S$^1$(3p)Cys & 0.03 & 0.02 \\
    Fe(3d)--S$^2$(3p)Cys & 0.17 & 0.07 \\
    Fe(4s)--S$^2$(3p)Cys & 0.02 & 0.02 \\
    Fe(3d)--S$^3$(3p)Cys & 0.21 & 0.12 \\
    Fe(4s)--S$^3$(3p)Cys & 0.02 & 0.03 \\
    Ni(3d)--S$^1$(3p)Cys & 0.26 & 0.31 \\
    Ni(4s)--S$^1$(3p)Cys & 0.07 & 0.07 \\
    Ni(3d)--S$^2$(3p)Cys & 0.34 & 0.14 \\
    Ni(4s)--S$^2$(3p)Cys & 0.07 & 0.07 \\
    Ni(3d)--S$^3$(3p)Cys & 0.35 & 0.08 \\
    Ni(4s)--S$^3$(3p)Cys & 0.06 & 0.06 \\
    Ni(3d)--S$^4$(3p)Cys & 0.33 & 0.32 \\
    Ni(4s)--S$^4$(3p)Cys & 0.08 & 0.08 \\
    Ni(3d)--O$^1$(2p) & 0.10 & 0.02 \\
    Ni(4s)--O$^1$(2p) & 0.12 & 0.09 \\
    Ni(3d)--O$^2$(2p) & 0.10 & 0.03 \\
    Ni(4s)--O$^2$(2p) & 0.11 & 0.08 \\
    \hline
  \end{tabular*}
  \label{tab:NiFe_bond_order1_full}
\end{table}

\begin{table*}
  \centering
  \caption{
  Bond orders for the \textit{Ht}-SH active site model system,
  computed at the DFT/TPSSh, MP2, and LNO-CCSD(T)/IAO levels
  with the def2-TZVP basis set.}
  \begin{tabular*}{1.0\textwidth}{@{\extracolsep{\fill}}lccccc}
    \hline\hline
    Bond & TPSSh & MP2 &\mc{3}{c}{LNO-CCSD(T)}\\
    \cline{4-6}
    & & & $\zeta = 1 \times 10^{-4}$
    & $\zeta = 5 \times 10^{-5}$
    & $\zeta = 2 \times 10^{-5}$ \\
    \hline
    Fe--C$^1$N & 0.39 & 0.42 & 0.28 & 0.24 & 0.23 \\
    Fe--C$^2$N & 0.39 & 0.43 & 0.27 & 0.24 & 0.22 \\
    Fe--CO & 0.69 & 0.77 & 0.67 & 0.90 & 0.92 \\
    Fe--S$^1$Cys & 0.25 & 0.29 & 0.15 & 0.12 & 0.11 \\
    Fe--S$^2$Cys & 0.20 & 0.19 & 0.13 & 0.10 & 0.10 \\
    Fe--S$^3$Cys & 0.24 & 0.18 & 0.16 & 0.15 & 0.16 \\
    Ni--O$^1$ & 0.23 & 0.30 & 0.15 & 0.12 & 0.12 \\
    Ni--O$^2$ & 0.22 & 0.27 & 0.15 & 0.12 & 0.12 \\
    Ni--S$^1$Cys & 0.34 & 0.66 & 0.30 & 0.39 & 0.38 \\
    Ni--S$^2$Cys & 0.41 & 1.39 & 0.29 & 0.22 & 0.22 \\
    Ni--S$^3$Cys & 0.41 & 1.73 & 0.28 & 0.15 & 0.15 \\
    Ni--S$^4$Cys & 0.43 & 0.82 & 0.35 & 0.42 & 0.41 \\
    \hline
  \end{tabular*}
  \label{tab:NiFe_bond_order}
\end{table*}

\subsubsection{Quadrupole Splitting}
The interaction of nuclear quadrupole moment with the electric-field gradient (EFG)
created by the charge density surrounding the nucleus splits the nuclear energy levels
$E_Q$ in the first order of the EFG tensor $\mathbf{V}$ for different magnetic spin quantum numbers $m_I$ according to\cite{Petrilli1998}
\begin{equation}
  E_Q = \frac{eQV_{zz}[3 m_I^2 - I(I-1)] (1 + \eta^2 / 3)^{1/2}}{4I(2I-1)} \;.
\label{eq:E_Q}
\end{equation}
In Eq.~\ref{eq:E_Q},
$e$ is the electron charge,
$Q$ represents the largest component of the nuclear quadrupole moment tensor
in the state characterized by $m_I = I$,
\begin{equation}
  \eta = \frac{|V_{xx} - V_{yy}|}{|V_{zz}|} \;,
\end{equation}
and $V_{xx}$, $V_{yy}$, and $V_{zz}$ are the eigenvalues of $\mathbf{V}$
following the convention $|V_{zz}| > |V_{xx}| \geqslant |V_{yy}|$.

In M\"{o}ssbauer spectroscopy, the most common probe nucleus is $^{57}$Fe,
for which the nuclear transition occurs between the $I=3/2$ excited state
and the $I=1/2$ ground state, with a $\gamma$-radiation energy ($E_{\gamma}$) of $14.4$ keV.
This leads to the quadrupole splitting between the $m_I = \pm 1/2$ states and the
$m_I = \pm 3/2$ states as
\begin{equation}
  \Delta = \frac{eQV_{zz} (1 + \eta^2 / 3)^{1/2}}{2} \;.
\end{equation}
In experiments, the splitting is usually expressed in terms of the velocity
of the source nucleus as
\begin{equation}
  \Delta_v = \frac{c}{E_\gamma} \Delta \;,
\label{eq:Delta_v}
\end{equation}
where $c$ is the speed of light.
Similarly, for a probe nucleus of $^{61}$Ni,
transition occurs between the $I=5/2$ excited state
and the $I=3/2$ ground state, with $E_\gamma = 67.4$ keV, resulting in
\begin{equation}
  \Delta = \frac{3 eQV_{zz} (1 + \eta^2 / 3)^{1/2}}{10} \;.
\end{equation}
The quadrupole moments of $^{57}$Fe and $^{61}$Ni are read from Ref.~\citenum{Stone2016}.

In Table~\ref{tab:NiFe_Mossbauer}, additional results of quadrupole splittings for
the two metal centers of the \textit{Ht}-SH active site model are presented.

\begin{table*}
  \centering
  \caption{Quadrupole splittings $\Delta_v$
  for the \textit{Ht}-SH active site model system,
  calculated at the DFT/TPSSh, MP2, and LNO-CCSD(T)/IAO
  levels with the def2-TZVP basis set.}
  \begin{tabular*}{1.0\textwidth}{@{\extracolsep{\fill}}lccccc}
    \hline\hline
     & TPSSh & MP2 & \mc{3}{c}{LNO-CCSD(T)} \\
    \cline{4-6}
    & & & $\zeta = 1 \times 10^{-4}$
    & $\zeta = 5 \times 10^{-5}$
    & $\zeta = 2 \times 10^{-5}$ \\
    \hline
    $^{57}$Fe $\Delta_v$ (mm/s) & 0.46 & -0.57 & 0.56 & 0.41 & 0.43\\
    $^{61}$Ni $\Delta_v$ (mm/s) & 0.17 & -1.64 & 0.22 & 0.33 & 0.35\\
    \hline
  \end{tabular*}
  \label{tab:NiFe_Mossbauer}
\end{table*}

\clearpage
\subsection{Additional Results for the AIMD Simulations}

\begin{figure}[h]
\includegraphics{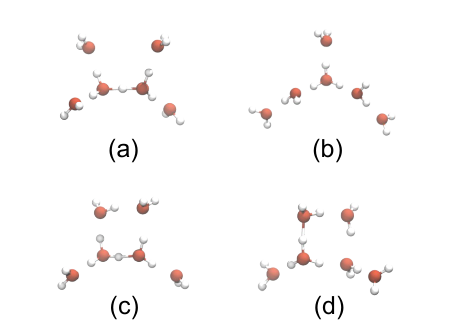}
\centering
\caption{
Initial structures for the AIMD simulations:
(a) Zundel-like,
(b) Eigen-like,
(c) and (d) four-water-ring-like.
}
\label{fig:water}
\end{figure}

\begin{figure}[h]
\includegraphics{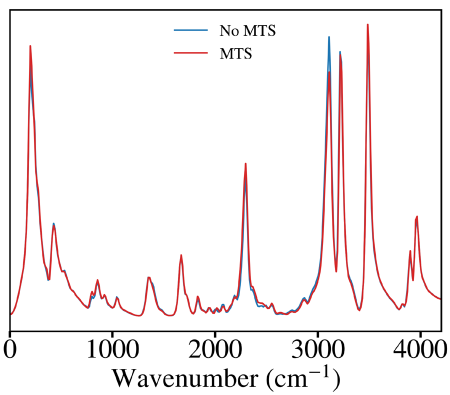}
\centering
\caption{
The comparison between the IR spectra of the four-water-ring-like conformer in the second row of Fig.~\ref{fig:ir_all} computed with and without MTS.
The dynamics without MTS was performed for 2.5 ps using a time step of 0.5 fs,
initiated from the same $NVT$-equilibrated configuration as that in the MTS dynamics.
The MTS dynamics of the first 2.5 ps was used to compute the spectrum here.
}
\label{fig:mts_compare}
\end{figure}

\begin{figure*}
\includegraphics{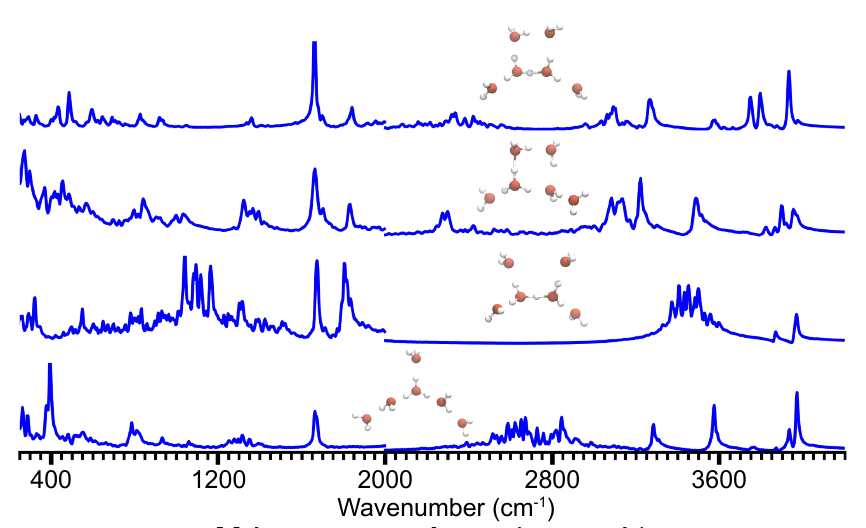}
\centering
\caption{
The computed IR spectra of the protonated water hexamer.
The intensity under 2000 cm$^{-1}$ is multiplied by 3 for clarity,
and the spectra are convoluted using a Gaussian filter with the width of 1 cm$^{-1}$.
}
\label{fig:ir_all}
\end{figure*}

\clearpage
\bibliography{ref}